\documentclass[11pt]{article}
  \usepackage{amsthm}
  \usepackage{amssymb}
  \usepackage{amsmath}
  \usepackage{eucal}
  \usepackage{mathrsfs}
  \usepackage[all]{xy}
  \usepackage{bbm}
\newcommand{\M}{\mathcal{M}} % manifold
\newcommand{\g}{\mathrm{g}} % metrica
\newcommand{\J}{{J}}%
\newcommand{\D}{{D}}% insiemi causali
\newcommand{\I}{{I}}%
\newcommand{\C}{\mathcal{C}} % superfici di Cauchy
\newcommand{\Mk}{\mathbb{M}^4} % spaziotempo di minkowski

\newcommand{\dc}{\mathcal{O}} % aperto
\newcommand{\K}{\mathcal{K}}  % insieme degli aperti
\newcommand{\Po}{\mathcal{P}} % Base per la topologia
\newcommand{\Si}{\Sigma} % Insieme simpliciale
\newcommand{\App}{\mathrm{App}} % approssimazione

\newcommand{\A}{\mathcal{A}} % algebre
\newcommand{\Al}{\mathscr{A}} % funtore causale
\newcommand{\Ss}{\mathcal{S}} % funtore superselezione
\newcommand{\Bh}{\mathfrak{B}(\mathcal{H})} % operatori limitati
\newcommand{\St}{\mathscr{S}} % spazio degli stati

\newcommand{\si}{\sigma}

\newcommand{\ga}{\gamma}
\newcommand{\be}{\beta}

\newcommand{\eps}{\varepsilon}
\newcommand{\io}{\iota}
\newcommand{\al}{\alpha}
\newcommand{\la}{\lambda}

\newcommand{\Z}{\mathcal{Z}}
\newcommand{\Loc}{\mathrm{\mathbf{Loc}}} % categoria degli spaziotempo
\newcommand{\Obs}{\mathrm{\mathbf{Obs}}} % categoria degli osservabili
\newcommand{\Sts}{\mathrm{\mathbf{Sts}}} % categoria degli spazi di stati 
\newcommand{\Sym}{\mathrm{\mathbf{Sym}}} % categorie delle categorie tensoriali

\newcommand {\nrestr}{\!\!\restriction}
\newcommand {\defi}{\equiv}
\newcommand {\norm}[1]{\Vert{#1}\Vert}
\newcommand {\con}[1]{\overline{{#1}}}
\newcommand{\R}{\mathbb{R}} % Insieme numeri reali
\newcommand{\Hil}{\mathcal{H}} % generico spazio di Hilbert
\newcommand{\Sub}{\mathrm{Sub}} %sottoposets
\newcommand{\Rep}{\mathrm{Rep}} %sottoposets

\newcommand{\Kcal}{\mathcal{K}}

\newcommand{\Scal}{\mathcal{S}}
\newcommand{\Zcal}{\mathcal{Z}}

\newcommand{\Rrm}{\mathscr{R}} 
 \newcommand{\Crm}{\mathrm{C}}
 \newcommand{\Erm}{\mathscr{E}}  
  
\newcommand{\Frm}{\mathscr{F}}  
\newcommand{\f}{\mathrm{f}}

 \author{\null\\Romeo Brunetti$^{(1)}$ and Giuseppe Ruzzi$^{(2)}$ \\
  \null\\
  \null\\
  \small{$^{(1)}$II Institute f\"ur Theoretische Physik, Universit\"at Hamburg,}\\
    \small{Luruper Chaussee 149, D-22761 Hamburg, Germany}\\[5pt] 
  \small{$^{(2)}$Dipartimento di Matematica, 
                 Universit\`a di Roma ``Tor Vergata,'' }\\
     \small{Via della Ricerca Scientifica, I-00133 Roma,  Italy}  \\[5pt]
           \small{\texttt{romeo.brunetti@desy.de, ruzzi@mat.uniroma2.it}}}

  \title{Superselection Sectors and General Covariance. ÷I}

\begin{document}

 \maketitle

\begin{abstract}
This paper is devoted to the analysis of 
charged superselection sectors in the framework
of the locally covariant quantum field theories. 
We shall analize sharply localizable charges, and use net-cohomology of J.E. Roberts
 as a main tool.
We show that to any 4-dimensional globally hyperbolic
spacetime it is attached a unique, up to equivalence, 
symmetric tensor $\Crm^*-$category with conjugates (in case of finite statistics); to any embedding
between different spacetimes, the corresponding categories can be 
embedded, contravariantly, in such a way that all the charged quantum
numbers of sectors are preserved. This entails that to 
any spacetime is associated a unique gauge group, up to isomorphisms,
and that to any embedding between two spacetimes there corresponds 
a group morphism between the related gauge groups. This form of covariance between sectors also brings to light the issue whether local and global sectors are the same. We conjecture this holds that at least on simply connected spacetimes. It is argued that the possible failure might be related to the presence of topological charges.
Our analysis seems to describe theories which have a well defined short-distance asymptotic behaviour. 
\end{abstract}

%\swapnumbers
  \theoremstyle{plain}
  \newtheorem{df}{Definition}[section]
  \newtheorem{teo}[df]{Theorem}
  \newtheorem{prop}[df]{Proposition}
  \newtheorem{cor}[df]{Corollary}
  \newtheorem{lemma}[df]{Lemma}
  
  \theoremstyle{plain}
  \newtheorem*{Main}{Main Theorem}
  \newtheorem*{MainT}{Main Technical Theorem}

  \theoremstyle{definition}
  \newtheorem{oss}[df]{Remark}

 \theoremstyle{definition}
  \newtheorem{ass}{\underline{\textit{Assumption}}}[section]
 
 \newpage

%*************************************************************

\tableofcontents
\markboth{Contents}{Contents}
\newpage
%*************************************************************

\section{Introduction}
The present paper represents a first step toward a new investigation of
superselection sectors based on the general framework of locally covariant 
quantum field theory \cite{BFV}.

One of the milestones of quantum field theory is the investigation of the superselection rules. 
They are, in essence, constraints to the nature and scope of possible measurements, proving that in quantum physics the superposition principle does not hold unrestrictedly. Stimulated by 
the preliminary investigation about the existence of superselection rules in 
quantum field theory by Wick, Wightman and Wigner \cite{WWW},  Haag and Kastler \cite{HK} found that the correct interpretation resides in the
inequivalent representations of abstract $\Crm^*-$algebras describing local observables, i.e. equivalence classes of local measurements. 
Some years later, Doplicher, Haag and Roberts \cite{DHR1, DHR2} undertook a awesome analysis which culminated 
at the beginning of the nineties with a long series of remarkable papers 
(Doplicher and Roberts, e.g. see \cite{DR1,DR2}).
The main result that has been proved is that besides any algebra of observables there is associated, canonically, a field algebra with normal Bose-Fermi commutation relations, on which a compact gauge group of the first kind acts. This group determines the superselection structure of the theory, in the sense that its irreducible representations are in one-to-one correspondence with the superselection sectors describing sharply localized charges. Well beyond the importance for the physical interpretation, 
they also obtained outstanding mathematical results, e.÷g. by founding an 
abstract theory of group duals extending that of Tannaka and Krein. 

Let us describe in more specific terms the basic ideas of superselection sectors in the 
algebraic approach of Haag and Kastler \cite{HK,Ha}. One assumes that all the physical information is contained in the a priori correspondence between (open, bounded) regions of Minkowski spacetime and algebras of observables that can
be measured inside it, i.÷e., we have an abstract net of \emph{local observables}, namely the correspondence
$$
\dc \to \mathscr{A}(\dc)
$$
which associates, for instance, to every element $\dc$ of the family $\K^{dc}$ of double cones of
Minkowski spacetime the (unital) $\Crm^{*}$-algebra $\mathscr{A}(\dc)$. The $\Crm^{*}-$algebra
generated by all local ones is termed the quasi-local algebra of observables 
and denoted by $\mathscr{A}_{\K^{dc}}$.
 
A few basic principles can be adopted, among which, causality and
Poin\-ca\-r\'e covariance. The claim is that this is all one needs for investigating (structural) properties of a 
physical theory. This claim is substantiated by many results, among which the theory of superselection
sectors is the leading example \cite{Ha}, as far as the intrinsic perspective is concerned, although one should not forget to cite that also the 
time-honored perturbation theory can be fruitfully formulated in the algebraic framework \cite{BF3,DF,HW}.  

An important point, where the physical interpretation enters heavily, is in the choice of the correct physical states for the theory at hand. As far as applications to elementary particle physics is concerned, one assumes the existence of a canonical state, that of the (pure) vacuum state $\omega_0$.
Doplicher, Haag and Roberts invoked the intuitive ideas that any particle (at least in the case in which no massless excitations are present in the physical theory) arises as an elementary and localized excitation of the vacuum state. Hence, they put forward that, if $\pi_{0}$ is the
representation of the algebra $\mathscr{A}_{\K^{dc}}$ associated with the vacuum state $\omega_0$, 
any representation
$\pi$ which describes an elementary and localized excitation of the vacuum should satisfy
\begin{equation}
\label{selection}
\pi\nrestr\!\mathscr{A}(\dc^{\perp})\cong
\pi_{0}\nrestr\!\mathscr{A}(\dc^{\perp})\ ,\qquad \dc\in\K^{dc}\ , 
\end{equation}
where the symbol $\cong$ means unitary equivalence, and $\perp$ denotes the causal complement.

This put the previous intuitive remark of particles as localized excitations of the vacuum on rigorous footing by declaring those representations to be ``physically relevant''
that are unitarily equivalent to the vacuum one in the causal complement of any double cone.

\emph{Superselection sectors} of the
quasi-local algebra $\mathscr{A}_{\K^{dc}}$ are then defined to be the unitary equivalence classes 
of those irreducible representations, satisfying the selection condition, and where the labels distinguishing them are called
\emph{charged quantum numbers}. One then considers the representations satisfying condition (\ref{selection}) as referring to ``localizable charges.''  The domain of validity of such intrinsic development is, for instance, that of the massive sector of hadronic theories.

Borchers \cite{Bor}, in a previous attempt, based the selection of physical sectors on the requirement of positivity of the energy, since also in this case one can exclude states for which the matter density does not vanish at infinity. On the base of his remarks, the analysis of Doplicher, Haag and Roberts was further refined by Buchholz and Fredenhagen \cite{BF}, at least for the case of gauge theories without massless excitations, by requiring the selection condition on unbounded regions called spacelike cones. These are regions extending to spacelike infinity. Their criterion is essentially that shown before, where one changes the double cones with the new regions. The so selected representations bear charges termed ``topological charges,'' and the results are similar to those of Doplicher, Haag and Roberts.

However, there are important theories for which the above requirements are not fulfilled, for instance in abelian gauge theories like QED. There the above localization is not pertinent since any gauge charge has attached, via Gauss' Law, its own flux line or string which forces a poorer localization. 
 QED itself is still awaiting a complete determination of its sectors, but see e.g. \cite{Buc} and \cite{Martino1, Martino2}.

In both localizable and topological cases, the analysis leads to the following remarkable results shedding light on deep characteristic aspects of elementary particle physics, namely;
\begin{enumerate}
\item \emph{Particle-Antiparticle symmetry}:
 To each charge corresponds a unique conjugated charge which entails the particle-antiparticle symmetry;
 \item \emph{Particle Statistics}: Each charge has a permutation symmetry to which there corresponds uniquely  a 
 sign and an integer $d$, its statistical dimension; the corresponding particles, if any, then satisfy parastatistics of order $d$, the sign distinguishes between para-Bose and para-Fermi statistics. Usual Bose and Fermi statistics correspond to $d=1$.
 \item \emph{Composition of Charges}: Two charges can be composed because they can be created in spacelike separated regions. 
 \end{enumerate}

The analysis developed so far has been fruitfully applied in other contexts, for instance in (chiral) 
conformal quantum field theory by Fredenhagen, Kawahigashi, Longo, Rehren, Schroer and various others 
collaborators \cite{FRS,KL,R}, for quantum field theory on curved spacetime in a paper 
by Guido, Longo, Roberts and Verch \cite{GLRV}, and initially for soliton physics and then 
towards more mathematical aims by M\"uger \cite{Mug, Mug1}. In a more mathematical direction one 
should cite the work done by Baumg\"artel and Lled\'o \cite{BL} and by Vasselli \cite{Zio}.

On the previous investigation of the superselection structure in the context of curved backgrounds 
Guido, Longo, Roberts and Verch \cite{GLRV} used as a major tool the results by Roberts on the 
connection between net cohomology and superselection sectors \cite{Rob1,Rob2}.  

Net cohomology has made clearer that it is the causal and topological structure of spacetime that are the basic relevant features when studying superselection sectors. In particular, we now understand that the main point is how these two properties are encoded in the structure of the index set of the net (e.g. the set of double cones $\K^{dc}$ for Minkowski spacetime) as a \emph{poset} (partially ordered set) ordered under inclusion. Let us then call $\K^{\bullet}$ the preferred choice of index set on the spacetime $M$, and call $\mathscr{A}_{\K^{\bullet}}$ the corresponding net of local observables. One proves, under some conditions, that representations satisfying the selection criterion (\ref{selection}) are, up to equivalence, in one-to-one correspondence with $1$-cocycles of the poset with values in the (vacuum) representation of the net of local algebras, $\mathscr{B}_{\K^{\bullet}}: \dc\to \mathscr{B}(\dc)$ where $\mathscr{B}(\dc)\equiv\pi_0(\mathscr{A}(\dc))^{\prime\prime}$, for any $\dc\in\K^{\bullet}$. They define a $\Crm^*-$category $\Z^{1}_{t}(\mathscr{B}_{\K^{\bullet}})$. In order to be a good starting point for the analysis of superselection structure, it should posses good features like being a \emph{tensor category}, having a \emph{permutation symmetry} and a \emph{conjugation} (in case of finite statistics). 

Recently, in \cite{Ruz3}, one of us completed the investigation of \cite{GLRV} in the case where
the spacetime is globally hyperbolic, relying on a refined form of net cohomology where a more precise investigation of the properties of indices was made clear, much stimulated by some previous results in \cite{Rob3}. Among several things, it is discussed why the choice made in \cite{GLRV} of basing the net structure on the so-called ``regular diamonds'' index set $\K^{rd}$ is unfortunate. There are two main problems with such a choice; the first is that this index set is not directed, causing problems with the definition of the tensor product structure of the category of $1$-cocycles; the second is that it has elements with non arcwise-connected causal complement, a topological feature that seems, at least, to forbid a straightforward application of Haag duality, if possible at all. In \cite{Ruz3} a better choice was made, that of using as a basic index set that of ``diamonds,'' noted $\K^{d}$, having only elements with arcwise-connected causal complements. On this basis the analysis was easier and more powerful, and one arrives at a category of $1$-cocycles $\Z^1_t(\mathscr{B}_{\K^{d}})$ that indeed possesses all the previously stated properties.

Our aims, as announced at the beginning, is to initiate a study of 
superselection sectors 
for the case of locally covariant quantum field theories \cite{BFV}, stepping forward from the previous developments \cite{GLRV,Ruz3,Verch-ab}.
This new framework deals with a 
non-trivial blend of two principles, 
locality and general covariance, in quantum physics. 
We refer to the review paper \cite{BF2} for further discussion. 
It contains the Haag-Kastler setting as a subcase, it found deep application to perturbation theory on curved spacetime \cite{HW} and appears to be well suited
 to developing a new look at perturbative quantum gravity \cite{Freddy}. Especially, it gives a new perspective in quantum field theory, for instance see  \cite{Ve2, Ruz2}. Hence, it is an important point to determine and  
study the features of the superselection structure. The basic features of the approach will be recalled in Section \ref{B}, here we just notice that it is a categorical approach in which a quantum field theory is considered as a covariant functor from a category $\Loc$ 
whose objects are $4$-dimensional globally hyperbolic spacetimes, with isometric embeddings as arrows, to a category of $\Crm^*-$algebras of observables, with injective $^*-$homomorphisms as arrows.

Aiming at founding superselection theory for this more general setting, we are then faced with the following list of issues: 
\begin{itemize}
\item[(a)] The good choice of the index sets for any globally hyperbolic spacetime and their stability under isometric embeddings.
\item[(b)] A pertinent choice of a reference state (space), with a minimal set of assumptions for nets of observable algebras.
\item[(c)] The choice of what kind of $1$-cocycles one wants to investigate, i.e., regarding their localization properties and the construction of their $\Crm^*-$categories, with the sought for properties, and especially their behaviour under isometric embeddings.
\end{itemize}

For the first problem, a ``canonical'' choice was made in \cite{BFV}. Namely, a family of regions 
of each $M$ with the properties of being relatively compact, causally convex, i.e., such that all 
causal curves starting and ending inside such a region would always remain inside it, 
and with non-empty causal complement. We denote it by $\K^{h}(M)$. One immediately faces a problem 
with such a family. Indeed, it shares with the regular diamonds the same annoying topological feature 
of having non arcwise-connected causal complements. However, we learned in \cite{Ruz3} how to overcome this
by passing to the diamonds' family $\K^{d}(M)$. Although for the first family stability under 
isometric embeddings is essentially obvious, our proof for the second relies on a nice recent result 
due to Bernal and S\'anchez \cite{BS3} on the extension of 
\emph{compact} hypersurfaces.

A choice of a reference state is more involved and delicate. Notice that, 
in the Minkowskian analysis of sectors \emph{\'a la}\  Doplicher-Haag-Roberts, 
a choice is made in terms of a \emph{single pure} 
reference state, typically the vacuum. In the locally covariant setting one can 
prove \cite{BFV} that there is no choice of single states on each $M$, pure or not, which is 
covariant under local diffeomorphisms. However, the covariance works well \cite{BFV}, in examples, if instead one chooses \emph{folia} of states for each manifold. This entails that it is appropriate \cite{BFV} to consider a functorial description of \emph{state space} that takes into account the covariance under local diffeomorphisms. 
It is then rather natural to choose, for our present purposes,
a \emph{locally quasi-equivalent} state space 
$\St_o (M)$ for any $M$, that satisfies some further technical conditions as Borchers property and such that there is at 
least one state for which the associated net of von Neumann algebras satisfies irreducibility and 
punctured Haag duality. Since no preferred choice of states is at hand, one of our main tasks would 
be therefore to exploit the relation between nets defined in terms of different states belonging to the 
reference state space. It is perhaps one of the virtues of our approach that such a relation can be 
fully discussed, and brings, under certain technical conditions, an isomorphism between the nets, 
which moreover behaves well under isometric embeddings. At the level of categories of $1$-cocycles, 
we have been able to prove that they \emph{do not} depend on the choice of the states inside the 
reference state space.
As far as assumptions for the net are concerned, we have chosen mainly to require irreducibility 
and punctured Haag duality. This is really a minimal requirement since a form of Haag duality seems 
to be really necessary for exploiting properties of superselection sectors. Innocent as it looks, 
this choice will turn out to be crucial for our construction.

However, the real crux of the matter comes with the last issue. Here, there are several possibilities, depending on the choice of the index set and the topological features of the spacetimes. We recall that according to the results in general relativity \cite{Wit}, Cauchy surfaces can have any topology as (closed) $3$-dimensional topological spaces. For instance,  they can be compact and/or non simply connected, e.g. as in de Sitter space and for the $\mathbb{RP}^3$ geon (see, e.g., \cite{FSW}), respectively. Hence, the problem lies in the large number of possibilities that we have for choosing the localization properties of the $1$-cocycles, namely what kind of index set we would like to use, and accordingly, what features of the charges do we want to highlight. This is additionally complicated by the fact that some of the index sets have the ability to recover the topology of the spacetime while others do not \cite{Ruz3}. 

It is in order to clarify another point, namely, although we are working within the locally 
covariant setting, this does not mean that the charges have to be necessarily localized in relatively 
compact regions. Local covariance only requires that the charges are local functionals of the metric, 
i.e., they depend only on local \emph{geometrical} data, nothing forbids the charges testing 
the topological structure of spacetime. Something like the Buchholz and Fredenhagen analysis of 
topological charges can be envisaged.

In this first paper, however, we restrict our attention to localizable charges, namely we fix as 
a main index set that of diamonds $\K^{d}(M)$ on any globally hyperbolic spacetime $M$, and we choose 
as $1$-cocycles those having the property of being \emph{path-independent}, in the precise sense of 
paths in the poset $\K^{d}(M)$. In this case the $1$-cocycles provide trivial representations of the 
first fundamental group of the manifold \cite{Ruz3}, hence they do not probe the topology of the 
spacetime. This choice corresponds to that of localizable charges in the Doplicher, Haag and Roberts 
sense on Minkowski spacetime. In the forthcoming paper \cite{BR} we will discuss how different choices 
of the posets are related to each other, still in the localizable case. We are working on the more 
complicated analysis of ``charges carrying topological structure'' i.e., on 
\emph{path-dependent} 1-cocycles.

Even in the easier case of localizable charges there is a subtlety. Although diamonds form a 
well behaved family for several of the questions underlying the mathematical development of the 
structure of sectors, one needs the family $\K^{h}(M)$ for some crucial results, since it contains 
nonsimply connected regions. For instance, 
under the isomorphism between nets for different choices of states, a $1$-cocycle of $\K^{d}(M)$
does not remain necessarily path-independent. The stability under net isomorphisms  
seems to hold only when the nets are extended to $\K^{h}(M)$ (see Section \ref{Caa}).
Nonetheless, these results involve only 
investigating the category of $1$-cocycles, not the tensor structure. Hence, one of the
main points of the analysis to prove that the \emph{categories} of $1$-cocycles associated to the  
different families are actually \emph{equivalent}. Granted that, one can proceed to the analysis of 
the \emph{tensor} structure of the categories. It is not clear to us, at the moment, whether the 
presence of both families underlies some further subtle point of more physical origin. 
A similar situation has been studied by Ciolli \cite{Cio} for massless scalar fields 
in 1+1 dimensions. 

The main result of the paper is the following: Calling $\Rrm$ the restriction operation of 
$1$-cocycles defined on spacetime $M$ to a subspacetime $N\subset M$, we have

\begin{MainT} The restriction $\Rrm$ lifts to a full and faithful covariant 
$^*$-functor between $\Z^1_t(\omega,\K^{d}(M))$ and $\Z^1_t(\omega,\K^{d}(N))$, for any 
choice of the state $\omega$ in the reference state space $\St_o$.
\end{MainT}

This result entails the good behaviour under isometric embeddings of spacetimes. Furthermore, 
it shows that the kind of theories we study behave well under the scaling limit. Indeed, the Main 
Technical Theorem seems to be the cohomological counterpart of the ``equivalence between local 
and global intertwiners'' that ``good'' theories posses in the scaling limit \cite{DMV}.

Furthermore, the main result of the paper is that, according to the expectations coming out
the locally covariant approach, the suitably defined superselection structure is also functorial, 
namely, one can define a map $\Scal$ from the category  of spacetimes $\Loc$ to the category $\Sym$ 
of symmetric tensor $\Crm^*-$categories, with full and faithful  symmetric tensor 
$^*-$functors as arrows, for which there holds

\begin{Main} The map $\Scal:\Loc\rightarrow \Sym$ is a  contravariant functor.
\end{Main}

Roughly speaking, the message coming out the Theorem is that the physical content of superselection 
sectors based on different spacetimes can be faithfully transported provided the spacetimes can be embedded.
Namely, \emph{all} charged quantum numbers are preserved in this embedding.

An important byproduct of the Main Theorem  is that to any spacetime there is associated a 
unique gauge group, up to isomorphism, and that to any embedding between spacetimes there 
corresponds a group morphism between the respective gauge groups.

Another consequence of the Main Theorem is that it makes possible 
a clear investigation of the possible relation between local 
and global superselection sectors of a spacetime.  
Because of the sharp localization, one expects  an equivalence between local and global 
superselection sectors. However there is no general evidence 
of this equivalence, that we call 
\emph{local completeness of superselection sectors}. Indeed, 
one can prove it only in models derived from  free quantum fields.
We  point out that a possible violation of local completeness might be 
related to the nontrivial topology of spacetimes and, in particular, 
to the existence of path-dependent 1-cocycles.

We now pass on to outline the content of the paper. In Section \ref{A} we start by elaborating geometrical features which are crucial for the following parts. We discuss some background results on Lorentzian spacetime mainly for fixing our notation. Then we pass to the discussion of families of subsets for each globally hyperbolic spacetime in $4$-dimensions and their stability properties under isometric embeddings. Some categorical notions are briefly outlined. In Section \ref{B} we start by recalling the basic definitions of the locally covariant quantum field theory as a covariant functor. 
There we introduce some crucial definitions and assumptions. Moreover we prove some results on state spaces. Section \ref{C} contains only a brief discussion of the basic definitions and properties of net cohomology. The heart of the paper is in Section \ref{D}. There we prove our Main Theorem and several related results. Eventually, these last results are applied, in subsection \ref{Cb}, to prove the generally covariant behaviour of the superselection structures. In Section \ref{Z} we point out the difficulties in proving the equivalence between local and global sectors.  Conclusions and Outlook follow. An Appendix is provided where some categorical notions and some results about Doplicher-Roberts Reconstruction Theorem are briefly recalled.

%********************************************************************
%********************************************************************
%********************************************************************
\section{Spacetime geometry}
\label{A}
In the first part of this section we review some basics notions 
of Lorentzian geometry, introduce the category of spacetimes and
define some families of sets of technical importance. We prove some results of 
independent interests.

\subsection{Lorentzian spacetimes}
\label{Aa}
We recall some basics on the causal
structure of spacetimes and establish our notation.
Standard references for this topic are \cite{BEE, One, EH, Wal}.
%********************************
\paragraph{Spacetimes:} A \textit{spacetime} $M$, in our framework, consists of
a Hausdorff, paracompact, connected, without boundary, smooth, oriented
4-dimensional manifold $M$ endowed
with  a smooth metric $\g$ with signature $(-,+,+,+)$, and with
a time-orientation, that is a smooth vector field $v$ satisfying 
the equation $\g_p(v_p,v_p) < 0$ for each $p\in M$.
(Throughout this paper smooth means $C^\infty$). \\
\indent A curve $\ga$ in $M$ is a continuous,  piecewise smooth, 
 regular function $\ga:I \longrightarrow  M$,  
 where $I$ is a connected subset of $\R$ with nonempty interior. It
 is called timelike, lightlike, spacelike 
 if respectively $\g(\dot{\ga},\dot{\ga})<0$, $=0$, $> 0$
 all along $\ga$, where $\dot{\ga}=\frac{d\ga}{dt}$.
 Assume now that $\ga$ is \textit{causal}, i.e.
 a nonspacelike curve; we can classify it according to the 
 time-orientation $v$ as future-directed or 
 past-directed if respectively 
 $\g(\dot{\ga}, v) < 0, >0$ all along $\ga$. When $\ga$ is future-directed 
and 
 there exists $\lim_{t\rightarrow\sup I}\ga(t)$ 
 ($\lim_{t\rightarrow\inf I}\ga(t)$), then it is said to have 
 a future (past) endpoint. In the negative case, it is said to be 
 future (past) endless; $\ga$ is said to be endless if none of them
 exist. Analogous definitions are assumed for past-directed causal curves.\\
%***********************************************************
\indent The \textit{chronological future} $\I^+(S)$,
the \textit{causal future} $\J^+(S)$
and the \textit{future domain of dependence } $\D^+(S)$
of a subset $S\subset M$ are defined as:
\begin{align*}
&\I^+(S)  \defi  \{ x\in M \ | \ 
\mbox{exists a future-directed timelike curve
    from $S$ to $x$ }\}\ ; \\
&\J^+(S)  \defi   S \cup \{ x\in M \ | \ 
\mbox{exists a future-directed  causal
    curve from $S$ to $x$ }\}\ ; \\
&\D^+(S) \defi   \{ x\in M \ | \ 
\mbox{any past-directed } \\ 
&   \qquad\qquad\qquad\qquad \ \ \mbox{endless causal
    curve through $x$ meets $S$}\}\ .
\end{align*}
These definitions
have duals in which ``future'' is replaced by ``past''
and the $+$ by $-$. By this,
we define $\I(S) \defi \I^+(S) \cup  \I^-(S)$,
$\J(S)\defi\J^+(S) \cup  \J^-(S)$ and $\D(S) \defi\D^+(S)\cup
\D^-(S)$.
%**************
\begin{oss}
\label{Aa:0}
It is worth recalling the following properties of causal sets\footnote{$cl(S)$ and 
 $int(S)$ denote respectively the closure and the internal part 
                              of the set $S$.}.\\[3pt]
(a) \  Let $S\subseteq M$. Then 
  $\I^+(S)$ is an open set and  $\I^+(cl(S)) = \I^+(S)$; 
   $cl(\J^+(S)) =cl(\I^+(S))$ and 
    $int(\J^+(S)) = \I^+(S)$. \\[3pt]
(b) \ Let  $S_1,S_2,S_3\subseteq M$. 
            If $S_1\subseteq\J^+(S_2)$ and $S_2\subseteq\I^+(S_3)$, then 
               $S_1\subseteq \I^+(S_3)$. \\[3pt]
By (a), we have that 
$cl\big(\J^+(cl(S))\big)= cl\big(\J^+(S)\big)$ for any
$S\subseteq M$.
\end{oss}
%*************
The \textit{causal disjointness relation}
is a symmetric binary relation $\perp$ on the subsets of $M$ 
defined as follows  
\begin{equation}
\label{Aa:1}
 S\perp V \ \iff \ V\subseteq M\setminus\J(S)\ .
\end{equation}
The \textit{causal complement} of a set $S$ is the open set
$S^\perp$ defined as 
\begin{equation}
\label{Aa:2}
S^\perp \defi M\setminus cl(\J(S))\ .
\end{equation}
A set $S$  is \textit{acausal}
if $\{p\}\perp \{q\}$ for each pair $p,q\in S$. A set $S$ is
\textit{achronal} if $\I^+(S)\cap S=\emptyset$. 
A (\textit{acausal}) \textit{Cauchy surface}
$\C$ of $M$ is an achronal (acausal) set verifying $\D(\C)= M$.
Any Cauchy surface is a closed, connected, Lipschitz hypersurface of
$M$. A \textit{spacelike} Cauchy surface is a smooth Cauchy surface
whose tangent space is everywhere spacelike. Any spacelike Cauchy
surface is acausal.
% For any relatively compact $S\subset M$
% the following properties hold: \textbf{1.}~$\J^+(\overline{S})=\overline{\J^+(S)}$;
% \textbf{2.}
% $\D^+(\overline{S})$ is compact; \textbf{3.} for each
% Cauchy surface $\C$ the set $\J^+(\overline{S})\cap \C$
% is either empty or compact; \textbf{4.}
% $\overline{\J^+(S\cup \{p\})} =  \J^+(\overline{S}\cup \{p\})$
% for any  $p\in M$ such that $\overline{S}\cap \{p\}=\emptyset$.\\[3pt]
%*****************************************************
\paragraph{Global hyperbolicity and the category $\Loc$:}
A spacetime $M$ is \textit{globally hyperbolic} 
if it admits a smooth \cite{BS1,BS2} foliation by spacelike Cauchy surfaces, 
namely, there is a 3-dimensional smooth manifold
$\Sigma$ and a diffeomorphism $F:\R \times \Sigma \longrightarrow  M$
such that:  for each $t\in\R$ the set $\C_t\defi \{F(t,y) \ | \ y\in\Sigma\}$
is a spacelike Cauchy surface of $M$; 
the curve $t\in\R\rightarrow F(t,y)\in M$ is a future-directed 
(by convention)
endless timelike curve for any $y\in\Sigma$. 
We recall some causality properties of a globally hyperbolic spacetimes.
%****************
\begin{oss}
\label{Aa:3}
Assume that  $M$ is
a globally hyperbolic spacetime. \\[3pt]
(a) \ Let $K$ be a relatively compact subset of $M$. 
        Then, $\J^+(cl(K))$ is closed, and  
         $\J^+(cl(K))=cl\big(\J^+(K) \big)$;
          $\D^+(cl(K))$ is compact.
         If $\C$ is  a spacelike Cauchy surface, then 
                 $\J(cl(K))\cap \C$ is a compact subset of $\C$. \\[3pt]
(b) \ The family of the open subsets 
   $\I^+(x)\cap \I^-(y)$, for any $x,y\in\M$, 
             is  a basis for the topology of $M$. 
       Hence, if $S$ is open set and   $x\in S$, then  there are 
         $y_1,y_2\in S$ such that $x\in \I^+(y_1)\cap\I^-(y_2)\subseteq S$.
\end{oss} 
%*****************
As an easy  consequence of the Remark \ref{Aa:3}.(b) we have the following  
\begin{lemma}
\label{Aa:4}
Let $M$ be a  globally hyperbolic spacetime. If $S$ 
is an open subset of $M$,  then the 
sets $\J^+(S),\J^-(S)$ and $\J(S)$ are open, hence 
$\J^+(S)=\I^+(S)$, $\J^-(S)=I^-(S)$ and $\J(S)=\I(S)$.
\end{lemma}
\begin{proof}
It is clear that  $\I^+(S)\subseteq \J^+(S)$. Now, let 
$x\in \J^+(S)$. Take  a point $x_0$ in $\J^-(x)\cap S\ne\emptyset$.  
Since $S$ is an open set, 
there is $y\in S$ such that $x_0\in \I^+(y)$ (see Remark \ref{Aa:3}.(b)). 
By Remark \ref{Aa:0}.(b), we have that 
\[ 
x\in\J^+(x_0), \  x_0\in \I^+(y) \ \ \Rightarrow \ \ x\in \I^+(y) \ \ 
\Rightarrow
\ \ x\in\I^+(S)\ .
\] 
Therefore $\J^+(S)= \I^+(S)$ and the set $\J^+(S)$ is open, because
so is $\I^+(S)$. 
Analogously, $\J^-(S)=\I^-(S)$. Hence  $\J(S)=\I(S)$.
\end{proof}
\begin{lemma}
\label{Aa:5}
Let $K, S$ be two subsets of a globally hyperbolic spacetime $M$. 
Assume  that $K$ is relatively compact with $K^\perp \ne\emptyset$, 
and that $S$ is open. If $cl(K)\subset S$, then $K^\perp\cap S\ne\emptyset$.
\end{lemma}
\begin{proof}
Assume that  $K^\perp\cap S =\emptyset$. 
This is equivalent to say that $S\subseteq 
(M\setminus K^\perp)= cl(\J(K))=\J(cl(K))$ 
(see Remark \ref{Aa:3}.(a)). Since $cl(K)\subset S$, we have that 
$\J(cl(K))\subseteq \J(S)\subseteq \J(cl(K))$. Hence
$\J(S)=\J(cl(K))$ and this leads to a 
contradiction. Indeed, by Lemma \ref{Aa:4}  $\J(S)$ is an open set
while $\J(cl(K))$ is closed and different from $M$ since
$K^\perp\ne\emptyset$. Therefore
$\J(S)$ is an open and  closed proper subset of $M$; 
this is not possible because 
$M$ is arcwise connected.
\end{proof}
Let $M$ and $ M_1$ be  globally hyperbolic spacetimes
with metrics $\g$ and $\g_1$ respectively.
A smooth function $\psi$ from $ M_1$ into $M$ is called an
\textit{isometric embedding}  if $\psi: M_1\rightarrow \psi( M_1)\subseteq M$
is a diffeomorphism and $\psi_*\g_1 = \g\nrestr_{\psi( M_1)}$.
The category $\Loc$ is the category whose objects are the
4-dimensional globally  hyperbolic spacetimes; the arrows
$(M_1,M)$ are the isometric embeddings
$\psi: M_1\longrightarrow M$ preserving the orientation
and the time-orientation of the embedded spacetime, and that satisfy
the property
\[
\forall x,x_1\in\psi(M_1), \ \J^+(x)\cap \J^-(x_1) \mbox{ is 
contained  in } \psi(M_1)\ .
\]
The composition law between two arrows $\psi$ and $\phi$, denoted by
$\psi\phi$, is given by the usual composition between
smooth functions; the identity
arrow  $\mathrm{id}_M$ is the identity  function of $M$.
%*************************************
\subsection{Stable families of indices} 
Given $M\in\Loc$, the aim is to introduce families of open subsets 
of $M$ which will be used as index set for nets of local algebras. We prove their stability under isometric embeddings.
\paragraph{Globally hyperbolic regions:} 
Let $\K^{h}(M)$ be the collection of subsets
$\dc\subseteq M$ satisfying the following properties: 
\begin{itemize}
\item[(i)] $\dc$ is open, arcwise connected, relatively compact set, and 
           $\dc^\perp\ne \emptyset$;
\item[(ii)] if $x_1,x_2\in\dc$, then $\J^+(x_1)\cap \J^-(x_2)$ 
 contained in $\dc$.
\end{itemize}
It turns out by this definition that $\Kcal^{h}(M)$ is a basis 
for the topology of $M$ and that any element $\dc$ of 
$\Kcal^{h}(M)$ -- with the metric
$\g\nrestr_{\dc}$ and with the induced orientation and time orientation -- is
a globally hyperbolic spacetime, hence $\dc\in\Loc$.
%***************************
We now show some straightforward geometrical results.
\begin{lemma}
\label{Aa:6}
Let $M\in\Loc$, then the following assertions hold:\\
(a) Let  $\dc,\dc_1\in\Kcal^{h}(M)$ be such that $cl(\dc_1)\perp\dc$. Then 
    there exists $\dc_2\in\Kcal^{h}(M)$ such that  
    $cl(\dc)\cup cl(\dc_1)$ $\subset \dc_2$;\\ 
(b) For any $\dc\in\Kcal^{h}(M)$ there are $\dc_1,\dc_2\in\Kcal^{h}(M)$ such that 
    $cl(\dc_1)\perp\dc$ and $cl(\dc)\cup cl(\dc_1)\subseteq \dc_2$;\\
(c) Let $\dc\in\Kcal^{h}(M)$ and  let $S\subset M$ be an open set, 
    with $cl(\dc)\subset S$, then 
   $\dc^\perp\cap S\ne\emptyset$.
\end{lemma}
\begin{proof}
$(a)$ Since $cl(\dc_1)\perp \dc$, then $cl(\dc_1)\subset 
M\setminus \J(cl(\dc))$. By Lemma \ref{Aa:5},  
$\dc^\perp_1\cap \dc^\perp= M\setminus (\J(cl(\dc))\cup\J(cl(\dc_1)))$
is open and nonempty. Fix $x\in \dc^\perp_1\cap \dc^\perp$ and let 
$\C$ be a spacelike Cauchy surface that meets $x$.  
Consider the sets $K\defi \J(cl(\dc))\cap \C$ and 
$K_1\defi\J(cl(\dc_1))\cap\C$.
In the relative topology of $\C$,  
both $K$ and $K_1$ are compact because $cl(\dc)$ and $cl(\dc_1)$ 
are compact  (see Remark \ref{Aa:3}.(a)), and they are arcwise
connected because so are $cl(\dc)$ and $cl(\dc_1)$. Moreover 
$x\not\in K\cup K_1$. It is clear that we can find two 
arcwise connected and relatively compact open sets $W$ and $W_1$
of $\C$ such that: $K\subset W$, $K_1\subset W_1$ and 
$x\not\in cl(W)\cup cl(W_1)$.  We now consider the following cases. 
First, assume that $W\cap W_1\ne\emptyset$. In this case we define 
$\dc_2\defi \D(W\cup W_1)$. $\dc_2$ verifies the properties 
written in the statement. In fact $W\cup W_1$ is an open and 
arcwise connected subset of $\C$, then $\dc_2$ is open and globally
hyperbolic \cite{One} (Lemma 43). Moreover, since $W\cup W_1$ is relatively
compact, $\dc_2$ is relatively compact (see Remark \ref{Aa:3}.(a)). 
Finally, since $\C\setminus cl(W)\cup cl(W_1)\ne\emptyset$, we have
that $\dc_2^\perp\ne\emptyset$. Hence $\dc_2\in\Kcal^{h}(M)$. 
It is also clear that $cl(\dc)\cup cl(\dc_1)\subset\dc_2$.
Secondly, assume now that $cl(W)\cap cl(W_1)=\emptyset$. 
Let $\ga:[0,1]\rightarrow \C$ be a curve that meets $x$ and 
such that $\ga(0)\in W$ 
and $\ga(1)\in W_1$. We can find a family  
$U_i, i=1,\ldots n$ of open, relatively compact and arcwise connected 
sets of $\C$ which cover the curve $\ga$ and such that, 
if  $G\defi W\cup W_1\cup U_n\cup \cdots \cup U_1$, then 
$\C\setminus cl(G)\ne\emptyset$. By setting  $\dc_2\defi \D(G)$,
as before we have that $\dc_2$ verifies the properties written in the 
statement. $(b)$ follows from $(a)$: it is enough to observe that, 
since $\dc^\perp\ne\emptyset$ and since $\Kcal^{h}(M)$ is a basis for
the topology of $M$, there is 
$\dc_1\in\Kcal^{h}(M)$, with $cl(\dc_1)\perp \dc$. 
$(c)$ follows from Lemma \ref{Aa:5}.
\end{proof}
%*******************************
We now use the previous Lemma to show that compactness of Cauchy surfaces is the only obstruction to the directedness of the
family $\Kcal^{h}(M)$, namely that for any $\dc_1,\dc_2\in\Kcal^{h}(M)$ 
there is $\dc\in\Kcal^{h}(M)$ such that $\dc_1\cup\dc_2\subseteq\dc$.
\begin{lemma}
\label{Aa:6a}
Let $M\in\Loc$ have noncompact Cauchy surfaces, then 
$\Kcal^{h}(M)$ is directed. 
\end{lemma}
\begin{proof}
Let $\C$ be a spacelike Cauchy surface of $M$. Note that 
the set $K\defi(\J(cl(\dc_1))\cap \C)\cup (\J(cl(\dc_2))\cap \C)$ is compact.
Since $\C$ is noncompact, then $\C\setminus K$ is open and nonempty. 
From this point on the proof proceeds as in Lemma \ref{Aa:6}a . 
\end{proof}
Consider $M_1,M\in\Loc$ and let $\psi\in(M_1,M)$. Define 
\begin{equation}
\label{Aa:7}
\begin{array}{rcl}
\psi(\K^{h}(M_1)) & \defi & \{ \psi(\dc)\subseteq M \ | \  \dc\in \K^{h}(M_1)\}\ ,\\
\K^{h}(M)\nrestr_{\psi(M_1)} & \defi & \{ \dc\in\K^{h}(M) \ | \ cl(\dc)\subseteq \psi(M_1)\}\ .
\end{array}
\end{equation}
\begin{lemma}
\label{Aa:7bis}
Given $M_1,M\in\Loc$ and  $\psi\in(M_1,M)$,  then 
\[
\K^{h}(M)\nrestr_{\psi(M_1)}=\psi(\K^{h}(M_1))\ .
\]
\end{lemma}
\begin{proof}
By the definition of $\Loc$ it is clear that if $\dc_1\in\Kcal^{h}(M_1)$,
then $\psi(\dc_1)\in\Kcal^{h}(M)\nrestr_{\psi(M_1)}$. Conversely, 
let $\dc\in\Kcal^{h}(M)\nrestr_{\psi(M_1)}$. By Lemma \ref{Aa:6}.(c),
$\dc^\perp\cap\psi(M_1)\ne\emptyset$. Then the causal complement 
of $\psi^{-1}(\dc)$ in $M_1$ is nonempty. By this and the definition 
of $\Loc$ we have that $\psi^{-1}(\dc)\in\Kcal^{h}(M_1)$.   
\end{proof}
Finally, we want to stress that $\Kcal^{h}(M)$ have elements which are 
nonsimply connected subsets of
$M$ and elements whose causal complement is nonarcwise connected.  
This fact creates some problems in studying superselection sectors of a
net indexed by $\Kcal^{h}(M)$. However, as we shall see in Section
\ref{C}, these problems will be overcome by basing the theory 
on a particular subfamily of $\Kcal^{h}(M)$ whose elements do not present 
the above topological features. To this end, define  
\[
\Sub(\Kcal^{h}(M))\defi \{ \Kcal_1\subseteq \Kcal^{h}(M) \ | \ 
           \Kcal_1 \mbox{ is a basis for the topology of } M\}\ .        
\]
It is clear that $\Kcal^{h}(M)\in\Sub(\Kcal^{h}(M))$. Further properties are discussed in Section \ref{C}.

\paragraph{Diamonds regions:}
A relevant  element of $\Sub(\Kcal^{h}(M))$ is  
the set $\Kcal^d(M)$ of diamonds of $M$ \cite{Ruz3}. 
An open subset $\dc$ of $M$ is called a \textit{diamond} if
there is a spacelike Cauchy surface $\C$, a chart $(U,\phi)$ 
of $\C$, and an open ball $B$ of $\mathbb{R}^3$ such that 
\[
 \dc = \D(\phi^{-1}(B)), \ \ \ cl(B)\subset \phi(U)\subset\R^3\ .
\]
We will say that $\dc$ is \emph{based} on $\C$ and call 
$\phi^{-1}(B)$ the \emph{base} of $\dc$.
Any diamond $\dc$ is an open, relatively compact, arcwise and simply
connected subset of $M$, and its causal complement 
$\dc^\perp$ is arcwise connected. Furthermore, for any 
diamond $\dc$ there exists a pair of diamonds $\dc_1,\dc_2$ such that 
\begin{equation}
\label{Aa:7a}
cl(\dc),cl(\dc_1)\subset\dc_2 \ , \ \ \  \dc\perp\dc_1\ .
\end{equation}  
The set of diamonds $\Kcal^d(M)$ is a basis for the topology of $M$  
and  $\Kcal^d(M)\in \Sub(\Kcal^{h}(M))$. Let 
\begin{equation}
\label{Aa:8}
\begin{array}{rcl}
\psi(\Kcal^d(M_1)) & \defi & \{ \psi(\dc)\subseteq M \ | \  \dc\in \Kcal^d(M_1)\}\ ,\\
\Kcal^d(M)\nrestr_{\psi(M_1)} & \defi & \{ \dc\in\K^{d}(M) \ | \ cl(\dc)\subseteq \psi(M_1)\}\ .
\end{array}
\end{equation}
%********************
\begin{lemma}
\label{Aa:9}
Given $M_1,M\in\Loc$ and  $\psi\in(M_1,M)$,  then 
\[
\Kcal^d(M)\nrestr_{\psi(M_1)}=\psi(\Kcal^d(M_1))\ .
\]
\end{lemma}
\begin{proof}
We prove the inclusion ($\supseteq$). Let 
$\dc$ be a diamond of $M_1$. This means that there exists 
a spacelike Cauchy surface $\C_1$ of $M_1$,  
a chart $(U,\phi_1)$ such that 
$\dc_1=\D(\phi_1^{-1}(B))$, where $B$ is a ball of $\mathbb{R}^{3}$
such that $cl(\phi_1^{-1}(B))\subseteq U$. Let $B_1$ be 
a ball of $\mathbb{R}^{3}$ such that
$cl(B)\subseteq cl(B_1)$ and $cl(\phi_1^{-1}(B_1))\subseteq U$. 
Observe that $\psi\phi_1^{-1}(B_1)$ is a relatively compact, spacelike 
acausal open set of $M$  with boundaries and with a nonempty complement.  
By \cite{BS3}, there exists in $M$ a spacelike Cauchy surface 
$\C$ such that $cl(\psi\phi_1^{-1}(B_1))\subseteq \C$. 
Define $V\defi \psi\phi_1^{-1}(B_1)$ and 
$\phi\defi \phi_{1}\psi^{-1}$. The pair 
$(V,\phi)$ is a chart of $\C$ and $cl(\phi^{-1}(B))\subset V$. 
Finally observe that by the properties 
of $\psi$ we have that 
$\psi(\D(\phi_1^{-1}(B)) = \D(\psi\phi_1^{-1}(B))= \D(\phi^{-1}(B))$,
namely $\Kcal^d(M)\nrestr_{\psi(M_1)}\supseteq\psi(\Kcal^d(M_1))$.
The proof of the reverse inclusion is very similar to the previous one 
and we omit it. 
\end{proof}

There are many other elements of $\Sub(\K^{h}(M))$ which may be of a certain interest. For instance, one such is the family of \emph{regular diamonds} used in \cite{GLRV} (see it for the exact definition, or \cite{Ve1}). It contains the family of diamonds as a subset, and has the same stability property as the others. We refrain  from giving details here,  since this family does not play a r\^ole hereafter. More discussion will be found in \cite{BR}.

%****************************************************************
%****************************************************************
\section{Locally covariant quantum field theory}
\label{B}
Locally covariant quantum field theory is a categorical
approach to the quantum theory of fields which incorporates 
the locality principle
of classical field theory in a generally covariant manner \cite{BFV,BF2}.
In order to introduce the axioms of
the theory, we give a preliminary definition.
Let us denote by $\Obs$ the category whose objects $\A$
are unital $\Crm^*$-algebras and whose
arrows  $(\A_1,\A_2)$ are the  unit-preserving injective
$\Crm^*$-morphisms from $\A_1$ into $\A_2$.
The composition law between the arrows $\al_1$ and $\al_2$,
denoted by $\al_1\al_2$, is given by the usual
composition between $\Crm^*$-morphisms; the unit
arrow  $\mathrm{id}_\A$ of $(\A,\A)$ is the identity morphism of $\A$. \\[3pt]
%***********************************
\indent A \textbf{locally covariant quantum field theory} is a
covariant functor $\Al$ from the category $\Loc$ (see Section \ref{A})  into
the category $\Obs$, that is, a diagram
\[
\xymatrix{
 M_1 \ar[d]_\Al \ar[r]^{\psi} & M_2 \ar[d]^\Al\\
\Al(M_1)  \ar[r]^{\alpha_\psi} &\Al(M_2)}
\]
where $\alpha_{\psi}\defi \Al(\psi)$, such that
$\al_{\mathrm{id}_{M}} = \mathrm{id}_{\Al(M)}$, and
$\al_\phi\al_\psi =\al_{\phi\psi}$
for each $\psi\in ( M_1, M)$ and
$\phi\in ( M, M_2)$.
%**************************************
The functor $\Al$ is said to be \textbf{causal}
if, given  $\psi_i\in (M_i, M)$
for $i=1,2$,
\[
\psi_1( M_1) \perp \psi_2( M_2) \ \Rightarrow \
\left[ \alpha_{\psi_1}(\Al( M_1)), \alpha_{\psi_2}(\Al( M_2))\right] = 0\ ,
\]
where $\psi_1( M_1)\perp\psi_2( M_2)$ means that $\psi_1( M_1)$ and
$\psi_2( M_2)$ are causally
disjoint in $M$. From now on $\Al$ will denote
a causal locally covariant quantum field theory.\\[3pt]
\indent In conclusion, let us see
how a net of local algebras over $M\in\Loc$
can be recovered from a locally covariant quantum
field theory $\Al$. To this end,
recall that any  $\dc\in \Kcal^{h}(M)$, considered as a spacetime
with the metric $\g\nrestr_{\dc}$,
belongs to $\Loc$.
The injection $\io_{M,\dc}$ of
$\dc$ into $M$ is an element of
$(\dc, M)$  because of the definition of $\Kcal^{h}(M)$.
Then, using  $\al_{\io_{M,\dc}}\in (\Al(\dc), \Al(M))$
to define $\A(\dc)\defi \al_{\io_{M,\dc}}\left(\Al(\dc)\right)$,
it turns out \cite{BFV} that the correspondence 
\begin{equation}
\label{Ab:1}
\Al_{\Kcal^{h}(M)}:\Kcal^{h}(M)\ni\dc\longrightarrow \A(\dc)\subset\Al(M)\ ,
\end{equation}
is a net of local algebras satisfying the Haag-Kastler axioms:
\[
\begin{array}{lcll}
\dc_1\subseteq\dc_2 & \Rightarrow & \A(\dc_1)\subseteq \A(\dc_2)\ , & 
\qquad isotony,\\
\null\\
 \dc_1\perp \dc_2   & \Rightarrow & [\A(\dc_1),\A(\dc_2)]=0\ , & 
\qquad causality.
\end{array}
\]
As for the local covariance of the theory, let $ M_1\in\Loc$ with
the metric $\g_1$, and let 
$\psi\in(M_1,M)$. Because of  Lemma \ref{Aa:7bis},
$\psi(\dc)\in \Kcal^{h}(M)$
for each $\dc\in\Kcal^{h}(M_1)$. Since
$\io^{-1}_{M,\psi(\dc)} \ \psi \ \io_{M_1,\dc}$ is an isometric
embedding of 
the spacetime $\dc$ onto the spacetime $\psi(\dc)$~---~the latter
equipped with the metric $\g\nrestr_{\psi(\dc)}$~---~one has that
\begin{equation}
\label{Ab:2}
\alpha_{\psi}:\A(\dc)\subset \Al(M_1)  \rightarrow
               \A(\psi(\dc))\subset \Al(M)\ .
\end{equation}
is a $\mathrm{C}^*$-isomorphism. As $\psi(\Kcal^{h}(M_1))\in
\Sub(\Kcal^{h}(M))$, denote by $ \Al_{\psi(\Kcal^{h}(M_1))}$ the net index by 
$\psi(\Kcal^{h}(M_1))$ obtained by restricting $\Al_{\Kcal^{h}(M)}$
to $\psi(\Kcal^{h}(M_1))$. Then the relation (\ref{Ab:2}) says that 
\begin{equation}
\label{Ab:3}
\alpha_{\psi}:\Al_{\Kcal^{h}(M_1)} \rightarrow \Al_{\psi(\Kcal^{h}(M_1))}.
\end{equation}
is a net-isomorphism.
%***********************************************************
%***********************************************************
\subsection{States and representations of nets}
\label{Ac}
Fix $M\in\Loc$, and consider 
the net $\Al_{\Kcal^{h}(M)}$. A state $\omega$ of the algebra $\Al(M)$, 
is defined to be a
positive ($\omega(A^\ast A)\ge 0, \ A\in\Al(M)$), 
and normalized ($\omega(\mathbbm{1})=1$) 
linear functional on it.
For any  state $\omega$ of the algebra $\Al(M)$
we will denote by $(\pi_\omega,\Hil_\omega, \Omega_\omega)$ the 
corresponding  GNS-construction; by 
$\omega^*\Al_{\K^{h}(M)}$ we will denote the net of von Neumann algebras 
defined as the correspondence 
\begin{equation}
\omega^*\Al_{\K^{h}(M)}: \K^{h}(M)\ni\dc\rightarrow 
\A_\omega(\dc)\subseteq  \mathfrak{B}(\Hil_\omega)\ ,
\end{equation}
where $\A_\omega(\dc)\defi \pi_\omega(\A(\dc))''$ for any 
$\dc\in\Kcal^{h}(M)$, and $\mathfrak{B}(\Hil_\omega)$ is the 
$\Crm^*$--algebra of linear bounded operators of $\Hil_\omega$.
Furthermore, we set 
$\Al_\omega(M)\defi\pi_\omega(\Al(M))$. \\
\indent Once $\omega^*\Al_{\K^{h}(M)}$ is given, to any element
$\Kcal_1\in \Sub(\Kcal^{h}(M))$ there corresponds a net 
of von Neumann algebras $\omega^*\Al_{\Kcal_1}$ defined as
\begin{equation}          
\omega^*\Al_{\Kcal_1} \defi \{\A_\omega(\dc) \ | \ \dc\in\Kcal_1 \}\ .
\end{equation}             
Properties of such a net 
of local algebras, which will be important for our purposes, 
are the following: 
%*****************
\begin{itemize} 
\item[P1.] $\omega^*\Al_{\Kcal_1}$ is said to be \textit{irreducible} whenever, 
given $T\in\Bh$ such that  $T\in\A_\omega(\dc)'$ for any 
$\dc\in \Kcal_1$, then  $T=c\cdot\mathbbm{1}$.
\item[P2.] $\omega^*\Al_{\K_1}$ satisfies 
the \textit{Borchers property} if given  $\dc\in \Kcal_1$
for any $\dc_1\in \Kcal_1$ with $cl(\dc)\subseteq \dc_1$ 
any nonzero orthogonal projection $E\in\A_\omega(\dc)$ 
is equivalent to $\mathbbm{1}$ in $\A_\omega(\dc_1)$.
\item[P3.]  $\omega^*\Al_{\Kcal_1}$ is \textit{locally definite} if
$\mathbb{C}\cdot \mathbbm{1} = \cap \{\A_\omega(\dc) \ | \
\dc\in\Kcal_1, \ x\in\dc\}$,  
for any point $x$ of $M$.
\item[P4.] $\omega^*\Al_{\Kcal_1}$ satisfies 
\textit{punctured Haag duality} 
if given $\dc\in\Kcal_1$, with $cl(\dc)\perp\{x\}$, then
\[
\A_\omega(\dc)= \cap \{\A_\omega(\dc_1)' \ | \ \dc_1\in\Kcal_1,  \  
   \dc_1\perp\dc, \ cl(\dc_1)\perp \{x\} \} \ ,
\]
for any point $x$ of $M$.
\end{itemize}
Some observations on these definitions are in order. 
First, given $\Kcal_1\in\Sub(\Kcal^{h}(M))$, the irreducibility of the net 
$\omega^*\Al_{\Kcal_1}$ is, in general, 
stronger than irreducibility of the representation $\pi_\omega$ of
$\Al(M)$. This is because the collection 
$\cup_{\dc\in\Kcal_1}\A_\omega(\dc)$ needs not to be dense in 
$\Al_\omega(M)$ (see also \cite{BFV}). Moreover,
it is clear that if $\omega^*\Al_{\Kcal_1}$ satisfies punctured
Haag  duality, then it satisfies \textit{Haag duality};
given $\dc\in\Kcal_1$, then 
$\A_\omega(\dc)$ $= \cap \{\A_\omega(\dc_1)' \ | \ \dc_1\in\Kcal_1, \ 
   \dc_1\perp\dc\}$. 
If $\omega^*\Al_{\Kcal_1}$ satisfies
punctured Haag duality and is irreducible, 
then it is locally definite \cite{Ruz2}.

From now on we will say that a state $\omega\in\St(M)$ 
\textit{satisfies punctured Haag duality} if $\omega^*\Al_{\Kcal^{h}(M)}$ 
is irreducible and satisfies punctured Haag duality.
We note the following straightforward results:
\begin{lemma}
\label{Ac:1}
The following assertions hold: \\
(a) If $\Al_{\Kcal_1}$ is irreducible, then 
    $\Al_{\Kcal_2}$ is irreducible, for any $\Kcal_2\in \Sub(\Kcal^{h}(M))$
    such that $\Kcal_1\subseteq \Kcal_2$;\\
(b) If $\Al_{\Kcal_1}$ satisfies the Borchers property, then 
       $\Al_{\Kcal_2}$ satisfies the Borchers property 
     for any $\Kcal_2\in\Sub(\Kcal^{h}(M))$ such that $\Kcal_2\subseteq \Kcal_1$; \\
(c) If there is $\K_1\in\Sub(\Kcal^{h}(M))$ such that $\Al_{\Kcal_1}$ is locally definite, then 
    $\Al_{\Kcal_2}$ is locally definite for any $\K_2\in \Sub(\Kcal^{h}(M))$.
\end{lemma}
\begin{proof}
$(a)$ and $(b)$ are obvious. $(c)$ derives from the fact 
that any element of $\Sub(\Kcal^{h})$ is a basis for the topology 
of $M$.
\end{proof}
%***********************************************************
%***********************************************************
%***********************************************************
\subsection{State  Space}
\label{Ad}
We now turn to the notion of a state space
of $\Al$ \cite{BFV}.  A \textit{state space} of a unital $\Crm^*$-algebra
$\A$ is a family of states $\St(\A)$ of $\A$ which is 
closed under finite convex combinations and 
operations $\omega(\cdot)\rightarrow \omega(A^*\cdot A)/\omega(A^*A)$ 
for $A\in\A$. We denote by $\Sts$  be the category
whose objects are the state spaces $\St(\A)$
of unital $\Crm^*$-algebras  $\A$ whose arrows 
are the positive
maps $\ga^*: \St(\A)\longrightarrow \St(\A')$, arising as dual maps
of injective morphisms of $\Crm^*$-algebras
$\ga:\A'\longrightarrow \A$, by
$\ga^*\omega(A) \defi \omega(\ga(A))$  for each $A\in\A'$.
The composition law between two arrows, as the definition of the identity
arrow of an object, are obvious. \\[3pt]
\indent A \textbf{state space} for
$\Al$ is a contravariant functor
$\mathscr{S}$ between $\Loc$ and $\Sts$, that is,
a diagram
\[
\xymatrix{
 M_1  \ar[d]_{\mathscr{S}} \ar[r]^{\psi} &
                          M_2 \ar[d]^{\mathscr{S}}\\
\mathscr{S}(M_1) & \ar[l]_{\alpha^*_\psi} \mathscr{S}( M_2)
}
\]
where $\mathscr{S}(M_1)$ is a state space of the algebra
$\Al(M_1)$, such that
$\alpha^*_{\mathrm{id}_{M}}= \mathrm{id}_{\St(M)}$, and
$\al^*_\psi\al^*_\phi =\al^*_{\phi\psi}$
for each $\psi\in(M_1,M)$ and
$\phi\in(M,M_2)$.\\[5pt]
\indent Some of the properties, involving nets and states, 
can be generalized to state spaces. First we recall, that a 
state space $\St$ is said to be 
\textit{locally quasi-equivalent}
if for any $M\in\Loc$ and for any pair $\omega,\si\in\St(M)$ we have
that 
\[
 F(\pi_\omega\nrestr\!\A(\dc)) =  F(\pi_\si\nrestr\!\A(\dc))\ , \qquad \forall\dc\in\Kcal^{h}(M)\ ,
\]
where $F(\pi_\omega\nrestr\!\A(\dc))$ is the local folium, i.e. the collection
of normal states of the algebras $\pi_\omega(\A(\dc))$. 
\begin{df}
We will say that a locally quasi-equivalent $\St$ satisfies the \textbf{Borchers property} 
(resp. \textbf{local definiteness}) if for any $M\in\Loc$ and for any 
$\omega\in\St(M)$ the net $\omega^*\Al_{\Kcal^{h}(M)}$ satisfies 
the Borchers property (resp. local definiteness). 
\end{df}
%*******************
\begin{lemma}
\label{Ad:1}
Let $\St$ be a locally quasi-equivalent state space. Given $M\in\Loc$,
for any pair $\si,\omega\in\St(M)$ the nets 
$\omega^*\Al_{\K^{h}(M)}$ and $\si^*\Al_{\K^{h}(M)}$ are isomorphic, namely,
there is a collection 
\[
\rho_{\omega,\si} \defi \{ \rho_\dc: \A_\omega(\dc)\longrightarrow 
                                     \A_\si(\dc) \  | \ \dc\in\Kcal^{h}(M)\}
\]
made of $^*$-isomorphisms of von Neumann algebras 
such that $\rho_\dc\nrestr\!\A_{\omega}(\dc_1) = \rho_{\dc_1}$ if 
$\dc_1\subseteq\dc$.
\end{lemma}
\begin{proof}
Local quasi-equivalence means that for any $\dc\in\K^h(M)$ there exists a 
\textit{unique} isomorphism 
$\rho_\dc:\A_\omega(\dc)\longrightarrow\A_\si(\dc)$ such that 
\[
 \rho_\dc\pi_\omega(A)=\pi_\si(A) \qquad A\in\A(\dc)\ .
\]
The  collection $\rho_{\omega,\si}\defi\{\rho_\dc \ |\ \dc\in\K^{h}(M)\}$ is compatible 
with the net structure. 
Indeed, note that given $\dc_1\subseteq \dc$, then 
$\rho_\dc\pi_\omega(A)=\pi_\si(A) = \rho_{\dc_1}\pi_\omega(A)$ for any 
$A\in\A_{\omega}(\dc_1)$.
By uniqueness we have $ \rho_\dc\nrestr\!\A_\omega(\dc_1) = 
 \rho_{\dc_1}$ if $\dc_1\subseteq \dc$.
\end{proof}
%***************
As an easy consequence of this lemma we have 
\begin{cor}
\label{Ad:2}
Let $\St$ be a locally quasi-equivalent state space.  Assume that for
any $M\in\Loc$ there is 
$\omega\in\St( M)$ such that $\omega^*\Al_{\Kcal^{h}(M)}$ satisfies 
the Borchers property (resp. local definiteness). Then 
$\St$ fulfills the Borchers property (resp. local definiteness).
\end{cor}
%***************
Now, consider $\psi\in(M_1,M)$ and 
the associated injective $\Crm^*-$morphism   
$\alpha_\psi: \Al(M_1)\rightarrow \Al(M)$.
By means of $\alpha_\psi$, a state $\omega\in\St(M)$ 
induces two different representations on $\Al(M_1)$. 
On the one hand, if $(\pi_\omega,\Hil_\omega,\Omega_\omega)$ is the 
GNS construction associated with $\omega$, then $\pi_\omega\alpha_\psi$
is a representation of $\Al(M_1)$. On the other hand, 
by local covaraince  $\omega\alpha_{\psi}\in\St(M_1)$. If 
$(\pi_{\omega\alpha_\psi},\mathcal{H}_{\omega\alpha_\psi},
\Omega_{\omega\alpha_\psi})$ is the GNS construction  associated with 
$\omega\alpha_{\psi}$, then $\pi_{\omega\alpha_\psi}$ is a representation
of $\Al(M_1)$. In order to understand the relation between 
$\pi_\omega\alpha_\psi$ and $\pi_{\omega\alpha_\psi}$, define 
\begin{equation}
\label{Ad:2a}
 V  \pi_\omega\alpha_{\psi}(A)  \Omega_\omega
 \defi \pi_{\omega\alpha_\psi}(A)  \Omega_{\omega\alpha_\psi}, 
 \qquad A\in\Al(M_1)\ .
\end{equation}
Since $\Omega_{\omega\alpha_\psi}$ is cyclic, $V$ is a partial 
isometry from $\Hil_\omega$ onto $\Hil_{\omega\alpha_\psi}$
such that $VV^* = \mathbbm{1}_{\mathcal{H}_{_{\omega\alpha_\psi}}}$  and 
$V^*V\in \pi_\omega(\alpha_\psi(\Al(M_1)))'$. Let 
\begin{equation}
\label{Ad:3}
\tau^\omega_\psi ( \pi_\omega\alpha_\psi(A)) \defi 
\pi_{\omega\alpha_\psi}(A) \ , 
 \qquad A\in\Al(M_1)\ .
\end{equation}
Observe that $\tau^\omega_\psi:  \pi_\omega\alpha_\psi(\Al(M_1))\rightarrow 
  \pi_{\omega\alpha_\psi}(\Al(M_1))$ is a $\Crm^*-$morphism. Furthermore, 
$\tau^\omega_\psi$ defines a net-morphism from the net 
$\omega^*\Al_{\psi(\Kcal^{h}(M_1))}$ into the net 
$(\omega\alpha_\psi)^*\Al_{\Kcal^{h}(M_1)}$ (see also (\ref{Ab:3})). 
%*********************
Collecting the definitions and properties just stated we have that
\begin{prop}
\label{Ad:4}
Assume that  $\omega^*\Al_{\Kcal^{h}(M)}$ satisfies the Borchers property.\\
(a) $\pi_\omega\alpha_\psi$ and $\pi_{\omega\alpha_\psi}$ 
    are locally quasi-equivalent 
     representations of $\Al(M_1)$.\\
% (b) $V^*  V$ is equivalent to $\mathbbm{1}$ on $\A_\omega(\dc)'$ 
%      for any $\dc\in\psi(\Kcal^{h}(M_1))$.\\
(b) $\tau^\omega_\psi$ is a net-isomorphism.
\end{prop}
\begin{proof}
$(a)$ Since $V^* V\in (\pi_\omega\alpha_\psi(\Al(M_1)))'$, we have 
\[
V^* V\in \A_\omega(\dc)'=\pi_\omega(\A(\dc))', \qquad 
\forall \dc\in\psi(\Kcal^{h}(M_1)). 
\]
We show that $V^*V$  has central support equal to 
$\mathbbm{1}$ on  $\A_\omega(\dc)'$ for $\dc\in\psi(\Kcal^{h}(M_1))$.
This is equivalent to showing that 
\[
E\cdot V^* V = 0  \ \ \ \iff \ \ \ E=0 \ ,
\]
for any orthogonal projection 
$E$ of $\A_\omega(\dc)$.
Assume that $E\ne 0$. By Lemma \ref{Aa:6}.(b), there is 
$\dc_1\in\psi(\Kcal^{h}(M_1))$ such that  $cl(\dc)\subseteq\dc_1$.
By the Borchers property there is  an  isometry 
$W\in \A_\omega(\dc_1)$ such that 
$WW^*=E$. Then 
\[
E\cdot V^* V = 0 \iff WW^*\cdot V^* V=0
\iff V^* V\cdot W^*=0\iff V^* V =0\ . 
\]
This leads to a contradiction. Hence $V^* V$ has central support 
$\mathbbm{1}$ on $\A_\omega(\dc)'$. % (b) By the Borchers property, 
% the algebras $\A_\omega(\dc)'$ with $\dc\in\psi(\Kcal^{h}(M_1))$
% are properly infinite. Then $V^*V\sim \mathbbm{1}$ on
% $\A_\omega(\dc)$. 
$(b)$ The proof follows from $(a)$ because 
by (\ref{Ad:2a}) $\tau^\omega_\psi ( \pi_\omega\alpha_\psi(A)) = V
             \pi_\omega\alpha_\psi(A) V^*$ for any 
   $A\in\Al(M_1)$.
\end{proof}
%**********************************
%**********************************
We now study the relations between the net-isomorphisms 
introduced in this section.
%**********************************
%**********************************
\begin{lemma}
\label{Ad:6}
Let $\St$ be a locally quasi-equivalent state space satisfying the
Borchers property. Given $\psi\in(M_1,M)$, then 
$\rho_{\omega\alpha_\psi,\si\alpha_\psi} \  \tau^\omega_\psi
= \tau^\si_\psi \ \rho_{\omega,\si}$,
for any pair  $\si,\omega\in\St(M)$.
\end{lemma}
\begin{proof}
Let $\dc\in\Kcal^{h}(M_1)$ and $A\in\A(\dc)$. By the definition of 
$\rho_{\omega,\si}$ (see Lemma \ref{Ad:1}) and by  
(\ref{Ad:3}), we have that 
\begin{align*}
\tau^\si_\psi(\pi_\si(\alpha_\psi(A))) & = \pi_{\si\alpha_\psi}(A)
  =  \rho_{\omega\alpha_\psi,\si\alpha_\psi}(\pi_{\omega\alpha_\psi}(A))\\
& =  \rho_{\omega\alpha_\psi,\si\alpha_\psi} \ \tau^\omega_\psi
       (\pi_\omega(\alpha_\psi(A)))  \\
&  = \rho_{\omega\alpha_\psi,\si\alpha_\psi} \ \tau^\omega_\psi \ 
  \rho_{\si,\omega}   (\pi_\si(\alpha_\psi(A)))\ .
\end{align*}
\end{proof}
%***************
\begin{lemma}
\label{Ad:7}
Let $\St$ be a locally quasi-equivalent state space satisfying the
Borchers property. Given 
$\psi\in(M_1,M)$, $\phi\in(M_2,M_1)$, then 
$\tau^\omega_{\psi\phi} = 
 \tau^{\omega\alpha_\psi}_\phi 
 \tau^\omega_\psi$,
for any $\omega\in\St(M)$.
\end{lemma}
\begin{proof}
First observe that $\omega\al_{\psi\phi}\in\St(M_2)$. By the
definition (\ref{Ad:3}) we have that  
\[
\begin{array}{rcll}
\tau^\omega_{\psi\phi}(\pi_\omega\alpha_{\psi\phi}(A)) & 
 = & \pi_{\omega\alpha_{\psi\phi}}(A) \ , & A\in \Al(M_2) \ , \\
\tau^{\omega}_\psi(\pi_\omega(\alpha_{\psi}(A))) & = & 
 \pi_{\omega\alpha_{\psi}}(A)\ ,  &  A\in \Al(M_1) \ , \\
\tau^{\omega\alpha_\psi}_\phi(\pi_{\omega\alpha_\psi}(\alpha_{\phi}(A)))
 & = &  
 \pi_{\omega\alpha_{\psi\phi}}(A) \ , &  A\in\Al(M_2) \ . 
\end{array} 
\]
By using these relations we have that 
\[
\tau^{\omega\alpha_\psi}_\phi   \tau^\omega_\psi 
(\pi_\omega\alpha_{\psi\phi}(A)) = 
\tau^{\omega\alpha_\psi}_\phi
(\pi_{\omega\alpha_\psi}(\alpha_\phi(A))) = 
\pi_{\omega\alpha_{\psi\phi}}(A) = 
 \tau^\omega_{\psi\phi}(\pi_\omega\alpha_{\psi\phi}(A))\ , 
\]
for any $A\in\Al(M_2)$.
\end{proof}

\section{Homotopy of posets and net cohomology}
\label{C}
The mathematical framework under which we will study 
superselection sectors is that of the net cohomology of 
posets. In the present section we give some notions and results 
referring to \cite{Rob2, Rob3, Ruz3}. Throughout this section, we fix 
$M\in\Loc$,  and denote by $\Po$ a basis for the topology 
of $M$ whose elements are open and arcwise connected subsets of $M$,
and have a nonempty causal complement.
Furthermore,  by the same symbol 
$\Po$,  we denote the partially ordered set (\emph{poset}) formed
by the basis $\Po$ ordered under \emph{inclusion} $\subseteq$. 
%*************************
%*************************
\subsection{Homotopy of posets}
\label{Ba}
%******************************
%******************************
\paragraph{The simplicial set of $\Po$:} We need a few  observations
on standard n-simplices. Consider the standard  n-simplex
$\Delta_n$, namely  the subset of $ \mathbb{R}^{n+1}$, defined as
\[
\Delta_n \defi \big\{ (\la_0,\ldots,\la_n)\in \mathbb{R}^{n+1} \ 
  | \ \la_0+\cdots +\la_n=1, \ \ \la_i\in[0,1]\big\}.
\]
Observe that $\Delta_n$  can be regarded as a partially ordered set 
with respect to the inclusion of its subsimplices, and that for any 
$n\geq 1$ there is  
a family $d^n_i:\Delta_{n-1}\rightarrow \Delta_n$, with $0\leq i\leq
n$, of inclusion preserving maps, defined as  
\[
d^n_i(\la_0,\ldots,\la_{n-1})= 
(\la_0,\la_1,\ldots,\la_{i-1}, 0, \la_{i},\ldots\la_{n-1}).
\]
Now, a \emph{singular n-simplex} of the  poset $\Po$ 
is an inclusion preserving map $f:\Delta_n\rightarrow \Po$. 
We denote by $\Si_n(\Po)$ the set of  singular n-simplices. 
The collection $\Si_*(\Po)$ of all singular simplices 
is called  the \emph{simplicial set} of $\Po$. 
The inclusion maps $d^n_i$ between 
standard simplices are extended  to maps
$\partial^n_i :\Sigma_n(\Po)\rightarrow \Sigma_{n-1}(\Po)$, called 
\emph{boundaries},
between singular simplices by setting
$\partial^n_i f  \defi f \circ d^n_i$. One can easily check, by 
the definition of $d^n_i$,  that the following 
relations
\begin{equation}
\label{Ba:0}
\partial^{n-1}_i\circ \partial^{n}_j = 
\partial^{n-1}_j \circ \partial^{n}_{i+1},\qquad  i\geq j\ , 
\end{equation}
hold. From now on, we will omit the superscripts from the symbol
$\partial^n_i$, and will denote: the composition 
$\partial_i\circ \partial_j$ by  the symbol $\partial_{ij}$;
0-simplices by the letter $a$; 1-simplices by $b$
and 2-simplices by $c$. Notice that a 0-simplex $a$ is nothing
but an element of $\Po$. A 1-simplex $b$ is formed by two 0-simplices 
$\partial_0b$, $\partial_1b$ and an element $|b|$ of $\Po$, called the 
\emph{support} of $b$, such that $\partial_0b$, 
$\partial_1b\subseteq |b|$; a 2-simplex $c$ is formed 
by three 1-simplices $\partial_0c,\partial_1c,\partial_2c$, whose 0-boundaries 
are chained between them according to (\ref{Ba:0}), 
and by a 0-simplex $|c|$, the \emph{support},
such that $\partial_0c,\partial_1c,\partial_2c\subseteq |c|$.
Given a 1-simplex $b$,
the \emph{reverse} of $b$ is  the  1-simplex $\overline{b}$ 
defined as 
\[
\partial_0\con{b} = \partial_1b\ , \ \ \  
\partial_1\con{b} = \partial_0b\ , \ \ \ 
|\con{b}| = |b|\ .
\] 
Finally, a 1-simplex $b$ is said 
 to be \emph{degenerate to} a 0-simplex $a_0$ whenever
 \[
  \partial_0 b =a_0 =\partial_1b, \ \ \ \  a_0=|b| \ .
 \]
We will denote by $b(a_0)$ the 1-simplex degenerate to $a_0$.
%***************************************************
%***************************************************
\paragraph{Paths:} Given $a_0,a_1\in\Si_0(\Po)$, 
a  \emph{path from $a_0$ to $a_1$} is a finite ordered 
sequence $p=\{b_n,\ldots,b_1\}$ of 1-simplices  enjoying  the relations
\[
 \partial_1b_1=a_0\ , \ \ \ \partial_0 b_{i} = \partial_1 b_{i+1} \  
            \mbox{ with } 
\ i\in\{1,\ldots,n-1\}\ , \ \ \  \partial_0b_n=a_1\ .
\]
The \emph{starting point} of $p$, written $\partial_1p$, 
is the 0-simplex $a_0$, while the \emph{end point} of $p$, 
written $\partial_0p$, is the 0-simplex  $a_1$. The \emph{support} $|p|$
of the path $p$ is the open set
\[
|p|\defi \cup^n_{i=1}|b_i|\ .
\]
We will denote 
by $\Po(a_0,a_1)$ the set of paths from 
$a_0$ to $a_1$, and by $\Po(a_0)$ the set of closed paths whose end point 
is $a_0$.  \\
\indent The set of paths is equipped with  the following 
operations. Consider 
a path $p=\{b_n,\ldots,b_1\}\in\Po(a_0,a_1)$. 
The \emph{reverse} $\overline{p}$ of $p$  is the path 
\[
\con{p}\defi \{ \overline{b}_1,\ldots,\overline{b}_n\}\in \Po(a_1,a_0)\ .
\]
The \emph{composition} of $p$ with a path 
$q=\{b'_k,\ldots b'_1\}$ of $\Po(a_1,a_2)$,  
is  defined as 
\[
q*p\defi \{b'_k,\ldots, b'_1,b_n,\ldots,b_1 \}\in \mathcal{P}(a_0,a_2)\ .
\]
Note that the reverse $-$ is involutive, while 
the composition  $*$ is associative. 
\indent An \emph{elementary deformation} of a path $p$
consists in replacing a 1-simplex $\partial_1c$ of the path
by a pair $\partial_0c,\partial_2c$, where $c\in\Si_2(\Po)$, or, conversely 
in replacing a consecutive pair $\partial_0c,\partial_2c$ of 1-simplices 
of $p$ by a single 1-simplex $\partial_1c$. Two paths with the same endpoints 
are \emph{homotopic} if they can be obtained from one other 
by a finite set of elementary deformations. Homotopy defines 
an equivalence relation  $\sim$  
on the set of paths with the same end points. The reverse and the 
composition are compatible with the homotopy equivalence relation,
namely 
\begin{equation}
\label{Ba:1}
\begin{array}{rclr}
 p\sim q   &   \iff   &    \con{p}\sim\con{q}\ , & 
    p,q\in \Po(a_0,a_1), \\
 p\sim q\ , \ p_1\sim q_1  &   \Rightarrow  & 
  p_1*p\sim q_1* q\ , \ \   &  p_1,q_1\in \Po(a_1,a_2)\ .
\end{array}
\end{equation}
Furthermore, for any $p\in \Po(a_0,a_1)$, the following relations hold:
\begin{equation}
\label{Ba:2}
\begin{array}{rcl} 
  p* b(a_0) \sim p & \mbox{ and } &  p \sim b(a_1)*p\ ,  \\
 \con{p}* p \sim b(a_0) & \mbox{ and }  & b(a_1)\sim p*\con{p}\ ,
\end{array}
\end{equation}
where $b(a_0)$ is the 1-simplex degenerate to $a_0$. 
The following lemma will be useful for later purposes.
%***********************************
\begin{lemma}
\label{Ba:3}
Let $p=\{b_n,\ldots,b_1\}$ be a path, and let  
$q=q_n*\cdots*q_1$ be a path such that  
\begin{equation}
\label{Ba:3a}
|q_i|\subseteq |b_i| \ ,  \ \ \ 
\partial_0q_i\subseteq \partial_0b_i \ , 
\ \ \ \partial_1q_i \subseteq\partial_1b_i \ ,
\end{equation}
for $i=1,\ldots,n$. The following assertions hold:\\
(a) there exist a pair of 1-simplices $\tilde{b},\hat{b}$ such that 
$p\sim \tilde{b}*q*\hat{b}$; \\
(b) if $p$ and $q$ are closed paths, there is a 1-simplex 
    $b$ such that $p\sim \overline{b}*q* b$, where $\overline{b}$ is
the reverse of $b$.
\end{lemma}
\begin{proof}
$(a)$ By (\ref{Ba:1}) we can assume, 
without loss of generality, $p$ to be a 1-simplex $b$.
So, let $q$ be a path such that $|q|\subseteq |b|$, 
$\partial_0q\subseteq\partial_0b$, and $\partial_1q\subseteq\partial_1b$.
Let $\hat{b}$ be the 1-simplex defined as 
\[ 
|\hat{b}|= |b|, \ \ \partial_1\hat{b}=\partial_1b, \ \  
\partial_0\hat{b}=\partial_1q,
\]
and let 
$\tilde{b}$ be the 1-simplex defined as 
\[
|\tilde{b}|= |b|, \ \ \partial_1\tilde{b}= \partial_0q, 
\ \ \partial_0\tilde{b}=\partial_0b.
\]
Note that  $b$ and $\tilde{b}*q*\hat{b}$ have the same endpoints. 
Since the poset formed by the collection $\dc\in\Po$ such that 
$\dc\subseteq|b|$ is directed under 
inclusion, these two paths are homotopic
\cite{Ruz3} (see also (\ref{Ba:5})). $(b)$ follows from $(a)$.
\end{proof}
\paragraph{Poset approximation and pathwise connectedness:} 
Let $S$ be an open subset 
of $M$. Let $\Po_S\defi \{\dc\in\Po \ | \ cl(\dc)\subseteq S\}$. 
Since the elements of $\Po$ are arcwise connected subsets 
of $M$, it turns out that $S$ is arcwise connected if, and only if, 
$\Po_S$ is \emph{pathwise connected}, namely 
\[
 \forall \dc_1,\dc_2\in\Po_S
 \mbox{ there exists a path } \ p \mbox{ from } \dc_1 \mbox{ to } \dc_2 
 \mbox{ with }  cl(|p|)\subseteq S.
\]
In particular, since $M$ is arcwise connected, 
then $\Po$ is pathwise connected.  
Given a curve $\ga:[0,1]\rightarrow M$, a path 
$p =\{b_n,\ldots,b_1\}$ is said to be an 
\emph{approximation} of $\ga$
if there is a partition $0=s_0 < s_1<\ldots <s_n=1$ of the interval $[0,1]$
such that 
\[
\ga([s_{i-1},s_i])\subseteq |b_i|, \qquad \ga(s_{i-1})\in 
\partial_1b_i, \ \  \ga(s_i)\in \partial_0b_i,
\]
for $i=1,\ldots n$. 
We denote by $\App(\ga)$ the set of approximations
of $\ga$. It turns out that $\App(\ga)\ne\emptyset$ for any curve
$\ga$.
%*********************************
%*********************************
\begin{lemma}[poset-approximation \cite{Ruz3}]
\label{Ba:4}
Assume that the elements of $\Po$ are simply connected.
Let $p,q\in\Po(a_0,a_1)$ be 
respectively two approximations of a pair of  curve $\ga$ and $\be$
with the same endpoints. Then 
$\ga$ and $\be$ are homotopic if, and only if,  $p$ and $q$ are homotopic.
\end{lemma}
%***********************************
%***********************************
\paragraph{Fundamental group of a poset:} Fix a base 0-simplex $a_0$,
and define 
\[
 \pi_1(\Po,a_0) \defi \Po(a_0) / \sim \ , 
\]
i.e. the quotient of the set $\Po(a_0)$ of closed paths with end point 
$a_0$ with respect to homotopy 
equivalence relation. Denote by $[p]$ the equivalence 
class of $p\in\Po(a_0)$ with respect to homotopy equivalence relation, 
and define
\[
 [p]\cdot [q] \defi [p*q] \ , \qquad [p],[q] \in   \pi_1(\Po,a_0)\ . 
\]
By the relations (\ref{Ba:1}), (\ref{Ba:2}), it turns out that 
$\pi_1(\Po,a_0)$ is, with respect to this composition law,
a group. The identity of the group is 
the equivalence class  $[b(a_0)]$  associated with the 1-simplex 
degenerate to $a_0$;  the inverse $[p]^{-1}$ of $[p]$ 
is the equivalence class $[\overline{p}]$ associated with 
the reverse $\overline{p}$ of the path $p$. This group is
\emph{the first homotopy group of $\Po$ with base 0-simplex} $a_0$. 
Moreover,  as $\Po$ is pathwise connected the  first homotopy group 
does not depend, up to isomorphism,  on the chosen base 0-simplex $a_0$.
This equivalence class of groups, denoted by $\pi_1(\Po)$, 
is called \emph{the fundamental group of} $\Po$.  
In the case that $\pi_1(\Po)$ is trivial we will say that 
$\Po$ is simply connected. We have 
% Finally the poset 
% $\Po$ is said to be \emph{simply connected} whenever 
% $\pi_1(\Po)$ is trivial. \\
% \indent We recall that $\Po$ is said to be \emph{directed} 
% if for any pair $\dc_1,\dc_2\in \Po$, there exists 
% $\dc_3\in\Po$ such that $\dc_1,\dc_2\subseteq \dc_3$. Then  we have 
\begin{equation}
\label{Ba:5}
  \Po\ \mbox{is directed} \ \ \Rightarrow \ \ \Po\ \mbox{is simply connected} \ .
\end{equation}
The link with the corresponding topological notions of the underlying 
spacetime is obtained by means of the poset approximation lemma. 
\begin{teo}[\cite{Ruz3}]
\label{Ba:6}
Assume that all the elements of  $\Po$ are simply connected. 
Then the fundamental group $\pi_1(\Po)$ of the poset 
and the fundamental group $\pi_1(M)$  of the spacetime $M$ 
are isomorphic.
% Let $x_0\in M$ and $a_0\in\Si_0(M)$ with $x_0\in a_0$. 
% Then the first homotopy  group 
% $\pi_1(\Po,a_0)$ of the poset with base 0-simplex $a_0$  
% is isomorphic  to the first homotopy group $\pi_1(M,x_0)$ 
% of $M$ with basepoint $x_0$. 
\end{teo}
Since the set $\Kcal^d(M)$ of diamonds of $M$ 
verifies the property written in the statement,  we have that 
$\pi_1(\Kcal^d(M))\simeq\pi_1(M)$. 
The statement fails  if $\Po$ admits nonsimply connected elements.
A counterexample is provided by  $\Kcal^{h}(M)$ which is simply 
connected irrespectively of the topology of $M$. To show this 
a preliminary result is necessary.
%******************** 
\begin{lemma}
\label{Ba:7}
Let $p$ be a closed path of $\Kcal^d(M)$ with end point $a_0$. 
There exists a closed path 
$q$ of $\Kcal^d(M)$, with end point $a_0$, and an element 
$\dc\in\Kcal^{h}(M)$ such that   
$p$ is homotopic to $q$ and $|q|\subseteq \dc$ .
\end{lemma}
\begin{proof}
Let $\ga$ be a closed curve with  $p\in\App(\ga)$.
Furthermore if $a_0$ is a diamond of the form $\D(G)$, with  
$G$ contained in a spacelike Cauchy surface $\C_0$, 
then we can assume that $\ga$ meets $G$ in a point $x_0$.
By \cite[Lemma 3.1]{Ruz3} $\ga$ is homotopic to  a closed curve $\be$
lying on $\C_0$, and with endpoint $x_0$. Let $q=\{b_n,\ldots b_1\}$ 
be a closed path, with endpoint $a_0$,  
formed by diamonds  $b_i=\D(G_i)$, 
with $G_i\subseteq\C_0$ for any $i$, 
and such that $q\in \App(\be)$.
The poset approximation Lemma \ref{Ba:4} entails that $p\sim q$. 
Let $K\defi \cup_i G_i$. $K\subset\C_{0}$ is open, relatively compact
and arcwise connected. Let $\dc\defi \D(K)$ and note that 
$|q|\subseteq \dc$. $\dc$ is open 
relatively compact, arcwise connected and, since it is contained in the
Cauchy surface $\C_{0}$, by a standard argument (see, e.g. \cite{One}) 
$\dc$ is globally hyperbolic. It is also clear that we can shrink 
the sets  $G_i$ in such a way that $\dc^\perp\ne\emptyset$. Hence 
$\dc\in\Kcal^{h}(M)$. 
\end{proof}
%***********************
\begin{prop}
\label{Ba:8}
The poset $\Kcal^{h}(M)$ is simply connected for any $M\in\Loc$.
\end{prop}
\begin{proof}
Let $p$ be a closed path of $\Kcal^{h}(M)$, and let 
$\ga$ be a closed curve with $p\in\App(\ga)$. Since $\Kcal^d(M)$
is a basis for the topology of $M$ we can find 
a closed path $q$ in $\Kcal^d(M)$, with endpoint $a_0$, such that:
$q\in\App(\ga)$; $q$ satisfies (\ref{Ba:3a})
with respect to $p$.
By Lemma \ref{Ba:3}.(b)  there is  $b\in\Si_1(\Kcal^{h}(M))$, with 
$\partial_1b=\partial_1p$, and $\partial_0b=a_0$,
such that $p\sim \overline{b}*q*b$, 
where  $\overline{b}$ is the reverse of $b$. By using the previous lemma,
there is a closed path $q_1$ in $\Kcal^d(M)$, with endpoint $a_0$,  
and $\dc\in\Kcal^{h}(M)$
such that $q_1\sim q$ and $|q_1|\subseteq \dc$. 
As the subposet of $\Kcal^{h}(M)$ formed by
$\dc$  and $\{\dc_1\in\Kcal^{h}(M) \ | \ 
 \dc_1\subset \dc\}$ is directed under inclusion, 
by (\ref{Ba:5}) we have that $q_1\sim b(a_0)$.  
Finally, by (\ref{Ba:1}) and (\ref{Ba:2})  we have
\[
p\sim \overline{b}*q*b\sim \overline{b}*b(a_0)*b\sim b(a_0) \ ,
\]
and this completes the proof.
\end{proof}
%***********************************************
%***********************************************
\subsection{Net cohomology}
\label{Bb}
\paragraph{The category of 1-cocycles:} Consider a net 
of von Neumann algebras over a Hilbert space $\Hil$, 
$\Al_\Po:\Po\ni\dc\rightarrow \Al(\dc) \subseteq \mathfrak{B}(\Hil)$ 
indexed by $\Po$. A \emph{1-cocycle} $z$ of the poset $\Po$, 
with values in $\Al_\Po$, is a field 
\[
 z:\Si_1(\Po)\ni b\rightarrow z(b)\in \Bh
\]
of unitary operators of $\mathfrak{B}(\Hil)$ satisfying the following properties: 
\[
\begin{array}{rll}
(i) & z(\partial_0 c) \cdot z(\partial_2 c) =  z(\partial_1 c)\ , \ \ \  &   c\in\Si_2(\Po) \ ;  \\ 
(ii) & z(b)\in\A(|b|) \ ,  & b\in\Si_1(\Po)  \ . 
\end{array}
\]
The property $(i)$ is  called the \emph{1-cocycle identity}, while $(ii)$ is called 
the \emph{locality condition} for cocycles. 
%\mbox{\emph{locality \ condition}}
%  \mbox{\emph{1-cocycle  identity}} \\
% \emph{1-cocycle identity},
% \[
% z(\partial_0 c) \cdot z(\partial_2 c) =  z(\partial_1 c),  
% \qquad  c\in\Si_2(\Po),
% \]
% and the \emph{locality condition},
% $z(b)\in\A(|b|)$ for any  1-simplex $b\in\Si_1(\Po)$.
Given a pair $z,z_1$ of 1-cocycles an \emph{intertwiner} 
$t$ between $z,z_1$ is a field 
\[
t:\Si_0(\Po)\ni a\rightarrow t_a\in\Bh  
\]
satisfying the following properties:
\[
\begin{array}{rll}
(iii)  &   t_{\partial_0b}\cdot z(b) =  z_1(b)\cdot t_{\partial_1b} \ , \ \ \  &   b\in\Si_1(\Po) \ ;  \\ 
(iv) &   t_a\in\A(a)\ ,  &  a\in\Si_0(\Po)\ . 
\end{array}
\]
$(iii)$ is the \textit{intertwining property} while  $(iv)$ is the \emph{locality condition} for intertwiners.
We will denote by $(z,z_1)$ the set of the intertwiners between $z$ and $z_1$.\\
\indent The \emph{category of 1-cocycles} $\Zcal^1(\Po)$  is the category 
whose objects are 1-cocycles and whose arrows are the corresponding 
set of intertwiners. 
It turns out that this is a $\Crm^*-$category, details 
can be found in \cite{Ruz3}. 
Two 1-cocycles $z,z_1$ are   \emph{equivalent} (or  \emph{cohomologous}) if there exists a
unitary arrow $t\in(z,z_1)$.  A 1-cocycle $z$ is 
\emph{trivial} if it is equivalent to the \emph{identity} cocycle 
$\io$ defined as $\io(b)=\mathbbm{1}$ for any 1-simplex $b$. 
\begin{lemma}
\label{Bb:1}
If $\Al_\Po$ is either irreducible or locally definite,  
then  the identity cocycle $\io$ is irreducible, that is 
$(\io,\io)=\mathbb{C}\cdot \mathbbm{1}$. 
\end{lemma}
\begin{proof}
If $t\in(\io,\io)$, then $t_{\partial_0b}=t_{\partial_1b}$ for any 
1-simplex $b\in\Si_1(\Po)$. This entails that $t$ is a constant field, 
that is, $t_a=t_{a_1}$ for any pair $a,a_1\in\Si_0(\Po)$, because 
$\Po$ is pathwise connected. Hence $t_a\in\A(\dc)$ for any
$\dc\in\Po$. This in turn entails that   $t_a$ belongs both to 
$\cap\{\A(\dc)' \ | \ \dc\in\Kcal^{h}(M) \}$ 
and to
$\cap\{\A(\dc) \ | \ \dc\in\Kcal^{h}(M) \ , \ \ x\in\dc\}$ for any   
$x\in M$.
Hence, the proof follows from the definition of irreducibility 
and local definiteness (see Section \ref{Ac}).
\end{proof}
From now on we will assume that $\Al_\Po$ is irreducible;
hence, by the above lemma, $\io$ is an irreducible object of 
$\Zcal^1(\Po)$.\\  
\indent The evaluation of a 1-cocycle $z\in\Zcal^1(\Po)$ 
on a path $p=\{b_n,\ldots,b_1\}$ in $\Po$ is defined as  
\[
z(p)\defi z(b_n)\cdots z(b_2)\cdot z(b_1).
\]
A  1-cocycle $z$ is said to be \emph{path-independent} if for any 
$a_0,a_1\in\Si_0(\Po)$ we have that 
\[
 z(p)=z(q) \ \ \mbox{ for any pair of paths }  p,q\in\Po(a_0,a_1). 
\]  
It turns out that path-independence 
is equivalent to \emph{triviality in} $\Bh$, namely,  there exists  a field 
$V:\Si_0(\Po)\ni a \rightarrow V_a\in\Bh$ of unitary operators 
such that 
\[
z(b) = V_{\partial_0b}\cdot  V^*_{\partial_1b}, \qquad 
 b\in\Si_1(\Po).
\]
We denote by $\Zcal^1_t(\Po)$  the set of path-independent 
1-cocycles $z$ of $\Po$, and 
with the same symbol  the full $\Crm^*-$subcategory of
$\Zcal^1(\Po)$  whose objects belong to $\Zcal^1_t(\Po)$. 
We will  refer to  $\Zcal^1_t(\Po)$ as 
the category of \emph{path-independent 1-cocycles}.
% %**********************************************************
\paragraph{Changing the index set:}  We now focus on path-independent 
1-cocycles and show  how they behave under a change of the index set. 
Given the poset $\Po$,
consider a subfamily $\Po_1\subseteq \Po$ which forms 
a basis for the topology of $M$. As we have already done in 
Section \ref{Ac}, we obtain  a net of von Neumann algebras
$\Al_{\Po_1}$, indexed by $\Po_1$,  by restricting the 
$\Al_\Po$  to $\Po_1$. 
We want to understand the relation between $\Zcal^1_t(\Po)$ 
and the category $\Zcal^1_t(\Po_1)$ of the path-independent 
1-cocycles of $\Po_1$ with values in $\Al_{\Po_1}$. 
This topic has been analyzed in great generality in \cite{Ruz3} 
in terms of abstract posets. According to this paper, for our aim,
it is enough to observe that the following two properties 
are verified. 
\begin{enumerate}
\item For any $\dc\in\Po$ there exists $\dc_1\in\Po_1$ such that 
      $\dc_1\subseteq \dc$.
\item Given $\dc\in\Po$, for any pair $\dc_1,\dc_2\in\Po_1$ 
      with $\dc_1,\dc_2\subseteq \dc$ there exists a path 
      $p\in\Po_1$ from $\dc_1$ to $\dc_2$ such that 
      $|p|\subseteq \dc$.  
\end{enumerate}
Both of them  derive from the fact that $\Po_1$ 
is a basis for the topology of $M$ (the second, in particular, is also  
a consequence of the fact that the elements of 
$\Po$ are, by assumption, arcwise connected).  
The properties 1 e 2 imply that $\Po_1$ is a locally relatively 
connected refinement of $\Po$ (\cite[Def. 2.9]{Ruz3}). This entails, 
first, that the identity cocycle $\io$ of  $\Zcal^1_t(\Po_1)$ 
is irreducible (\cite[Lemma 2.11]{Ruz3}). Secondly, 
for any $z,z_1\in\Zcal^1_t(\Po)$
and for any $t\in(z,z_1)$, define
\begin{equation}
\label{Bb:1a}
\begin{array}{ll}
 \Rrm(z)(b)\defi z(b) \ , & b\in\Si_1(\Po_1)\ ,\\ 
 \Rrm(t)_a \defi t_a \ , & a\in\Si_0(\Po_1)  \ .
\end{array}
\end{equation}
It can be easily seen  that $\Rrm: \Zcal^1_t(\Po)\rightarrow\Zcal^1_t(\Po_1)$
is a covariant functor (the restriction to $\Po_1$). 
By  \cite[Def. 2.11]{Ruz3} we have the following
\begin{lemma}
\label{Bb:2}
The  categories $\Zcal^1_t(\Po_1)$ and  $\Zcal^1_t(\Po)$
are equivalent, in particular there exists a covariant functor
$\Erm:\Zcal^1_t(\Po_1)\rightarrow \Zcal^1_t(\Po)$ such that 
\[
\Erm\circ \Rrm \simeq \mathrm{id}_{\Zcal^1_t(\Po)}, \ \ \ 
\Rrm\circ \Erm =  \mathrm{id}_{\Zcal^1_t(\Po_1)} \ , 
\]
where the symbol $\simeq$ stands for natural equivalence. 
\end{lemma}
% The relation $\Rrm\circ \Erm =  \mathrm{id}_{\Zcal^1_t(\Po_1)}$
% means that $\Erm$ provides an extension of path-independent 1-cocycles 
% of $\Po_1$ to path-independent 1-cocycles 
% of $\Po$ since  $\Rrm(\Erm(z)) =z$ for any $z\in \Zcal^1_t(\Po_1)$.
% %**********************************************************
%**********************************************************
%**********************************************************
\paragraph{Connection between homotopy and  net cohomology:} 
To begin with we recall some properties of 1-cocycles 
which will be useful also for later applications.  
First,  any  1-cocycle $z$ is  
\emph{invariant for homotopic} paths, i.÷e.
\begin{equation}
\label{Bb:3}
p\sim q \ \ \Rightarrow \ \ z(p) = z(q) \ .
\end{equation}
Secondly, 
\begin{equation}
\label{Bb:4}
\begin{array}{rll}
(a) & z(\overline{p}) =z(p)^* & \mbox{ for any path } p  \ , \\ 
(b) & z(b(a_0)) = \mathbbm{1} & \mbox{ for any 0-simplex } a_0 \ . 
\end{array}
\end{equation}
Following \cite{Ruz3},  by using homotopic invariance 
of 1-cocycles and the relations  (\ref{Bb:4}), 
one can easily deduce the following result.
% Let $a_0$ be  0-simplex. Given  $z$  a 1-cocycle of  $\Zcal^1(\Po)$, 
% define 
% \begin{equation}
% \label{}
% \si_z([p])\defi z(p) \ , \qquad [p]\in\pi_1(\Po,a_0) \ .
% \end{equation}
% This definition is well posed 
% by homotopic invariance of 1-cocycles. Furthermore  
\begin{teo}[\cite{Ruz3}]
\label{Bb:5}
To  any  $z\in\Zcal^1(\Po)$  there corresponds  a unitary representation  
$\si_z$ of the fundamental group $\pi_1(\Po)$  
such that $z$ is path-independent if, and only if, $\si_z$ 
is trivial.
% Let $a_0$ be  0-simplex. Then 
% to any  $z\in\Zcal^1(\Po)$  
% there corresponds  a unitary representation  $\si_z$ of $\pi_1(\Po,a_0)$
% such that $z$ is path-independent if, and only if, $\si_z$ 
% is trivial.
\end{teo}
%************
\indent As a consequence of this result we have that 
if $\Po$ is simply connected then any 1-cocycle of $\Po$ 
is path-independent. Hence by Lemma \ref{Ba:8} we have
\begin{equation}
\label{Bb:6}
  \Zcal^1(\Kcal^{h}(M))=\Zcal^1_t(\Kcal^{h}(M)), \qquad M\in\Loc\ .
\end{equation}
So, any  1-cocycle of the poset $\Kcal^{h}(M)$ is 
path-independent. \\
\indent The relation between the net cohomology of $\Po$ 
and the topology of the underlying spacetime 
is obtained by means of the theorems \ref{Ba:6} and \ref{Bb:5}. 
%*************
\begin{teo}[\cite{Ruz3}]
\label{Bb:7}
Assume that all the elements of $\Po$ 
are simply connected. To any  1-cocycle $z\in\Zcal^1(\Po)$, 
there corresponds a unitary representation 
$\tilde{\si}_z$ of $\pi_1(M)$
such that $z$ is path-independent if, and only if, $\tilde{\si}_z$ 
is trivial.
% Assume that all the elements of $\Po$ 
% are simply connected. Let $x_0\in M$ and let $a_0$ be a 0-simplex, 
% with $x_0\in a_0$. To any  1-cocycle $z\in\Zcal^1(\Po)$, 
% there corresponds a unitary representation 
% $\tilde{\si}_z$ of $\pi_1(M,x_0)$ 
% such that $z$ is path-independent if, and only if, $\tilde{\si}_z$ 
% is trivial.
\end{teo}
%*************
The set $\Kcal^d(M)$ of diamonds of $M$ verifies the hypotheses 
of Theorem \ref{Bb:7}. This allows us to give a 
topological characterization of the set  $\Zcal^1(\Kcal^d(M))$ 
of 1-cocycles of $\Kcal^d(M)$. \\[3pt]
\indent (1) The 1-cocycles  of $\Kcal^d(M)$ which are path-independent
      \emph{do not carry} any information about the topology of 
      the spacetime since they provide trivial representations 
      of the fundamental group. On the other hand,  this type 
      of 1-cocycles has a direct physical interpretation;
      they represent sharply localized sectors of the net 
      of local observables (see Section \ref{1Cad}). The inspection 
      of their charge structure and of their local covariance   
      is the subject of the present paper, 
      and it will be carried out throughout Section \ref{D}.  \\[3pt] 
\indent (2)  The 1-cocycles of $\Kcal^d(M)$ which are path-dependent
      \emph{are of a topological nature} since
      they provide nontrivial representations 
      of the fundamental group of the spacetime $M$ 
      (clearly, if $M$ is nonsimply connected).
      In our opinion, path-dependent 1-cocycles might be 
      charged sectors induced by the nontrivial topology of 
      the spacetime, a phenomenon  predicted  and studied 
      in the literature \cite{FWh, AS, So}. However, 
      until now, there is no interpretation of these
      1-cocycles in terms of superselection sectors, namely 
      representations of the net of local observables. 
      We investigate in  Section \ref{Z} a notion that possibly 
      points towards the correct interpretation.
%*******************************************************
%*******************************************************

%**************************************************
%**************************************************
%**************************************************
\section{Charged superselection sectors}
\label{D}
By charged superselection sectors
in the Minkowski space $\Mk$, it is meant the unitary equivalence classes 
of the irreducible representations of a net of local observables 
which are local excitations of the vacuum representation.
We can distinguish two types of charged sectors  according to the regions 
of the spacetime used as index set of the net of local observables.
Charged sectors of Doplicher-Haag-Roberts type, when one considers double cones of 
$\Mk$, and the charges of Buchholz-Fredenhagen type  associated with  a particular class 
of nonrelatively compact regions like spacelike cones. 
In both cases,  sectors
define a $\Crm^*-$category in which the charge structure 
manifests itself by the existence of a tensor product,
a permutation symmetry, and a conjugation 
(Doplicher-Haag-Roberts analysis \cite{DHR1,DHR2}, and Buchholz-Fredenhagen analysis \cite{BF}). 
Furthermore, it is possible to reconstruct the (unobservable) fields and 
the gauge group  underlying the theory \cite{DR2}.\\[5pt]
\indent Our purpose is to study superselection sectors 
of Doplicher-Haag-Roberts type in
the framework of a locally covariant quantum field theory $\Al$.  
The first step is to introduce the notion of reference state space, 
which will play for the theory the same r\^ole played by the 
vacuum representation. To this aim, it is worth observing that  
the vacuum representation plays the r\^ole 
of a reference representation that singles out charged sectors, and that 
in both cited analysis it is enough to take as vacuum representation one  
satisfying the Borchers property and 
Haag duality\footnote{It is possible to make do with less than the
Borchers property, \cite{Ruz1}. On the contrary, up to now, Haag duality
or a weaker form of it \cite{Rob2}, seems to be an essential requirement
on the vacuum representation for the theory of superselection sectors.} 
\cite{DR2}. 
%************************
\begin{df}  
\label{C:1}
We call  a \textbf{reference state space} for $\Al$ 
a locally quasi-equivalent state space $\St_o$   
satisfying the Borchers property, and
such that for any $M\in\Loc$ there is at least one state 
$\omega\in\St_o(M)$ satisfying punctured Haag duality. 
\end{df}
%*************************
An example of a locally covariant quantum field theory
with a state space verifying the properties of the Definition 
\ref{C:1}, is provided by the Klein-Gordon scalar field 
and by the space of  quasi-free states satisfying the microlocal 
spectrum condition \cite{BFV, Ruz2} (see also in Section \ref{Z}).
We stress that we require the existence for any $M\in\Loc$ of at least one state
$\omega\in\St_o(M)$  satisfying 
punctured Haag duality.  This property seems 
to be the right generalization of Haag duality,  
which apparently seems to deal 
well with the nontrivial topology of arbitrary globally hyperbolic 
spacetimes \cite{Ruz3}. The reason why punctured Haag duality 
is assumed on the net indexed by  $\Kcal^d(M)$, is that 
$\Kcal^{h}(M)$ contains elements which are not simply connected, and 
elements whose causal complement is not arcwise connected. So,
punctured Haag duality (and also Haag duality) might not hold 
for a net indexed  by $\Kcal^{h}(M)$ (see \cite{Ruz2}). 
Finally, recall that with our convention a state $\omega$ satisfying 
punctured Haag duality means that the net $\omega^*\Al_{\Kcal^d(M)}$
is irreducible and satisfies punctured Haag duality. These two properties
entail local definiteness (see Section \ref{Ac}).
Then, by Corollary \ref{Ad:2},  $\St_o$ is locally definite. 
%*************************
\begin{df}
\label{C:2}
The \textbf{charged superselection sectors} of $\Al$, with 
respect to the reference state space $\St_o$,   
are the unitary equivalence classes of the irreducible 
elements of the categories $\Zcal^1_t(\omega, \Kcal^d(M))$
of path-independent 1-cocycles of $\Kcal^d(M)$ with values
in $\omega^*\Al_{\Kcal^d(M)}$,  as $\omega$
varies in $\St_o(M)$ and as $M$ varies in $\Loc$. 
\end{df}
%*************************
Our program for investigating superselection sectors is divided in two parts.
Our \emph{first}  aim is to understand  the charge 
structure of the categories $\Zcal^1_t(\omega, \Kcal^d(M))$ 
on a fixed spacetime background $M$ and  how these categories are related 
as $\omega$ varies in $\St_o(M)$. 
\emph{Secondly}, we will inspect the locally covariant structure of 
sectors. This means
that we will study the connection of sectors associated 
with different isometrically embedded spacetimes backgrounds.\\
\indent Now, some observations concerning the definition of superselection 
sectors in a locally covariant quantum field theory are in order.\\[3pt]
%**************
\indent (1) Our definition of superselection sectors in terms 
of net cohomology is equivalent to the usual one given in terms
of representations of the net of local observables which are 
sharp excitations of a reference representation. In particular,
we will show in Section \ref{1Cad}, that for any spacetime $M$
and for any $\si\in\St_o(M)$, to any  1-cocycle $z\in\Zcal^1_t(\si,\Kcal^d(M))$
there corresponds, up to equivalence, a unique
representation $\pi^z$ of the net of local 
observables which is a sharp excitation of a representation 
$\pi_\omega$ associated with a state $\omega\in\St_o(M)$ satisfying 
punctured Haag duality. \\[3pt]
%*********************************
\indent (2) There are several reasons why we choose to study 1-cocycles 
of the poset $\Kcal^d(M)$ instead of 1-cocycles of $\Kcal^{h}(M)$. 
On the one hand, $\Kcal^d(M)$ reflects the topological and causal 
properties of $M$ better than $\Kcal^{h}(M)$: 
The fundamental group of $\Kcal^d(M)$ is the same as that of $M$
and any diamond  has an  arcwise connected causal 
complement. These two properties have been one of the keys 
of the paper \cite{Ruz3} where, in the Haag-Kastler framework,  
the charge structure of sharply localized sectors in a 
fixed background spacetime has been investigated.  
On the contrary, 
$\Kcal^{h}(M)$ is simply connected irrespective of the topology of $M$
(Proposition \ref{Ba:8}), and it has elements with a 
nonarcwise connected causal complement.
On the other hand, Lemma \ref{Bb:2} states 
that there is no loss of generality in studying 
path-independent 1-cocycles of $\Kcal^d(M)$ instead of those 
of $\Kcal^{h}(M)$, because the corresponding categories are 
equivalent. We have to say, however, that 
the cited result provides an equivalence of $\Crm^*-$categories, but  
ignores the tensorial structure of the categories.
This topic will be analyzed in \cite{BR} where we will provide 
a symmetric tensor equivalence between the categories 
associated with $\Kcal^{d}(M)$ and $\Kcal^{h}(M)$.\\[3pt]
%*********************
%*********************
\indent (3) Since $\St_o$ satisfies the Borchers property,  by a routine
calculation (see \cite{Rob3}), it turns out  that 
the  category  $\Zcal^1_t(\omega, \Kcal^d(M))$  is closed under direct
sums and subobjects for any $\omega\in\St_o(M)$ and any $M\in\Loc$.
As observed above $\St_o$ is locally definite. 
Then, by Lemma \ref{Bb:1}, the identity cocycle of 
$\Zcal^1_t(\omega, \Kcal^d(M))$ is irreducible for any  
$\omega\in\St_o(M)$ and any $M\in\Loc$.
%***********************************************
%***********************************************
%***********************************************
%***********************************************
%**********************************************
%**********************************************
\subsection{Fixed spacetime background} 
\label{Ca}
In the present section we investigate the charge 
structure of superselection sectors in a fixed spacetime background 
$M\in\Loc$. We will start by  noticing that 
for a state $\omega\in\St_o(M)$ satisfying 
punctured Haag duality, the corresponding category 
has a tensor product, a permutation symmetry,  
and the objects with a finite statistics have conjugates. 
Afterwards, we will show that all the constructions can be coherently 
extended to the categories  $\Zcal^1_t(\si,\Kcal^{d}(M))$ 
for any $\si\in\St_o(M)$.  We conclude this section by studying 
the behaviour of these categories under restriction to subregions of $M$. 
%***************************************************************
%***************************************************************
\subsubsection{A preferred reference states}
\label{Cax}
As a starting point of our investigation  we apply  
to our framework  the results of the analysis \cite{Ruz3} of 
sharply localized sectors on a fixed spacetime background $M$, 
carried out in the Haag-Kastler framework.\\[5pt]
\indent Given $M\in\Loc$, let $\omega\in\St_o(M)$. Consider the category 
$\Zcal^1_t(\omega,\Kcal^d(M))$, and for any tuple
$z,z_1,z_2,z_3\in\Zcal^1_t(\omega,\Kcal^d(M))$
and $t\in(z,z_1)$, $s\in(z_2,z_3)$, define
\begin{equation}
\label{Cax:1}
\begin{array}{lcll}
(z\otimes_\omega z_1)(b) &\defi &z(b)\cdot \mathrm{ad}_{z(p)}(z_1(b))\ ,  & 
 b\in\Si_1(\Kcal^d(M))\ ,\\
(t\otimes_\omega s)_a &\defi & t_a\cdot \mathrm{ad}_{z(q)}(s_a)\ ,  & 
 a\in\Si_0(\Kcal^d(M))\ ,
\end{array}
\end{equation}
where $p$ is a path  with $\partial_1p\perp |b|$
and $\partial_0p=\partial_1b$;  $q$ is a path with 
$\partial_1q\perp a$ and $\partial_0p=a$; $\mathrm{ad}_{z(p)}$
denotes the adjoint action. Since the elements 
of $\Kcal^d(M)$ have arcwise connected causal complements,  
path-independence of 1-cocycles implies that 
these definitions do not depend on the choice 
of the paths $p$ and $q$. Now,  
given $z,z_1\in\Zcal^1_t(\omega,\Kcal^d(M))$, 
for any 0-simplex $a$ define 
\begin{equation}
\label{Cax:2}
\eps_\omega(z,z_1)_a \defi z_1(b)^*\cdot 
                       \mathrm{ad}_{z_1(p)}(z(b_1)^*) 
                       \cdot 
                      z(b_1)\cdot \mathrm{ad}_{z(p_1)}(z_1(b)) \ ,
\end{equation}
where $b_1,b$ are 1-simplices such that 
$\partial_0b_1\perp\partial_0b$
and $\partial_1b_1=\partial_1b=a$;   
$p$ is a path from the causal complement
of $|b_1|$ to $\partial_1b$; $p_1$ is a path from the causal
complement of $|b|$ to $\partial_0b_1$. This
expression is independent of the choices of 
$b_1,b,p_1,p$. We will refer to the pair $(\otimes_\omega,\eps_\omega)$ 
as \emph{the tensor structure of}  $\Zcal^1_t(\omega,\Kcal^d(M))$,
although, up until now, we cannot affirm either that  
$\otimes_\omega$ is a tensor product or that $\eps_\omega$ 
is a permutation symmetry.  One can easily see, for instance, 
that $z\otimes_\omega z_1$ satisfies the 1-cocycle identity, 
but it is not clear if it satisfies the locality condition
nor if it is path-independent. However, for a particular 
choice of $\omega$ we have the following result 
%************************
\begin{teo}[\cite{Ruz3}]
\label{Cax:3}
Let $\omega\in\St_o(M)$ satisfy  punctured Haag duality. Then  
relations (\ref{Cax:1}) and (\ref{Cax:2}) define, 
respectively, a tensor product $\otimes_\omega$ and a permutation 
symmetry $\eps_\omega$
of $\Zcal^1_t(\omega,\Kcal^d(M))$;  the category has left inverses and 
a notion of statistics of objects;
the objects with finite statistics have conjugates.
\end{teo}
%************************
By using this result, in the next two sections we will be able to prove 
that $(\otimes_\si,\eps_\si)$ define  a tensor product and a permutation 
symmetry of $\Zcal^1_t(\si,\Kcal^d(M))$ for any $\si\in\St_o(M)$.
Thus,  the existence, for any $M\in\Loc$,   
of at least one state  $\omega\in\St_o(M)$  satisfying punctured 
Haag duality is a corner stone for our analysis. 
\begin{oss}
\label{Cax:4}
It is worth observing that 
it is possible to choose the paths $b,b_1,p,p_1$ 
involved in the definition (\ref{Cax:2}) in such a way that all their supports 
are contained in a unique diamond $\dc$. In particular it is possible 
to replace the paths $p$ and $p_1$ by two 1-simplices 
$\tilde{b}$ and $\tilde{b}_1$ which have the same 
endpoints of $p$ and $p_1$, respectively, and whose support is $\dc$.   
This is an easy  consequence 
of the definition of diamonds and of the property (\ref{Aa:7a}). 
The same holds true for the paths involved in the definition (\ref{Cax:1})
\end{oss}
%***************************************************************
%***************************************************************
%***************************************************************
\subsubsection{Independence of the choice of states I}

\label{Caa}
The aim of this section is to show that for any pair
$\si,\omega\in\St_o(M)$ the corresponding categories 
$\Zcal^1_t(\omega,\Kcal^d(M))$ and $\Zcal^1_t(\si,\Kcal^d(M))$ are 
$^*-$isomorphic. Let  
\[
\rho_{\omega,\si}:\omega^*\Al_{\Kcal^{h}(M)} \longrightarrow
\si^*\Al_{\Kcal^{h}(M)},
\]
be the net isomorphism defined in Section \ref{Ad} (see Lemma \ref{Ad:1}). 
We stress here that despite considering the categories 
associated with the set $\Kcal^d(M)$, the fact that $\rho_{\omega,\si}$
is a net isomorphism of the nets indexed by  $\Kcal^{h}(M)$ is of a crucial 
importance when stating the claimed isomorphism 
(see proof of the Lemma \ref{Caa:2}). 
Now, for any pair  $z,z_1\in\Zcal^1_t(\omega,\Kcal^d(M))$ and $t\in(z,z_1)$ 
define
\begin{equation}
\label{Caa:1}
 \begin{array}{lcll}
 \Frm_{\omega,\si}(z)(b) & \defi & \rho_{|b|}(z(b)) \ , & 
   b\in\Si_1(\Kcal^d(M)) \ , \\
  \Frm_{\omega,\si}(t)_a & \defi  & \rho_{a}(t_a)  \ , & 
   a\in\Si_0(\Kcal^d(M)) \ .
\end{array}
\end{equation}
Clearly, $\Frm_{\omega,\si}(z)(b)\in \A_\si(|b|)$ and 
$\Frm_{\omega,\si}(t)_a\in\A_\si(a)$. For any $c\in\Si_2(\Kcal^d(M))$ 
we have that 
\begin{align*}
 \Frm_{\omega,\si}(z)(\partial_0c)\cdot  & \Frm_{\omega,\si}(z)(\partial_2c)  = \\
& = \rho_{|\partial_0c|}(z(\partial_0c))\cdot 
        \rho_{|\partial_2c|}(z(\partial_2c))  = 
\rho_{|c|}(z(\partial_0c))\cdot 
        \rho_{|c|}(z(\partial_2c)) \\
 & =  \rho_{|c|}(z(\partial_0c)\cdot z(\partial_2c)) 
   = \rho_{|\partial_1c|}(z(\partial_1c)) \\ 
 & =  \Frm_{\omega,\si}(z)(\partial_1c)\ . 
\end{align*}
Hence $\Frm_{\omega,\si}(z)$ is a 1-cocycle 
of $\Kcal^d(M)$. Observe that if $M$ is simply connected, 
then, by Theorem \ref{Bb:7},  $\Frm_{\omega,\si}(z)$ is a path-independent 
1-cocycle. For the general case we have the following 
%*****************
\begin{lemma}
\label{Caa:2}
For any $M\in\Loc$ then $\Frm_{\omega,\si}(z)$ is a path-independent 1-cocycle of 
$\Zcal^1_t(\si,\Kcal^d(M))$.
\end{lemma}
\begin{proof}
Let $p$ be a closed path of $\Kcal^d(M)$ with endpoint $a_0$. 
By Lemma \ref{Ba:7}, there exists a closed path $q$ of $\Kcal^d(M)$,
with endpoint $a_0$, and an element $\dc\in\Kcal^{h}(M)$ such that 
$p\sim q$ and $|q|\subseteq\dc$. Assume $q=\{b_n,\ldots,b_1\}$. 
By  homotopic invariance of 1-cocycles  we have 
\begin{align*}
 \Frm_{\omega,\si}(z)(p) &= 
  \Frm_{\omega,\si}(z)(q) 
    = \rho_{|b_n|}(z(b_n))\cdots 
  \rho_{|b_1|}(z(b_1))\\
  &  = \rho_{\dc}(z(b_n))\cdots 
  \rho_{\dc}(z(b_1))
   =  \rho_{\dc}(z(b_n)\cdots 
       z(b_1))\\
 & =\rho_{\dc}(z(q)) = \rho_{\dc}(\mathbbm{1}) = \mathbbm{1} \ ,    
\end{align*}
where the path-independence of $z$ has been used. 
\end{proof}
It is clear that 
$\Frm_{\omega,\si}(t)\in(\Frm_{\omega,\si}(z),\Frm_{\omega,\si}(z_1))$. 
Moreover, since $\rho_{\omega,\si}$ is net isomorphism, 
$\Frm_{\omega,\si}$ is an isomorphism. Indeed, given 
the functor $\Frm_{\si,\omega}$ which is associated with the net isomorphism
$\rho_{\si,\omega}$, one can easily see that 
\[
 \Frm_{\si,\omega}\circ \Frm_{\omega,\si} = 
\mathrm{id}_{\Z^1_t(\omega,\Kcal^d(M))} \ , \ \ \  
\Frm_{\omega,\si}\circ \Frm_{\si, \omega} = 
\mathrm{id}_{\Z^1_t(\si,\Kcal^d(M))} \ .
\]
In conclusion we have the following
%*******************
\begin{prop}
\label{Caa:3}
$\Frm_{\omega,\si}: \Z^1_t(\omega,\Kcal^d(M))\rightarrow 
\Z^1_t(\si,\Kcal^d(M))$ is an  isomorphism of $\Crm^*$-categories
for any pair $\omega,\si\in\St_o(M)$.
\end{prop}
We will refer to the functor $\Frm_{\omega,\si}$ as the \emph{flip functor}.
%************%************%************%************%*****
%************%************%************%************%*****
%************%************%************%************%*****
\subsubsection{Independence of the choice of states II} 
\label{Cab}
The aim of this section is to show that the superselection sectors 
of the category $\Zcal^1_t(\si,\Kcal^d(M))$, 
for any $\si\in\St_o(M)$, have a charge structure,  
and that all these categories carry  the same physical 
information. In other words we want to show that 
for any $\si\in\St_o(M)$ the category 
$\Zcal^1_t(\si,\Kcal^d(M))$ has a tensor product, a permutation symmetry, 
and that the objects with finite statistics 
have conjugates; furthermore, that all these categories are 
symmetric tensor $^*-$isomorphic.  
To this end, we will first note that, thanks to the flip functor and 
Theorem \ref{Cax:3}, 
the category associated with a state satisfying punctured Haag duality,
induces a tensor structure  on  
$\Zcal^1_t(\si,\Kcal^d(M))$ for any $\si\in\St_o(M)$.
Secondly,  we will show that the induced structure 
coincide with the ambient one, defined by 
(\ref{Cax:1}) and (\ref{Cax:2}). This, in turn,  
will entail that the flip functor
is a symmetric tensor $^*-$isomorphism.\\[5pt]
%*****************************************
\indent Consider a state $\omega\in\St_o(M)$ satisfying  
punctured Haag duality. Recall that by Theorem \ref{Cax:3}, 
$\otimes_\omega$ and $\eps_\omega$  are, respectively, 
a tensor product and a permutation symmetry of 
$\Zcal^1_t(\omega,\Kcal^d(M))$. Now, 
given  $\si\in\St_o(M)$, for any $z,z\in \Zcal^1_t(\si,\Kcal^d(M))$
and for any pair of arrows $t,s$ of the category 
$\Zcal^1_t(\si,\Kcal^d(M))$, define
\begin{equation}
\label{Cab:1a}
\begin{array}{lcl}
 z \otimes^\omega_\si z_1 & \defi & 
  \Frm_{\omega,\si}\big(\Frm_{\si,\omega}(z)\otimes_\omega 
  \Frm_{\si,\omega}(z_1)\big) \ , \\
 t \otimes^\omega_\si s  & \defi & 
  \Frm_{\omega,\si}\big(\Frm_{\si,\omega}(t)\otimes_\omega 
  \Frm_{\si,\omega}(s)\big) \ , 
\end{array}
\end{equation}
and 
\begin{equation}
\label{Cab:1b}
\eps^\omega_\si(z,z_1) \defi \Frm_{\omega,\si}(
           \eps_\omega(\Frm_{\si,\omega}(z),\Frm_{\si,\omega}(z_1))) \ .
\end{equation}
Since the flip functor is an $^*-$isomorphism
the above formulas define, respectively, a tensor product 
and a permutation symmetry of 
$\Zcal^1_t(\si,\Kcal^d(M))$. We will refer  to the pair  
$(\otimes_\si^\omega, \eps_\si^\omega)$ as \emph{the tensor structure 
of $\Zcal^1_t(\si,\Kcal^d(M))$ induced by $\omega$}. 
%**************************
\begin{lemma}
\label{Cab:2}
Let $\omega\in\St_o(M)$ satisfy  punctured Haag duality. 
Then, the following assertions hold for any $\si\in\St_o(M)$.\\
(a) $\otimes^\omega_\si = \otimes_\si$  and  $\eps^\omega_\si =\eps_\si$; \\
(b) The pair $(\otimes_\si,\eps_\si)$ define a tensor product 
    and a permutation 
    symmetry of the category $\Zcal^1_t(\si,\Kcal^d(M))$.
\end{lemma}
\begin{proof}
(a) Given a 1-simplex $b\in\Si_1(\Kcal^d(M))$, by Remark \ref{Cax:4} 
there is a 1-simplex $b_1$ and a diamond $\dc$ such that 
$\partial_0b_1=\partial_1b$, 
$\partial_0b_1\perp |b|$,  and $|b|,|b_1|\subseteq \dc$.
Now recalling the definition 
(\ref{Cax:1}), we have that
\begin{align*}
\Frm_{\si,\omega}(z)\otimes_\omega 
  \Frm_{\si,\omega}(z_1) (b)& = 
   \Frm_{\si,\omega}(z)(b)\cdot \mathrm{ad}_{\Frm_{\si,\omega}(z)(b_1)}\big(
  \Frm_{\si,\omega}(z_1)(b))\big) \\
& = 
  \rho_{|b|}(z(b))\cdot \mathrm{ad}_{\rho_{|b_1|}(z(b_1))}\big((
  \rho_{|b|}(z_1(b))\big) \\
& = \rho_{\dc}(z(b))\cdot \mathrm{ad}_{\rho_{\dc}(z(b_1))}\big(
     \rho_{\dc}(z_1(b))\big) \\
& = \rho_{\dc}\big(z(b)\cdot \mathrm{ad}_{z(b_1)}(z_1(b))\big)\ . 
\end{align*}
Using this equation and the fact that 
$\Frm_{\si,\omega}(z)\otimes_\omega \Frm_{\si,\omega}(z_1) (b)\in
       \A_\omega(|b|)$, we have that 
\begin{align*}
(z\otimes^\omega_\si  & z_1)(b)  = \\ 
& = \Frm_{\omega,\si} \big(\Frm_{\si,\omega}(z)\otimes_\omega 
  \Frm_{\si,\omega}(z_1)\big) (b)  = \rho^{-1}_{|b|}
  \big(\Frm_{\si,\omega}(z)\otimes_\omega 
  \Frm_{\si,\omega}(z_1) (b)\big) \\
 &  = 
 \rho^{-1}_{\dc}
  \big(\Frm_{\si,\omega}(z)\otimes_\omega 
  \Frm_{\si,\omega}(z_1) (b)\big) =  \rho^{-1}_{\dc}\big( 
  \rho_{\dc}\big(z(b)\cdot \mathrm{ad}_{z(b_1)}(z_1(b))\big)\big) \\
 & = z(b)\cdot \mathrm{ad}_{z(b_1)}(z_1(b)) \\
 &  = (z\otimes_\si z_1)(b)\ . 
\end{align*}
The same argument leads to the identity
$t\otimes^\omega_\si s =  t\otimes_\si s$ for any pair of arrows 
$t,s$ of $\Zcal^1_t(\si,\Kcal^d(M))$. So we have 
$\otimes^\omega_\si=\otimes_\si$. We now  apply a similar 
argument to the permutation symmetry $\eps^\omega_\si$ 
(see the definition (\ref{Cax:2})). 
Consider a 0-simplex $a$.
By the Remark \ref{Cax:4},  there are 
four 1-simplices $b,b_1,\tilde{b},\tilde{b}_1$ which fulfil,
with respect to $a$,  the properties 
of the definition (\ref{Cax:2}), and such that all the supports
$|b|,|b_1|,|\tilde{b}|,|\tilde{b}_1|$ are contained in a diamond $\dc$. So 
\begin{align*}
\eps_\omega(\Frm_{\si,\omega}(z),& \Frm_{\si,\omega}(z_1))_a  = \\
& =   \Frm_{\si,\omega}(z_1)(b_1)^*\cdot 
  \mathrm{ad}_{\Frm_{\si,\omega}(z_1)(\tilde{b}_1)}
   (\Frm_{\si,\omega}(z)(b)^*) \ \cdot \\
& \phantom{aaaaaaaaaaaa}  \cdot \ 
    \Frm_{\si,\omega}(z)(b)\cdot 
      \mathrm{ad}_{\Frm_{\si,\omega}(z)(\tilde{b})}
    (\Frm_{\si,\omega}(z_1)(b_1)) \\
& =   \rho_{|b_1|}(z_1(b_1)^*)\cdot 
  \mathrm{ad}_{\rho_{|\tilde{b}_1|}(z_1(\tilde{b}_1))}
   (\rho_{|b|}(z(b)^*) \ \cdot \\
& \phantom{aaaaaaaaaaaa}  \cdot \ 
      \rho_{|b|}(z(b))\cdot 
      \mathrm{ad}_{\rho_{|\tilde{b}|}(z(\tilde{b}))}
     \cdot (\rho_{|b_1|}(z_1(b_1)) \\
& =   \rho_{\dc}\big(z_1(b_1)^*\cdot 
  \mathrm{ad}_{z_1(\tilde{b}_1)}
   (z(b)^*) \big) \cdot 
      \rho_{\dc}\big( z(b)\cdot 
      \mathrm{ad}_{z(\tilde{b})}
     \cdot z_1(b_1)\big) \\
& =   \rho_{\dc}\big( z_1(b_1)^*\cdot 
  \mathrm{ad}_{z_1(\tilde{b}_1)}
   (z(b)^*) \cdot  z(b)\cdot 
      \mathrm{ad}_{z(\tilde{b})}
     \cdot z_1(b_1)\big) \\
& =   \rho_{\dc}(\eps_\si(z,z_1)_a )\ . 
\end{align*}
Using this equation and the fact that 
$\eps_\omega(\Frm_{\si,\omega}(z),\Frm_{\si,\omega}(z_1))_a\in\A_\omega(a)$
we have
\begin{align*}
\eps^\omega_\si(z,z_1)_a & = 
\Frm_{\omega,\si}(\eps_\omega(\Frm_{\si,\omega}(z), 
\Frm_{\si,\omega}(z_1)))_a  
 =  \rho^{-1}_{a}(\eps_\omega(\Frm_{\si,\omega}(z), 
                                 \Frm_{\si,\omega}(z_1))_a ) \\
& =  \rho^{-1}_{\dc}(\eps_\omega(\Frm_{\si,\omega}(z), 
                                 \Frm_{\si,\omega}(z_1))_a )
 = \rho^{-1}_{\dc}(\rho_{\dc}(\eps_\si(z,z_1)_a )) \\ 
& = 
\eps_\si(z,z_1)_a\ ,  
\end{align*}
which completes the proof. (b) follows from (a) 
\end{proof}
%**********************************************
The following relations are a consequence of 
the previous lemma and of the invertibility 
of the flip functor.
Consider $\omega,\si\in\St_o(M)$  such that $\omega$ 
satisfies   punctured Haag duality. For any 
$z,z_1\in \Zcal^1_t(\si,\Kcal^d(M))$ and for any 
pair $t,s$ of arrows of $\Zcal^1_t(\si,\Kcal^d(M))$, we have 
\begin{equation}
\label{Cab:2a}
\begin{array}{l}
\Frm_{\si,\omega}(z)\otimes_\omega 
  \Frm_{\si,\omega}(z_1) = \Frm_{\si,\omega}(z\otimes_{\si} z_1) \ , \\ 
\Frm_{\si,\omega}(t)\otimes_\omega 
  \Frm_{\si,\omega}(s) = \Frm_{\si,\omega}(t\otimes_{\si} s) \ ,  
\end{array}
\end{equation} 
and 
\begin{equation}
\label{Cab:2b}
\eps_\omega(\Frm_{\si,\omega}(z),\Frm_{\si,\omega}(z_1)) =  
\Frm_{\si,\omega}(\eps_\si(z,z_1))\ . 
\end{equation} 
In conclusion we have the following 
%*********************
\begin{teo}
\label{Cab:3}
For any  $\omega\in\St_o(M)$ the category
$\Zcal^1_t(\omega, \Kcal^d(M))$ equipped,  with 
$\otimes_\omega$ and $\eps_\omega$,  
is a symmetric tensor $\Crm^*-$category with left inverses. 
Any object with finite statistics have  conjugates.  
For any $\si\in\St_o(M)$, 
\[
\Frm_{\omega,\si}: \Zcal^1_t(\omega,\Kcal^d(M))
\rightarrow \Zcal^1_t(\si, \Kcal^d(M))
\]
is a covariant symmetric tensor  $^*-$isomorphism.
\end{teo}
\begin{proof}
Consider a pair of states $\omega,\si\in\St_o(M)$. In order to 
prove that $\Frm_{\omega,\si}: \Zcal^1_t(\omega,\Kcal^d(M))
\rightarrow \Zcal^1_t(\si, \Kcal^d(M))$ is a symmetric tensor 
$^*-$isomorphism, we consider a state $\omega_o\in\St_o(M)$  
satisfying punctured Haag duality
and use the fact that the tensor structures of 
$\Zcal^1_t(\omega,\Kcal^d(M))$ and 
$\Zcal^1_t(\si, \Kcal^d(M))$ are equal to the tensor structure induced by 
$\omega_o$. Given $z,z_1\in \Zcal^1_t(\omega,\Kcal^d(M))$, 
we have 
\begin{align*} 
\Frm_{\omega,\si}(z\otimes_\omega  z_1)  & = 
\Frm_{\omega,\si}(z\otimes_\omega^{\omega_o}  z_1)  \\
& = \Frm_{\omega,\si} \circ \Frm_{\omega_o,\omega} 
\big(\Frm_{\omega,\omega_o}(z)\otimes_{\omega_o}  
 \Frm_{\omega,\omega_o}(z_1)\big)  \\
& = \Frm_{\omega_o,\si}
\big(\Frm_{\omega,\omega_o}(z)\otimes_{\omega_o}  
 \Frm_{\omega,\omega_o}(z_1)\big)  \\
& = \Frm_{\omega_o,\si}
\big( \Frm_{\si,\omega_o}\circ\Frm_{\omega,\si}(z)\otimes_{\omega_o}  
      \Frm_{\si,\omega_o}\circ\Frm_{\omega,\si}(z_1)\big)  \\
\mbox{by } (\ref{Cab:2a}) \ \  & = \Frm_{\omega_o,\si}\circ \Frm_{\si,\omega_o}
\big(\Frm_{\omega,\si}(z)\otimes_{\si}  
     \Frm_{\omega,\si}(z_1)\big)  \\
& =  \Frm_{\omega,\si}(z)\otimes_{\si}  
     \Frm_{\omega,\si}(z_1) \ . 
\end{align*} 
The same reasoning applied to arrows and to the permutation symmetry 
shows that the flip functor is a covariant symmetric tensor 
$^*-$isomorphism. As 
the category associated with a state
satisfying punctured Haag duality, has left inverses and the object 
with finite statistics have conjugates, the same holds 
for the category associated with any state of $\St_o(M)$. 
This is because the flip functor is a covariant symmetric tensor 
$^*-$isomorphism (see in Appendix).
\end{proof}
%********************************************
%********************************************
%********************************************
%********************************************
\subsubsection{Restriction to  subregions}
\label{Cac}
Let $N\subset M$ be an open arcwise connected subset of $M$, such that
for any pair $x_1,x_2\in N$ then $\J^+(x_1)\cap \J^-(x_2)$ is contained in $N$. This property says that $N$ is a globally 
hyperbolic spacetime. As $N$ is isometrically embedded in 
$M$ and as diamonds are stable under isometric embeddings (Lemma \ref{Aa:9}) 
we have 
\[
 \Kcal^d(M)\nrestr_N \defi \{ \dc\in\Kcal^d(M) \ | \ \overline{\dc}\subset N \} =
 \Kcal^d(N)\ . 
\]
Let $\Al_{\Kcal^d(N)}$ be the net of local algebras indexed by $\Kcal^d(N)$,
obtained by restricting $\Al_{\Kcal^d(M)}$ to $\Kcal^d(N)$.
Let  $\omega\in\St_o(M)$, $\omega^*\Al_{\Kcal^d(N)}$ inherits from 
$\omega^*\Al_{\Kcal^d(M)}$ the Borchers property and local definiteness, as follows from Lemma \ref{Ac:1}.
However, it needs not be irreducible.  
Let $\Zcal^1_t(\omega, \Kcal^d(N))$ be the category of
path-independent 1-cocycles of $\Kcal^d(N)$ with values in 
$\omega^*\Al_{\Kcal^d(N)}$. This is a $\Crm^*-$category closed under
direct sums and subobjects, and, by local definiteness, the identity 
cocycle is irreducible. The aim now is to show that 
the restriction functor 
$\Rrm: \Zcal^1_t(\omega,\Kcal^d(M))\rightarrow\Zcal^1_t(\omega, \Kcal^d(N))$,
defined by (\ref{Bb:1a}),
is a full and faithful covariant 
$^*-$functor.\\[5pt]
\indent Now, let $\omega\in\St_o(M)$ and recall  that 
the restriction functor (\ref{Bb:1a}) is defined, 
for any $z,z_1\in\Zcal^1_t(\omega, \Kcal^d(M))$ and $t\in(z,z_1)$, by
\[
\begin{array}{ll}
 \Rrm(z)(b)\defi z(b)\ ,  & b\in\Si_1(\Kcal^d(N)) \ , \\
 \Rrm(t)(a)\defi t_a \ ,  & a\in\Si_0(\Kcal^d(N))  \ .\\
\end{array}
\]
$\Rrm$ is a covariant $^*-$functor  from $\Zcal^1_t(\omega,\Kcal^d(M))$ into 
                  $\Zcal^1_t(\omega,\Kcal^d(N))$.
Note that if we take $\si\in\St_o(M)$,
then it can be easily shown that the following diagram is commutative
\[
\xymatrix{
  \Z^1_t(\omega,\Kcal^d(M)) \ar[d]_{\Rrm}
  \ar[r]^{\Frm_{\omega,\si}} & \Z^1_t(\si,\Kcal^d(M)) 
   \ar[d]^{\Rrm}\\
  \Z^1_t(\omega,\Kcal^d(N)) \ar[r]^{\Frm_{\omega,\si}}
   & \Z^1_t(\si,\Kcal^d(N))\ }
\]
Therefore if we prove that $\Rrm$ is full an faithful
for a particular choice of $\omega$ then it would be full and faithful 
for any other element $\si\in\St_o(M)$.
%**************** 
\begin{teo}
\label{Cac:1}
$\Rrm$ is a full and faithful $^*-$functor.
\end{teo}
\begin{proof}
In the first part of the proof we follow \cite[Theorem 30.2]{Rob3}
As observed above it is enough to prove the assertion 
when $\omega$  satisfies punctured Haag duality. 
Given $z,z_1\in \Zcal^1_t(\omega, \Kcal^d(M))$,  let 
$t\in (\Rrm(z),\Rrm(z_1))$. This means that 
\[
t_{\partial_0b}\cdot z(b) = z_1(b)\cdot t_{\partial_1b} \qquad b\in
\Si_1(\Kcal^d(N))\ .
\]
We want prove that there exists $\hat{t}\in(z,z_1)$ such that 
$\hat{t}_a= t_a$ whenever $a\in\Si_0(\Kcal^d(N))$. 
Fix $a_0\in\Kcal^d(N)$, define 
\[
 \hat{t}_a\defi z_1(p_a)\cdot t_{a_0}\cdot z(p_a)^* \qquad 
 a\in\Si_0(\Kcal^d(M))\ ,
\]
where $p_a$ is a path in $\Kcal^d(M)$ from $a_0$ to $a$. 
This definition does not depend on the 
chosen $a_0\in\Si_0(\Kcal^d(N))$ nor on the chosen path $p_a$. 
Moreover 
$\hat{t}_{\partial_0b}\cdot z(b) = z_1(b)\cdot \hat{t}_{\partial_1b}$
for any  $b\in\Si_1(\Kcal^d(M))$, and  
\[
 \hat{t}_{a} = t_{a} \qquad a\in\Si_0(\Kcal^d(N))\ .
\]
What remains to be shown is that $\hat{t}$ satisfies the locality 
condition, i.e. $\hat{t}_{a}\in\A_{\omega}(a)$ for any
$a\in\Si_0(\Kcal^d(M))$.  From now on the proof 
is very similar to  the proof of \cite[Proposition 4.19]{Ruz3}. 
Let $x_0\in N$, we prove that $\hat{t}_a\in\A_{\omega}(a)$  for any
$a\in\Si_0(\Kcal^d(M))$ 
whose closure $cl(a)$ is causally disjoint from $\{x_0\}$.
Fix  a 0-simplex $a_1$  of $\Kcal^d(M)$ to be such that 
$cl(a_1)\perp \{x_0\}$ and $a_1\perp a$. Recall  that the
definition of $\hat{t}$ does not depend both on the choice of $a_0$
and on the choice of the path.  
First observe that we can always find 
$a_0\in\Si_0(\Kcal^d(N))$ such that 
$a_0\perp a_1$ and $cl(a_0)\perp \{x_0\}$. Furthermore, 
since the causal complement of $a_1$
is arcwise connected,  there is a path $p_a$
which lies in the causal complement of $a_1$. Therefore:
\[
 \hat{t}_{a}\cdot A =  
 z_1(p_a)\cdot t_{a_0}\cdot z(p_a)^*\cdot A = 
 z_1(p_a)\cdot t_{a_0}\cdot  A\cdot z(p_a)^*=
 A\cdot  \hat{t}_{a}\ ,
\]
for any $A\in\A_{\omega}(a_1)$. Hence $\hat{t}_a\in \A_{\omega}(a_1)'$ for any 
$a_1\perp a$ and $cl(a_1)\perp x$. By punctured Haag duality 
$\hat{t}_a\in\A_{\omega}(a)$. Thus, we have shown 
that $\hat{t}_a\in\A_{\omega}(a)$ for any 0-simplex $a$ such that $cl(a)\perp \{x_0\}$.
By \cite[Proposition 4.19]{Ruz3} the proof follows.
\end{proof}
%************************
%************************
% \begin{prop}
% \label{Cac:2}
% $\Rrm$ is, up to equivalence, injective. 
% \end{prop}
% \begin{proof}
% Let  $\omega\in\St_o(M)$ be such that $\omega^*\Al_{\Kcal^d(M)}$
% is irreducible and satisfies punctured Haag duality.
% Let $z,z_1\in\Z^1_t(\omega, \Kcal^d(M))$ be such that 
% $\Rrm(z)(b) = \Rrm(z_1)(b)$ for any  $b\in\Si_1(\Kcal^d(N))$.
% Fix $a_0\in \Si_0(\Kcal^d(N))$, and define 
% \[
%  t_a \defi z(p_a)\cdot z_1(p_a)^*  \qquad a\in\Si_0(\Kcal^d(M))\ ,
% \]
% where $p_a$ is a path from $a_0$ to $a$. The definition 
% is independent of the choice of $p_a$ and $a_0$. Moreover
% $t_a=\mathbbm{1}$ whenever  $a\in\Si_0(\Kcal^d(N))$, and 
% $t_{\partial_0b}\cdot z_1(b) = z(b)\cdot t_{\partial_1b}$ 
% for any 1-simplex $b\in\Si_1(\Kcal^d(M))$.
% As in the proof of the previous lemma, 
% $t$ satisfies the locality condition. Hence $t$ is a unitary 
% arrow.
% \end{proof}
%*********************************************************
Two comments about Theorem \ref{Cac:1}  are in order.\\[3pt]       
\indent (1) This is a key result for our aims. It will entail that   
the embedding of a sector 
into a different spacetime preserves the statistical properties 
(see Remark \ref{Cbb:5}), this being the genesis of  
the local covariance of gauge groups (see Section \ref{Cbc}).\\[3pt]
\indent (2) Theorem \ref{Cac:1} is nothing but 
the cohomological version  of the ``equivalence 
between local and global intertwiners,'' a property  
that the superselection sectors which are preserved in the scaling limit
fulfills \cite{DMV} (see also \cite{Rob1}). We emphasize 
that in the present paper this equivalence arises as   a natural
consequence of punctured Haag duality.\\[3pt]
\indent (3) It can be easily shown that the restriction 
functor $\Rrm$ is a symmetric tensor functor. The proof is
contained, implicitely, in the proof of Proposition \ref{Cba:2}. 
%*********************************************************
%*********************************************************
%*********************************************************
\subsection{Net cohomology and sharply localized sectors}
\label{1Cad}
The purpose of this section  is to provide the interpretation 
of our definition of superselection sectors 
in terms of representations of the net of local 
observables which are sharply localized with respect 
to a reference representation.\\[5pt] 
\indent A representation $\pi$ on a Hilbert space 
$\Hil_\pi$ of the net $\Al_{\Kcal^d(M)}$ 
is  a collection $\{\pi_\dc\}_{\dc\in\Kcal^d(M)}$ of 
representations $\pi_\dc$ of the algebras $\A(\dc)$ on $\Hil_\pi$, 
which is compatible with the net structure, i.e. 
$\pi_{\dc_1}\nrestr_{\A(\dc)}=\pi_\dc$ if $\dc\subseteq \dc_1$.     
Given $\omega\in\St_o(M)$ let $\pi_\omega$ be the GNS representation 
of the algebra $\Al(M)$, on the Hilbert space $\Hil_\omega$, which 
is associated with $\omega$.  
A representation 
$\pi=\{\pi_\dc\}_{\dc\in\Kcal^d(M)}$ 
is a \emph{sharp excitation} of  $\pi_\omega$ if there exists 
a family  $\{V_\dc\}_{\dc\in\Kcal^d(M)}$ of unitary 
operators from $\Hil_\pi$ onto $\Hil_\omega$ such that 
\begin{equation}
\label{1Cad:1}
 V_{\dc_1}\pi_\dc(A)=\pi_\omega(A) V_{\dc_1} \qquad A\in\A(\dc)\ , \ 
 \dc\perp\dc_1\ .  
\end{equation}
Let  $(\pi,\pi_1)$ be the collection of linear bounded 
operators $T:\Hil_\pi\rightarrow \Hil_{\pi_1}$ such that 
$T\pi_\dc(A) = {\pi_1}_\dc(A)T$ for $A\in\A(\dc)$ and for
$\dc\in\Kcal^d(M)$.  We denote by $\Rep(\omega)$ the category 
whose objects are those representations of $\Al_{\Kcal^d(M)}$ which are a local 
excitation of $\pi_\omega$ and with arrows the corresponding
intertwiner operators. 
%**************
\begin{prop}
\label{1Cad:2}
For any $\si\in\St_o(M)$, the category $\Zcal^1_t(\si,\Kcal^d(M))$
is equivalent to the category $\Rep(\omega)$  for any 
$\omega\in\St_o(M)$ satisfying punctured Haag duality. 
\end{prop}
\begin{proof}
We first prove that $\Zcal^1_t(\omega,\Kcal^d(M))$
is equivalent to the category $\Rep(\omega)$ for 
a state $\omega$ as in the statement. We give only a sketch  
of the  the proof  because 
is very similar to  the proof
of \cite[Lemma 3A.6]{GLRV}. 
Consider $z\in\Zcal^1_t(\omega,\Kcal^d(M))$.  Fix  
$\dc_1\in\Kcal^d(M)$ and define
\[
 \pi^z_\dc(A)\defi  z(p_{\dc_1})\cdot \pi_{\omega}(A)\cdot 
  z(p_{\dc_1})^*\qquad A\in\A(\dc)\ ,
\]
where $p_{\dc_1}$ is a path with $\partial_0p_{\dc_1} = \dc_1$ 
and $\partial_1p_{\dc_1} \perp \dc$. This definition is well posed 
as diamonds have arcwise connected causal complements, and  it 
can be easily shown 
that  $\pi^z$ is a representation of $\Al_{\Kcal^d(M)}$. 
For any $\dc$ let $q_{\dc}$ be a path from 
$\dc_1$ to $\dc$, and let $V_{\dc}\defi z(q_{\dc})$. One can easily check 
that $V_{\dc}\pi_{\dc_2}(A)  =\pi_{\omega}(A)V_\dc$, 
for any $A\in\A(\dc_2)$ with $\dc_2\perp \dc$. Hence  $\pi^z\in
\Rep(\omega)$. Conversely,  given $\pi\in\Rep(\omega)$, 
let $V_{\dc}, \ \dc\in\Kcal^d(M)$, 
be the collection of unitary operators associated with $\pi$ by 
(\ref{1Cad:1}). Define 
\[
 z^\pi(b)\defi V_{\partial_0b}\cdot V^*_{\partial_1b}\qquad b\in\Si_1(\Kcal^d(M))\ .
\]
$z^\pi$ satisfies the 1-cocycle identity and is path-independent. 
Punctured Haag duality (hence, Haag duality) entails that  $z^\pi$ fulfills the locality condition,  
hence  $z^\pi\in\Zcal^1_t(\omega,\Kcal^d(M))$. Following the cited reference,
one arrives to the categorical equivalence 
between $\Zcal^1_t(\omega,\Kcal^d(M))$ and  $\Rep(\omega)$. 
Now the proof follows from 
the Proposition  \ref{Caa:3}.
\end{proof}
%******************************
It is now clear that, in a fixed spacetime 
background $M$ \emph{all} the categories $\Zcal^1_t(\si,\Kcal^d(M))$  
carry the \emph{same physical information} for any choice of 
$\si\in\St_o(M)$. Indeed, they are associated with 
representations of the net of local observables
which are a sharp excitation of the representation  
$\pi_{\omega}$ associated with a state $\omega\in\St_o(M)$ 
satisfying punctured Haag duality. 
%*********************************************************
%*********************************************************
%*********************************************************
%*********************************************************
%*********************************************************
%*********************************************************
%*********************************************************
%*********************************************************
%*********************************************************
%*********************************************************
\subsection{Locally covariant structure of sectors}
\label{Cb}

We show how the locally covariant structure 
of superselection sectors arises. 
We  introduce the embedding functor which gives a first important
information on the covariant structure of sectors. Such a structure 
is encoded in the superselection functor   analyzed 
in the subsequent section. Finally, by applying the
Doplicher-Roberts duality theory of compact groups to the superselection 
functor, we investigate  the locally covariant properties  of the associated gauge 
groups.
%*******************************************
%*******************************************
\subsubsection{The Embedding functor}
\label{Cba}
Consider $M_1,M\in\Loc$, with $\psi\in(M_1,M)$,
let  $\alpha_\psi:\Al(M_1)\rightarrow \Al(M)$ be the $\Crm^*-$morphism
associated with $\psi$. Given $\omega\in\St_o(M)$, let   
$\tau^\omega_\psi:\omega^*\Al_{\psi(\Kcal^d(M_1))}\rightarrow
                (\omega\alpha_\psi)^*\Al_{\Kcal^d(M_1)}$ the corresponding 
net-isomorphism introduced  in Section \ref{Ad}.  
%******************
\begin{df}
\label{Cba:1}  
Given $\psi\in(M_1,M)$ and $\omega\in\St_o(M)$. 
We call \textbf{embedding} the map of categories 
$\Erm^\omega_\psi:  \Zcal^1_t(\omega, \Kcal^d(M))
        \rightarrow 
                 \Zcal^1_t(\omega\alpha_\psi,\Kcal^d(M_1))$
defined, for $z,z_1\in\Zcal^1_t(\omega,\Kcal^d(M))$ and
$t\in(z,z_1)$, as 
\begin{align*}
 \Erm^\omega_\psi(z)(b) & \defi \tau_\psi^\omega(z(\psi(b)))\ , && 
                b\in\Si_1(\Kcal^d(M_1))\ , \\
 \Erm^\omega_\psi(t)_a & \defi \tau_\psi^\omega(t_{\psi(a)})\ , && 
                   a\in\Si_0(\Kcal^d(M_1)) \ .
\end{align*}
\end{df} 
%**********
In this  definition $\psi(b)$ is the 1-simplex of $\Kcal^d(M)$ 
defined as 
\[
|\psi(b)|=\psi(|b|) \ ,  \ \ \partial_0\psi(b)=
\psi(\partial_0b) \ ,  \ \ \partial_1\psi(b)= \psi(\partial_1b) \ . 
\]
%************************
\begin{prop}
\label{Cba:2}
Given $\psi\in(M_1,M)$ and $\omega\in\St_o(M)$. 
The embedding 
\[
\Erm^\omega_\psi: \Zcal^1_t(\omega, \Kcal^d(M))
        \rightarrow 
                 \Zcal^1_t(\omega\alpha_\psi,\Kcal^d(M_1))
\]
is a covariant symmetric tensor $^*-$functor which is full and faithful.
\end{prop}
\begin{proof}
That $\Erm^\omega_\psi$ is a covariant $^*-$functor 
is obvious from the fact that $\tau^\omega_\psi$ 
is a net-isomorphism. Given $z,z_1\in \Zcal^1_t(\omega, \Kcal^d(M))$
and $t\in(\Erm^\omega_\psi(z),\Erm^\omega_\psi(z_1))$, define
\[
s_a\defi (\tau^\omega_\psi)^{-1}(t_{\psi^{-1}(a)}) \ , \qquad 
 a\in \psi(\Kcal^d(M_1))) \ ,
\]
and observe that $s\in (\Rrm(z),\Rrm(z_1))$ where  
$\Rrm$ is the restriction functor from 
$\Zcal^1_t(\omega,\Kcal^d(M))$  into $\Zcal^1_t(\omega,\Kcal^d(\psi(M_1)))$. 
Since $\Rrm$ is full, there is $t'\in(z,z_1)$ such that 
\[
(\tau^\omega_\psi)^{-1}(t_{\psi^{-1}(a)})=\Rrm(t')_a = t'_a \ , 
\]
for any $a\in \Si_0(\psi(\Kcal^d(M_1)))$
which is equivalent to 
$t_a=  \tau^\omega_\psi(t'_{\psi(a)})=\Erm^\omega_\psi(t')_a$ 
for any $a\in\Si_0(\Kcal^d(M_1))$. This proves that 
$\Erm^\omega_\psi$ is full. If $\Erm^\omega_\psi$ were not-faithful, 
there would be $t_1,t_2\in(z,z_1)$ such that 
$\Erm^\omega_\psi(t_1)=\Erm^\omega_\psi(t_2)$. Therefore 
\[
\tau_\psi({t_1}_{\psi(a)})= \tau_\psi({t_2}_{\psi(a)}) \ \iff \ 
 {t_1}_{\psi(a)}= {t_2}_{\psi(a)}  
\]
for any $a\in\Si_0(\Kcal^d(M_1))$. This, in turns,  is equivalent
to the identity 
\[
\Rrm(t_1)_a = \Rrm(t_2)_a \ , \qquad a\in \Si_0(\psi(\Kcal^d(M_1))) \ ,  
\] 
and this leads to a contradiction because $\Rrm$ is faithful. 
What remains to be shown is that the embedding is a symmetric and tensor
$^*-$functor. To this end let  $z,z_1\in \Zcal^1_t(\omega, \Kcal^d(M))$. 
Recalling the definition (\ref{Cax:1}), 
for any 
$b\in\Si_1(\Kcal^d(M_1))$ we have 
\begin{align*}
\Erm^{\omega}_\psi(z\otimes_\omega  z_1) (b)  &   = 
  \tau^\omega_\psi(z\otimes_\omega z_1) (\psi(b))  \\ 
& =  \tau^\omega_\psi \big( z(\psi(b))\cdot \mathrm{ad}_{z(p)}
              \big(z_1(\psi(b))\big)\big) \\
& =  \tau^\omega_\psi(z(\psi(b))) \cdot \mathrm{ad}_{\tau^\omega_\psi(z(p))}
              \big(\tau^\omega_\psi(z_1(\psi(b)))\big)\\
&  =  (\Erm^{\omega}_\psi(z)\otimes_{\omega\alpha_\psi} \Erm^{\omega}_\psi(z_{1}))(b) \ ,
\end{align*}
where the fact that $\tau^\omega_\psi$ is a morphism 
of $\Crm^*-$algebras has been used (see Section \ref{Ad}).  
The same reasoning leads to the identity 
\[
\Erm^{\omega}_\psi(t\otimes_\omega s)_a 
 =  (\Erm^{\omega}_\psi(t)\otimes_{\omega\alpha_\psi} 
 \Erm^{\omega}_\psi(s))_a, \qquad 
a\in\Si_0(\Kcal^d(M_1))
\]
for any pair $t,s$ of arrows of $\Zcal^1_t(\omega, \Kcal^d(M))$. 
Now, let $z,z_1\in\Zcal^1_t(\omega,\Kcal^d(M))$. 
By recalling definition (\ref{Cax:2}), 
for any $a\in\Si_0(\Kcal^d(M_1))$ we have 
\begin{align*}
\Erm^{\omega}_\psi& (\eps(z,z_1))_a  =  
  \tau^\omega_\psi\big(z_1(b_1)^*\cdot 
               \mathrm{ad}_{z_1(\tilde{q})}(z(b)^*)\big) 
                       \cdot 
     \tau^\omega_\psi\big(z(b)\cdot \mathrm{ad}_{z(\tilde{p})}(z_1(b_1))\big),
\end{align*}
where $b_1,b$ are 1-simplices in $\Kcal^d(M)$ such that 
$\partial_1b = \partial_1b_1 = \psi(a)$ and 
$\partial_0b\perp \partial_0b_1$.  Let us consider 
the first term of the product.
We can take 
in $\Kcal^d(M_1)$ two 1-simplices 
$b'_1, b'$ of the following form: 
\[
 \partial_1b'_1= \partial_1b'= a, \ \ \ \partial_0b'_1\perp\partial_0b'  
\]
so the  1-simplices in $\Kcal^d(M)$ defined as 
$\psi(b'_1)$ and $\psi(b')$ fulfill the hypotheses of the
definition of a permutation symmetry. Furthermore, 
let $q'$ be a path in $\Kcal^d(M_1)$ such that
$\partial_0q'=\partial_1b'_1$ and $\partial_1q'\perp |b'|$. 
Then $\psi(q')$ has the same properties of $q'$. Therefore 
\begin{align*}
\tau^\omega_\psi\big(z_1(b_1)^*\cdot & \mathrm{ad}_{z_1(\tilde{q})}(z(b)^*)\big) =\\
& = \tau^\omega_\psi\big(z_1(\psi(b'_1))^*\cdot 
    \mathrm{ad}_{z_1(\psi(q'))}(z(\psi(b_1'))^*)\big)\\
& = \tau^\omega_\psi(z_1(\psi(b'_1)))^*\cdot 
    \mathrm{ad}_{\tau^\omega_\psi(z_1(\psi(q')))}
    \big(\tau^\omega_\psi(z(\psi(b_1')))^*\big)\\
& = \Erm^{\omega}_\psi(z_1)(b_1')^* \cdot 
    \mathrm{ad}_{\Erm^{\omega}_\psi(z_1)(q')}
    \big(\Erm^{\omega}_\psi(z)(b')\big)^* \ .
\end{align*}
Applying the same reasoning to the other term of the product
we arrive at  
$\Erm^{\omega}_\psi(\eps(z,z_1))_a  =
\eps(\Erm^{\omega}_\psi(z), 
      \Erm^{\omega}_\psi(z_1))_a$ for any $a\in\Si_0(\Kcal^d(M_1))$.
\end{proof}
%***********************************
%***********************************
\begin{lemma}
\label{Cba:3}
Given $\psi\in (M_1,M)$ let $\omega,\si\in\St_o(M)$.Then 
\[
 \Erm^\omega_\psi \circ \Frm_{\si,\omega} = 
 \Frm_{\si\alpha_\psi,\omega\alpha_\psi}\circ  \Erm^\si_\psi.
\]  
\end{lemma}
\begin{proof}
Given $z\in\Zcal^1_t(\si,\Kcal^d(M))$ and 
$b\in\Si_1(\Kcal^d(M_1))$, by Lemma \ref{Ad:6} we have 
\begin{align*}
\Frm_{\si\alpha_\psi,\omega\alpha_\psi}\circ\Erm^\si_\psi(z)(b) & 
 = \rho_{\si\alpha_\psi,\omega\alpha_\psi}(\tau^\si_\psi(z(\psi(b))))\\
& = \tau^\omega_\psi(\rho_{\si,\omega}(z(\psi(b)))
 = \Erm^\omega_\psi \circ \Frm_{\si,\omega}(z)(b)\ .
\end{align*}
\end{proof}
%******************
%****************** 
%******************************************************
%******************************************************
%******************************************************
\subsubsection{The Superselection Functor}
\label{Cbb}
We now are in the position to 
show  the covariant structure of superselection sectors. 
Let $\Sym$  be the category whose objects are symmetric 
tensor $\mathrm{C}^*-$categories, 
and whose arrows are the full and faithful, 
symmetric tensor $^*-$functors. According to 
the philosophy of the locally covariant quantum field 
theories, we expect that the superselection sectors 
can exhibits a structure of functor from the category 
$\Loc$ into the category  $\Sym$. We know  that 
the superselection sectors of any spacetime $M\in\Loc$  
identify a family of categories within the same isomorphism
class, any such category $\Zcal^1_t(\omega,\Kcal^d(M))$ 
is labeled by an element $\omega\in\St_o(M)$. Since there is no natural 
way to associate an element of this isomorphism class 
to the spacetime $M$, as $M$ varies in $\Loc$, 
we will be forced to make a choice. \\[3pt]
% Let $\Sym$  be the category whose 
% objects are symmetric tensor $\mathrm{C}^*-$categories, 
% and whose arrows are the full and faithul, symmetric tensor $^*-$functors. 
\indent Given a locally covariant quantum field theory $\Al$
and a reference state space $\St_o$, let 
\begin{equation}
\label{Cbb:1}
\underline{\omega}\defi\{\omega_M\in\St_o(M) \ | \ M\in\Loc \}
\end{equation}
be a \emph{choice of states}.  \\
\indent We call the \textbf{superselection 
functor} associated with the choice $\underline{\omega}$, 
the categories map $\Ss_{\underline{\omega}}:\Loc\rightarrow \Sym$ 
defined as            
\begin{equation}      
\label{Cbb:2}         
\left\{               
\begin{array}{lll}     
\Ss_{\underline{\omega}} (M)   \defi    
 \Zcal^1_t(\omega_M, \Kcal^d(M))  ,  &  M\in\Loc\ , \\ 
\\                      
\Ss_{\underline{\omega}}(\psi)    \defi   
\Frm_{\omega_M\alpha_\psi,\omega_{M_1}} 
\circ \Erm^{\omega_M}_\psi\ ,
&   \psi\in (M_1,M)\ . 
\end{array}\right.    
\end{equation}        
%***********************
\begin{teo}           
\label{Cbb:3}        
$\Ss_{\underline{\omega}}:\Loc\longrightarrow \Sym$ 
is a  contravariant functor.
\end{teo}            
\begin{proof}        
Let $\psi\in (M_1,M)$.
Since $\Ss_{\underline{\omega}}(\psi)$ 
is defined as the composition of the flip and of the 
embedding functor, Proposition \ref{Cba:2} and Theorem \ref{Cab:3} 
imply that           
$\Ss_{\underline{\omega}}(\psi): \Ss_{\underline{\omega}} (M) 
\rightarrow \Ss_{\underline{\omega}} (M_1)$ 
is a full, faithful symmetric tensor  $^*-$functor.
Given  $\phi\in(M_2,M_1)$, by Lemma \ref{Cba:3}, we have  
\begin{align*}      
\Ss_{\underline{\omega}}(\phi)\circ 
\Ss_{\underline{\omega}}(\psi)  & = 
 \Frm_{\omega_{M_1}\alpha_\phi,\omega_{M_2}}  
    \circ \Erm^{\omega_{M_1}}_\phi\circ 
   \Frm_{\omega_M\alpha_\psi,\omega_{M_1}}
   \circ \Erm^{\omega_M}_\psi\\
& = \Frm_{\omega_{M_1}\alpha_\phi,\omega_{M_2}}  \circ 
    \Frm_{\omega_M\alpha_{\psi\phi},\omega_{M_1}\alpha_\phi}\circ 
    \Erm^{\omega_{M}\alpha_\psi}_\phi\circ 
    \Erm^{\omega_M}_\psi\\
& =  \Frm_{\omega_{M}\alpha_{\psi\phi},\omega_{M_2}} \circ
     \Erm^{\omega_{M}}_{\psi\phi}\\
& =  \Ss_{\underline{\omega}}(\psi\phi)\ .
\end{align*}       
Finally, by the definitions of the flip and of the embedding
functors we have   
\[                 
\Ss_{\underline{\omega}}(\mathrm{id}_{M}) =
  \Frm_{\omega_{M},\omega_{M}} \circ
     \Erm^{\omega_{M}}_{\mathrm{id}_{M}} =
    \mathrm{id}_{\Zcal^1_t(\omega_M,\Kcal^d(M))} =
   \mathrm{id}_{\Ss_{\underline{\omega}}(M)}\ , 
\]                 
and the proof is now completed.
\end{proof}        
%*************     
\begin{prop}       
\label{Cbb:4}
If $\underline{\omega}$ and 
$\underline{\si}$ is a pair of choice of states, then  the functors  
$\Ss_{\underline{\omega}}$
and $\Ss_{\underline{\si}}$ are isomorphic.
\end{prop}
\begin{proof}
Define 
\[
u_{\underline{\omega},\underline{\si}}(M) =
\Frm_{\omega_M,\si_M} \ , \qquad M\in\Loc \ .
\]
By Theorem \ref{Cab:3}, it follows that 
$u_{\underline{\omega},\underline{\si}}(M):
\Ss_{\underline{\omega}}(M)\rightarrow \Ss_{\underline{\si}}(M)$
is an symmetric tensor isomorphism.
% \begin{align*}
%  u_{\underline{\omega},\underline{\si}}(M)   
%  (\Ss_{\underline{\omega}}(M))  
%   &  = F_{\omega_M,\si_M}(\Zcal^1_t(\omega_M, \Kcal^d(M))) \\
%   &  =  \Zcal^1_t(\si_M, \Kcal^d(M)) \\ 
%   & = \Ss_{\underline{\si}}(M).
% \end{align*} 
Given $\psi\in(M_1,M)$.  By Lemma \ref{Cba:3} we have that 
\begin{align*}
u_{\underline{\omega},\underline{\si}}& (M_1) \circ 
 \Ss_{\underline{\omega}}(\psi)  = \\
& =  \Frm_{\omega_{M_1},\si_{M_1}} \circ 
\Frm_{\omega_M\alpha_\psi,\omega_{M_1}} 
\circ \Erm^{\omega_M}_\psi
 =  \Frm_{\omega_M\alpha_\psi,\si_{M_1}} 
  \circ \Erm^{\omega_M}_\psi \\
& =  \Frm_{\si_M\alpha_\psi,\si_{M_1}} \circ 
 \Frm_{\omega_M\alpha_\psi,\si_M\alpha_\psi} 
  \circ \Erm^{\omega_M}_\psi 
 = \Frm_{\si_M\alpha_\psi,\si_{M_1}} \circ \Erm^{\si_M}_\psi
 \circ \Frm_{\omega_M,\si_M} \\
& = \Ss_{\underline{\si}}(\psi)
 \circ u_{\underline{\omega},\underline{\si}}(M).
\end{align*} 
Hence, $u_{\underline{\omega},\underline{\si}}:\Ss_{\underline{\omega}}\to 
\Ss_{\underline{\si}}$ is a natural isomorphism. 
\end{proof}
%******************************************************
%*********************
\begin{oss}
\label{Cbb:5}
This theorem  is the main result  of this paper because
it shows the covariance of charged superselection sectors:
if $\psi\in(M_1,M)$, then  to any   sector
of $M$ there correspond a unique  sector 
of $M_1$ with the same charged quantum numbers. 
To be precise, let
$z\in\Ss_{\underline{\omega}}(M)$ 
be an irreducible object with statistical parameter
$\la([z])=\chi([z])\cdot d([z])$, where $[z]$ denotes the equivalence 
class of $z$. Let $\overline{z}$
be the conjugate of $z$ (see Appendix \ref{X}). Then 
$\Ss_{\underline{\omega}}(\psi)(z)$ 
is an irreducible object of
$\Ss_{\underline{\omega}}(M_1)$ such that 
\[
\big[\Ss_{\underline{\omega}}(\psi)(z)\big]= 
\Ss_{\underline{\omega}}(\psi)([z]) \ .  
\]
Furthermore $z$ and $\Ss_{\underline{\omega}}(\psi)(z)$ 
have  the \emph{same} statistics, i.e.
\[
\chi([z])=
\chi\big(\big[\Ss_{\underline{\omega}}(\psi)(z)\big]\big)
 \ , \ \ \ \ 
d([z])=d\big(\big[\Ss_{\underline{\omega}}(\psi)(z)\big]\big) \ .
\]
Moreover 
\[
\Ss_{\underline{\omega}}(\psi)([\overline{z}]) 
=
\overline{\big[\Ss_{\underline{\omega}}(\psi)(z)\big]}
\ , 
\]
that means that 
$\Ss_{\underline{\omega}}(\psi)([\overline{z}])$ is
the conjugate sector of $\Ss_{\underline{\omega}}(\psi)(z)$. 
\end{oss}
%********************************************************
%********************************************************
%********************************************************
%********************************************************
%********************************************************
\subsubsection{The Gauge Weak Functor}
\label{Cbc}
The next natural step of the present investigation 
should be the application of  Doplicher-Roberts reconstruction theorem 
\cite{DR2}  to the pair of functors 
$(\Al,\Ss_{\underline{\omega}})$ in order to analyze the 
local covariance  of fields and of the gauge groups 
underlying the theory. 
In the present section we will follow partially this investigation line
by showing the covariant structure of  gauge groups 
associated with the superselection sectors. 
Because of some technical problems, that will be exposed below, 
we will not deal with the reconstruction of fields. Conversely, 
gauge groups can be easily reconstructed  by using 
the Doplicher-Roberts duality theorem (hereafter, DR-theorem)
for compact groups \cite{DR1} (note that in the reconstruction 
theorem \cite{DR2} the gauge group of fields is  the dual 
of the category of superselection sectors). \\[5pt]
% We apply the Doplicher-Roberts duality 
% theorem of compact groups\cite{DR1}, to show the covariant structure  
% of gauge groups associted with  superselection sectors. \\[5pt] 
\indent Before beginning the analysis  we need a preliminary observation 
on the DR-theorem  whose 
functorial properties are described in some detail in the Appendix \ref{X}. 
The DR-theorem states that 
any tensor $\Crm^*-$category with a permutation symmetry, 
and conjugates, is the abstract dual 
of a compact group which  is uniquely associated 
with the category only up to isomorphism. 
To any full and faithful, symmetric tensor $^*-$functor 
between categories there corresponds, 
in a contravariant fashion, a  group morphism. 
Also in this case the correspondence between functors and 
group morphisms is not injective. These degrees of freedom 
will reflect in the weakening of the local covariance 
of the gauge groups (see below). \\[5pt] 
\indent Consider a choice of states $\underline{\omega}$.
For any spacetime $M$, we \emph{choose} a compact group 
$G_{\underline{\omega}}(M)$ among the isomorphism
class of compact groups associated with 
the full subcategory of 
$\Ss_{\underline{\omega}}(M)$ whose objects have conjugates. 
Furthermore, for any $\psi\in(M_1,M)$, we \emph{choose} 
a group morphism
\[
\alpha_{\underline{\omega}}(\psi): G_{\underline{\omega}}(M_1)\rightarrow 
                                   G_{\underline{\omega}}(M) \ , 
\]
among the set of group morphisms 
associated with the functor 
\[
\Ss_{\underline{\omega}}(\psi): 
\Ss_{\underline{\omega}}(M)\rightarrow 
                             \Ss_{\underline{\omega}}(M_1) \ . 
\]
Let $\mathbf{Grp}$ be the category whose objects are 
compact groups and whose arrows are the corresponding set of 
group morphisms. Now, for any choice of states 
$\underline{\omega}$ we denote by $\mathcal{G}_{\underline{\omega}}$ 
the categories map 
$\mathcal{G}_{\underline{\omega}}: \Loc\rightarrow \mathbf{Grp}$
defined as 
\begin{equation}
\label{Cbc:1}
\left\{
\begin{array}{lll}
 \mathcal{G}_{\underline{\omega}}(M) \defi   
   G_{\underline{\omega}}(M)  &   M\in\Loc \ , \\ 
\\
\mathcal{G}_{\underline{\omega}}(\psi) \defi  
 \alpha_{\underline{\omega}}(\psi) & 
   \psi\in (M_1,M) \ ,
\end{array}\right.
\end{equation}
We have the following 
%**************
\begin{teo}
\label{Cbc:2}
The map $\mathcal{G}_{\underline{\omega}}$ 
satisfies the following properties: \\
(a) given $M\in\Loc$, there exists  
$g_{\underline{\omega}}(M) \in\mathcal{G}_{\underline{\omega}}(M)$ 
such that 
\[
\mathcal{G}_{\underline{\omega}}(\mathrm{id}_M) = 
\mathrm{ad}_{g_{\underline{\omega}}(M)}\ ;
\]
(b) given $\phi\in(M_2,M_1)$ and $\psi\in(M_1,M)$,
there exists 
$g_{\underline{\omega}}(\phi,\psi)\in\mathcal{G}_{\underline{\omega}}(M_2)$ such that 
\[
\mathcal{G}_{\underline{\omega}}(\phi)\circ 
\mathcal{G}_{\underline{\omega}}(\psi) =
\mathrm{ad}_{g_{\underline{\omega}}(\phi,\psi)} \circ 
\mathcal{G}_{\underline{\omega}}(\phi\psi) \ .
\]
\end{teo}
\begin{proof}
The proof is an easy  application of 
the DR-theorem  to the Theorem \ref{Cbb:3}.
\end{proof}
%********************
The Theorem \ref{Cbc:2}b says that the map 
$\mathcal{G}_{\underline{\omega}}$ is not,
in general, a covariant functor because the defining 
properties of functors are verified in a \emph{weak} sense, 
namely up to isomorphisms of 
the set of arrows. We will refer to the map 
$\mathcal{G}_{\underline{\omega}}: \Loc\rightarrow \mathbf{Grp}$
as the \textbf{gauge weak functor}
\footnote{
Note that the notion of a weak covariant 
(or contravariant)  functor can be given 
in terms of a 2-category, see for instance \cite{LR}.}.
% The word  \emph{gauge}  is due to the Doplicher-Roberts' 
% reconstruction theorem \cite{DR1} where 
% it is shown that the compact group associated with the category of 
% superselection sectors by the duality theorem, 
% is the global gauge group, or gauge group of the first kind, 
% of a net of local fields.\marginpar{(Completare ?)} 
%****************
\begin{teo}
\label{Cbc:3}
Given a pair $\underline{\omega}$, $\underline{\si}$ of choices
of states. Then for any $M\in\Loc$ there exists  a  group isomorphism 
\[
\alpha_{\underline{\omega},\underline{\si}}(M): 
\mathcal{G}_{\underline{\si}}(M)\rightarrow 
       \mathcal{G}_{\underline{\omega}}(M) \ , 
\] 
such that:  if $\psi\in(M_1,M)$  there exists 
$g_{\underline{\omega},\underline{\si}}(\psi)\in\mathcal{G}_{\underline{\omega}}(M)$  such that 
\[
\alpha_{\underline{\omega},\underline{\si}}(M)  
\circ \mathcal{G}_{\underline{\si}}(\psi)
= \mathrm{ad}_{g_{\underline{\omega},\underline{\si}}(\psi)}\circ 
\mathcal{G}_{\underline{\omega}}(\psi)\circ 
\alpha_{\underline{\omega},\underline{\si}}(M_1) \ .
\]
\end{teo}
\begin{proof}
Consider the natural isomorphism 
$u_{\underline{\omega},\underline{\si}}: 
\Ss_{\underline{\omega}}\rightarrow \Ss_{\underline{\si}}$ defined 
in the proof of the  Theorem \ref{Cbb:4}. Recall that 
$u_{\underline{\omega},\underline{\si}}$ satisfies the following 
properties: for any $M\in\Loc$ we have that 
\[
 u_{\underline{\omega},\underline{\si}}(M): 
 \Ss_{\underline{\omega}}(M)\rightarrow \Ss_{\underline{\si}}(M)\ ,
 \qquad (*)
\] 
is a covariant symmetric tensor $^*-$isomorphism which satisfies,
for any $\psi\in(M_1,M)$ the following equation: 
\[
u_{\underline{\omega},\underline{\si}}(M)\circ \Ss_{\underline{\omega}}(\psi)= 
 \Ss_{\underline{\si}}(\psi)\circ 
u_{\underline{\omega},\underline{\si}}(M_1) \ . 
 \qquad  (**)
\]
Consider the equation $(*)$. Since 
$u_{\underline{\omega},\underline{\si}}(M)$ is an isomorphism,
by the DR-theorem  there corresponds group isomorphism 
$\alpha_{\underline{\omega},\underline{\si}}(M): 
      \mathcal{G}_{\underline{\si}}(M)\rightarrow 
       \mathcal{G}_{\underline{\omega}}(M)$.
Consider now the equation $(**)$, and note that 
$u_{\underline{\omega},\underline{\si}}(M_1)\circ 
\Ss_{\underline{\omega}}(\psi)$ 
is full faithful symmetric tensor functor from 
$\Ss_{\underline{\omega}}(M)$
into $\Ss_{\underline{\si}}(M_1)$. Therefore we can find 
an $g\in \mathcal{G}_{\underline{\omega}}(M)$ such that 
\[
\mathcal{G}_{\underline{\omega}}(\psi)\circ 
\alpha_{\underline{\omega},\underline{\si}}(M_1) =
\mathrm{ad}_g(
\alpha_{u_{\underline{\omega},\underline{\si}}(M_1)\circ
        \Ss_{\underline{\omega}}(\psi)}) \ . 
\]
In an analog fashion, there exists 
$g_1\in \mathcal{G}_{\underline{\omega}}(M)$  such that 
\[
\alpha_{\underline{\omega},\underline{\si}}(M)  
\circ \mathcal{G}_{\underline{\si}}(\psi) = 
\mathrm{ad}_{g_1}(\alpha_{\Ss_{\underline{\si}}(\psi) \circ u_{\underline{\omega},\underline{\si}}(M_1)})
\]
therefore by $(**)$ we have
\[
\mathrm{ad}_{g^{-1}_1}\circ 
\alpha_{\underline{\omega},\underline{\si}}(M)  
\circ \mathcal{G}_{\underline{\si}}(\psi)
= \mathrm{ad}_{g^{-1}}\circ 
\mathcal{G}_{\underline{\omega}}(\psi)\circ 
\alpha_{\underline{\omega},\underline{\si}}(M_1)
\]
Namely 
\[
\alpha_{\underline{\omega},\underline{\si}}(M)  
\circ \mathcal{G}_{\underline{\si}}(\psi)
= \mathrm{ad}_{g_1\cdot g^{-1}}\circ 
\mathcal{G}_{\underline{\omega}}(\psi)\circ 
\alpha_{\underline{\omega},\underline{\si}}(M_1),
\]
and this completes the proof.
\end{proof}
In conclusion, this theorem shows that the gauge weak functor 
does not depends on the choice of states $\underline{\omega}$. 
In particular the groups $\mathcal{G}_{\underline{\omega}}(M)$ 
belong to the same isomorphism class for any possible 
choice $\underline{\omega}$. 
Therefore we call $\mathcal{G}_{\underline{\omega}}(M)$, for some choice 
$\underline{\omega}$, the \textbf{gauge group associated with}
$M\in\Loc$.
%****************************
\begin{oss}
\label{Cbc:4}
We point out two problems connected to the study of the local covariance 
of fields. \emph{First}, in the case that the set $\Kcal^{h}(M)$
is nondirected, the Doplicher-Roberts reconstruction theorem 
does not apply straightforwardly.  This happen for instance when either 
$M$ is  nonsimply connected or  $M$ has compact Cauchy surfaces. 
\emph{Secondly}, consider $M,M_1\in\Loc$ such that there is 
$\psi\in(M_1,M)$. Assume that 
$\Kcal^{h}(M)$ and $\Kcal^{h}(M_1)$ are directed.\footnote{  
For instance we can take as $M$ the Minkowski space, and as
$M_1$ a diamond of the Minkowski space. One can easily check 
that the set of diamonds of these two spaces are directed.}
In this case one can apply  the reconstruction 
theorem and obtain the  algebras of fields, 
say $\mathscr{F}(M)$ and $\mathscr{F}(M_1)$.  One expects that 
$\mathscr{F}(M_1)$ is isomorphic to the subalgebra 
$\mathscr{F}(\psi(M_1))$ of  $\mathscr{F}(M)$. It is not clear how to show
this because in the definition of $\mathscr{F}(M_1)$ intervenes 
the category $\Ss_{\underline{\omega}}(M_1)$, while in that 
of $\mathscr{F}(\psi(M_1))$ intervenes $\Ss_{\underline{\omega}}(M)$, 
and we do not know whether these categories are equivalent.
\end{oss}

%%%%%%%%%%%%%%%%%%%%%%%%%%%%%%%%%%%%%%%%%
%%%%%%%%%%%%%%%%%%%%%%%%%%%%%%%%%%%%%%%%%
%%%%%%%%%%%%%%%%%%%%%%%%%%%%%%%%%%%%%%%%%

\section{Local completeness}
\label{Z} 
The local covariance of the theory makes possible 
the analysis of the relation between local and global superselection 
sectors. This section is devoted to a preliminary analysis of this topic. 
In particular we will discuss how the possible nonequivalence 
between local and global superselection sectors might be related 
to the nontrivial topology of spacetimes and in particular 
to the  existence of path-dependent 1-cocycles.\\[5pt]
\indent Fix a spacetime $M\in\Loc$ and consider the 
set of diamonds $\Kcal^d(M)$. Any diamond 
$\dc$ is a globally hyperbolic spacetime, and the injection 
$\io_{M,\dc}:\dc\rightarrow M$ provides an embedding of $\dc$ into $M$.
Given  a choice $\underline{\omega}$ of states, we focus  on 
the superselection sectors $\Ss_{\underline{\omega}}(M)$ associated 
with $M$ and to those associated with any diamond $\dc$, that is 
$\Ss_{\underline{\omega}}(\dc)$. We know that 
\begin{equation}
\label{Z:1}
 \Ss_{\underline{\omega}}(\io_{M,\dc}): 
 \Ss_{\underline{\omega}}(M)\rightarrow \Ss_{\underline{\omega}}(\dc)
\end{equation}
is a  covariant symmetric tensor $^*-$functor which is full and faithful.
From the physical point of view, we expect that this functor 
is an equivalence.  The sectors under investigation 
are sharply localized, hence they should not be affected by the nontrivial 
topology of the spacetime and there should be no difference between 
local and global behaviour. However, 
up until now,  we have no argument to establish this  equivalence.
We explain what is the 
technical problem, to which we will refer as the 
\emph{extension problem}. \\
\indent Given $\dc\in\Kcal^d(M)$, let 
$\Kcal^d(\dc)$ be the set of diamonds of $\dc$ considered as a globally 
hyperbolic spacetime. Given $\omega\in\St_o(M)$, 
let $\Zcal^1_t(\omega,\Kcal^d(\dc))$ be the category 
of path-independent 1-cocycles of $\Kcal^d(\dc)$ with values 
in the net $\omega^*\Al_{\Kcal^d(\dc)}$. This category 
is a tensor $\Crm^*-$category with a permutation symmetry and its 
objects with  finite statistics have conjugates. 
One  can  easily see  that the functor (\ref{Z:1}) 
is an equivalence  if, an only if, for any  diamond 
$\dc$ and for any 1-cocycle $z_\dc$ of $\Zcal^1_t(\omega,\Kcal^d(\dc))$,
having finite statistics, there exists 
$z\in \Zcal^1_t(\omega,\Kcal^d(M))$ such that   
\begin{equation}
\label{Z:2}
z\nrestr_{\Kcal^d(\dc)} = z_\dc \ .  
\end{equation}
Now, it is convenient to introduce a new  terminology.
\begin{df}
\label{Z:3}
Given $M\in\Loc$. We will say that the superselection sectors 
of $(\Al(M),\St_o(M))$ are \textbf{locally complete} whenever 
$\Ss_{\underline{\omega}}(M)$ 
is equivalent to $\Ss_{\underline{\omega}}(\dc)$ for any $\dc\in\Kcal^d(M)$; 
conversely, we will say that they are \textbf{locally incomplete}. 
\end{df}
Examples of theories with locally complete sectors  can be easily 
provided (see below), while it seems to be a very hard task to prove 
the converse.  
On the other hand, we have no argument to prove that this 
holds true in general, and the only attempts, known by the authors, 
to solve this problem in the Haag-Kastler framework
are due to Roberts \cite{Rob1} and 
Longo\footnote{Private communication.}. However,
only partial results have been achieved.
Concerning the failure of local completeness,
it seems to be related in a subtle way to the nontrivial 
topology of the spacetime: in particular to the nonsimply connectedness, 
as we explain by means of the following reasonings.\\[3pt]
\indent (1) The 
first example of a locally covariant quantum field theory 
with a state space has been provided in \cite{BFV}. 
The correspondence which associates 
to any $M\in\Loc$ the CCR algebra  $\mathscr{F}(M)$ 
of the scalar Klein-Gordon field over $M$ is a locally covariant 
quantum field theory $\mathscr{F}$. 
The correspondence  which associates 
the space $\St_\mu(M)$ of the quasi-free states of $\mathscr{F}(M)$ 
satisfying the microlocal spectrum condition 
\cite{BFK} is a locally quasi-equivalent 
states space $\St_\mu$ of
$\mathscr{F}$. The pure states of $\St_\mu(M)$ fulfill the split property,
and satisfies punctured Haag duality (see \cite{Ruz2}). 
So the pair $(\mathscr{F},\St_\mu)$ 
satisfies all the properties that we have assumed 
in this analysis. Fix $M\in\Loc$ and consider a pure state $\omega$ of 
$\St_\mu(M)$. By means of the same reasoning used in 
\cite{BDLR} it turns out that any 1-cocycle of the category 
$\Zcal^1_t(\omega,\Kcal^d(M))$ is a  finite direct 
sums of the trivial 1-cocycle (the proof is relies 
on \cite[Theorem 3.5]{Rob1a}, which provides sufficient conditions 
for the absence of nontrivial superselection sectors,  and on the 
properties of the nets of free fields \cite{Ara}). 
The Proposition  \ref{Caa:3}  entails that this holds for any 
$\si\in\St_\mu(M)$. So for any choice $\underline{\omega}$ and for any 
$M$ the category $\Ss_{\underline{\omega}}(M)$ is trivial. Hence 
the functor 
$\Ss_{\underline{\omega}}(\psi): 
 \Ss_{\underline{\omega}}(M_1)\rightarrow 
 \Ss_{\underline{\omega}}(M)$ 
is an equivalence for any 
$\psi\in(M_1,M)$ and   the theory 
is locally complete. 
This easy\footnote{Modifying this example one can construct 
examples of locally complete theories  having a nontrivial 
superselection content.
It is enough to enlarge the symplectic 
space associated with any $M\in\Loc$ in a such way that it is possible to introduce an action of a compact group.} example says  that 
\textit{the spacetime topology, in particular the nonsimply connectedness, 
does not represent an obstruction to local completeness}. \\[3pt]  
% , because if it 
% were so, then, for a nonsimply connected spacetime $M$, 
% the functor $\Ss_{\underline{\omega}}(\io_{M,\dc}):
% \Ss_{\underline{\omega}}(M)\rightarrow 
% \Ss_{\underline{\omega}}(\dc)$ should not be  an equivalence 
% for some $\dc\in\Kcal^d(M)$. 
\indent (2) The existence of a possible relation 
between topology and local incompleteness, can be seen by 
analyzing  a situation easier than 
the extension problem.
Consider a spacetime $M$, and let 
$\omega$ be a state of $\St_o(M)$. 
Assume that there is a family $\{z_\dc\}$  
with $z_\dc\in\Zcal^1_t(\omega,\Kcal^d(\dc))$ for any diamond $\dc$, 
satisfying the following 
property:  given  $b\in\Si_1(\Kcal^d(M))$, then 
\begin{equation}
\label{Z:3bis}
 z_\dc(b) = z_{\dc_1}(b)\ .
\end{equation}
for any pair $\dc,\dc_1$ of diamonds of $M$ such that 
$cl(|b|)\subseteq \dc\cap \dc_1$. Now define   
\[
z(b) \defi  z_\dc(b)\ , \qquad b\in\Si_1(\Kcal^d(M)) \ , 
\]
where $\dc$ is a diamond which contains $cl(|b|)$. 
By (\ref{Z:3bis}), this definition does not depend
on the chosen diamond $\dc$.  It is clear that 
for any 1-simplex $b$ we have that 
$z(b)\in\A_{\omega}(|b|)$ because so does $z_\dc(b)$.   
Furthermore, as for any $c\in\Si_2(\Kcal^d(M))$,  by (\ref{Aa:7a}),
there is $\dc_1\in\Kcal^d(M)$, with $cl(|c|)\subseteq\dc_1$, 
we have that 
\[
z(\partial_0c)\cdot z(\partial_2c) = 
z_{\dc_1}(\partial_0c)\cdot z_{\dc_1}(\partial_2c) = 
z_{\dc_1}(\partial_1c) =z(\partial_1c)\ .
\]
Therefore,  $z$ is a 1-cocycle of $\Kcal^d(M)$ which extends
the family $\{z_\dc\}$. Notice, however,
that in the case that $M$ is nonsimply connected,  
then $z$ might be a path-dependent 1-cocycle.  
Hence \textit{the local incompleteness might be related to the existence 
of path-dependent 1-cocycles}.
Recall that this type of 1-cocycles are of a topological nature
as they provide nontrivial representations of the fundamental group 
of the manifold (see Section \ref{Bb}).\\[3pt]
\indent (3) In order to strengthen the idea in (2), 
consider a nonsimply connected spacetime  
$M$, and let $\omega$ be a state of $\St_o(M)$. Assume that 
there exists a path-dependent 1-cocycle $z$ of $\Kcal^d(M)$ with values 
in $\omega^*\Al_{\Kcal^d(M)}$. For any $\dc\in\Kcal^d(M)$ define 
\[
z_\dc(b)\defi z(b)\ , \qquad b\in\Si_1(\Kcal^d(\dc)) \ .
\]
Since $\dc$ is simply connected,  $z_\dc$ is a path-independent
1-cocycle of $\Kcal^d(\dc)$. Hence,  the family 
$\{z_\dc\}$ satisfies the condition 
(\ref{Z:3}) and its extension to $\Kcal^d(M)$ is a path-dependent 
1-cocycle. One should be cautious at this point. The above
does not imply the violation of local completeness because 
one should first ensure that $z_\dc$ has finite statistics. Unfortunately,
at the moment we are unable to prove this.
%*************************
\begin{oss} 
\label{Z:5}
According to the above discussion,
the obstruction to local completeness of sectors
seems to be the presence of topological 1-cocycles. So, we expect
that, in a simply connected spacetime,  the sectors are locally  
complete.
\end{oss}
\begin{oss} 
\label{Z:4}
To avoid confusion, we want  to stress that 
there is no relation between path-dependent 1-cocycle 
of a poset and the topological charged sectors  discovered by 
Buchholz and Fredenhagen in \cite{BF}. 
As observed the formers are of a topological nature 
because they provide nontrivial representation of the 
fundamental group of the poset. The latter are 
called topological because of their localization properties. 
In this case the poset underlying the theory is that formed by 
the set of spacelike cones of the Minkowski space. 
One can easily see  that the 1-cocycles 
associated with this type of charges 
are path-independent. Thus, they provide only trivial representations  
of the fundamental group of the poset. 
\end{oss}

%%%%%%%%%%%%%%%%%%%%%%%%%%%%%%%%%%%%%%
%%%%%%%%%%%%%%%%%%%%%%%%%%%%%%%%%%%%%%
%%%%%%%%%%%%%%%%%%%%%%%%%%%%%%%%%%%%%%

\section{Conclusions and Outlook}

The paper concerned with the analysis of superselection sectors 
in the framework of a locally covariant quantum field theory $\Al$.  
The main purpose  was the  understanding of the covariance 
behaviour of those sectors which describe sharply localized charges,
namely those type of sectors that on a fixed spacetime background 
are a sharp excitation of a reference representation,
the vacuum in the Minkowski space, of the observable net.\\
\indent As the present paper is the first investigation 
on this topic, it is worth recalling in some detail 
the basic assumptions and results. The first, very useful result, 
is of geometrical nature:
\begin{itemize}
\item[1.] The set of diamonds $\Kcal^d(M)$ 
of globally hyperbolic spacetimes $M$ is stable 
under isometric embeddings (Lemma \ref{Aa:9}). 
\end{itemize}
This allows to express the locally covariant principle in terms of  nets
of local observables $\Al_{\K^d(M)}$  indexed by the set of diamonds 
$\K^d(M)$ of $M$. The set $\Kcal^h(M)$  introduced in \cite{BFV} 
does not fit the topological and causal properties of the spacetime $M$, 
compared to $\K^d(M)$. \\
\indent The first needed step to introduce the superselection sectors 
of the quantum field theory functor $\Al$ has been the definition of a reference 
state space $\St_o$ for the theory. We required that 
$\St_o$ is locally quasi equivalent,  that it satisfies the Borchers 
property and punctured Haag duality (see Definition \ref{C:1}). 
In particular, we emphasize that the local quasi equivalence is required 
not only on diamonds but also on the larger family $\Kcal^h(M)$.  
This was an important assumption, as  we are requiring 
that the elements $\St_o(M)$ behave locally in the same way 
also on non simply connected regions. Given $\St_o$ 
we have defined the sectors as follows: The superselection sectors 
associated with a spacetime $M$ are the families 
$\Zcal^1_t(\omega,\Kcal^d(M))$, as $\omega$ varies in $\St_o(M)$,
of path-independent 1-cocycles of the poset  
obtained by ordering under inclusion $\Kcal^d(M)$, which take 
their values in the net $\omega^*\Al_{\Kcal^d(M)}$ 
(see Definition \ref{C:2}). 
\begin{itemize}
\item[2.] For any spacetime $M$ the sets 
      $\Zcal^1_t(\si,\Kcal^d(M))$, as $\si$ varies in $\St_o(M)$,   
      \emph{carry the same physical information}: 1-cocycles 
      of $\Zcal^1_t(\si,\Kcal^d(M))$  are, up to equivalence, 
      in bijective correspondence with representations 
      of the observable net which are sharp excitations of 
      the representation $\pi_\omega$ induced by a state $\omega$ in 
      $\St_o(M)$ 
      satisfying punctured Haag duality (see Proposition \ref{1Cad:2}).
\item[3.]  Sectors manifest a \emph{charge structure}:  
       Their quantum numbers have a composition law, a statistics and 
       a charge conjugation symmetry. This structure is independent 
       from the choice of the state $\omega$ in $\St_o(M)$. 
       This is expressed by the fact 
       that $\Zcal^1_t(\omega,\Kcal^d(M))$ is a symmetric tensor 
       $\Crm^*-$category whose objects with finite statistics have 
       conjugates; all the categories 
       $\Zcal^1_t(\omega,\Kcal^d(M))$ belong to the same 
       isomorphism class (see Theorem \ref{Cab:3}). 
\item[4.]  The \emph{charge structure is contravariant}: 
       If a spacetime $M_1$ can be isometrically embedded 
       in $M$, then to any sector of $M$ there corresponds a unique 
       sector of $M_1$ with the same charged quantum numbers;
       the correspondence associating to any spacetime 
       $M$ the category $\Zcal^1_t(\omega_M,\Kcal^d(M))$, 
       with $\omega_M\in\St_o(M)$, is a contravariant functor. 
       (see Theorem \ref{Cbb:3} and  Remark \ref{Cbb:5})
       \footnote{We point out that a similar  functorial structure 
                 arises in  the theory of subsystems \cite{CDR}.}.        
\end{itemize}

These results imply that \emph{the physical content of superselection 
sectors carried by each spacetime remains stable when the spacetimes can 
be coherently embedded}.

Notice that in this first paper we took a conservative 
point of view, namely, we work at a level that is as faithful as possible to 
the tradition in superselection theory. Indeed, one might have even started 
in a more abstract way by \emph{defining} representations in the generally 
covariant sense from the beginning.This amounts to view them as natural 
transformations. 
However, at the moment it is not clear how to generalize some 
of the important technical features that we used.

We now pass to briefly outline a prospect for future works.

An important point that will be tackled in a forthcoming contribution 
is how various indices are related to each other, 
in the sense of tensor categories. Here, one make use of more machinery 
from algebraic topology, in the sense of injections and retractions, 
always tailored to the poset situation.

New directions of research can be envisaged. For instance, 
an important step would be to clarify
whether the reconstruction procedure of Doplicher and Roberts can 
be fully implemented, in general, by reconstructing the field nets, and proving them to be locally covariant. 
In case the theory is locally complete and the family index is directed, 
the reconstruction can be done, and it will appear in the third 
paper of the series.

A most important direction would be the incorporation of the case when the 
$1$-cocycles are \emph{not} path-independent. Here the full topological 
structure of the spacetimes enters the game. This is strongly related 
also to the problem of proving, or disproving, local completeness in 
the sense of Section \ref{Z}. It looks like a difficult task but worth exploring.
We hope to come back to this point in the near future.

We conclude with a couple of possible further directions of research. The first deals with the gauge groups.
One may imagine that the gauge groups associated with any local region act as 
\emph{local} gauge groups, thus opening a fresh look at this problem 
in the algebraic setting. Here, a more geometrical setting might prove helpful.
The second direction deals with the fact that the theories described in this paper seems to have from the outset a well-defined ultraviolet behaviour, a fact which is exemplified by the properties of the functor of restriction $\Rrm$ (see Theorem \ref{Cac:1}).
A closer connection with the work done by D'Antoni, Morsella and Verch \cite{DMV}, possibly within the framework outlined in \cite{BrR}, would be desirable.

\section*{Acknowledgements}
We are particularly thankful to J.E. Roberts and K. Fredenhagen for their kind remarks during this investigation. We thank as well E. Vasselli for his precious help with the intricacies of the Doplicher-Roberts reconstruction Theorem. We are grateful to the DFG and to the European Network ``Quantum Spaces -- Noncommutative Geometry'' for financial support.

%%%%%%%%%%%%%%%%%%%%%%%%%%%%%%%%%%%%%%
%%%%%%%%%%%%%%%%%%%%%%%%%%%%%%%%%%%%%%
%%%%%%%%%%%%%%%%%%%%%%%%%%%%%%%%%%%%%%

%**************************************************
%**************************************************
\appendix
\numberwithin{equation}{section}

\section{Tensor $\Crm^*-$categories} 
\label{X}
For easy of notation we will denote the set of objects 
of a category $\C$ by the same symbol $\C$. 
We denote by $z,z_1,z_2,\ldots $  the objects of the category
and the set of the arrows between $z,z_1$ by $(z,z_1)$.
The composition of arrows is indicated by ``$\cdot$'' and 
the identity  arrow of $(z,z)$ by $1_z$. Recall that 
an arrow $t\in(z_1,z_2)$ is an \textit{isomorphism} if there exists an 
 arrow $s\in(z_2,z_1)$ such that 
\[
  s\cdot t = 1_{z_1}, \ \  t\cdot s = 1_{z_2}.
\] 
The objects $z_{1}$ and $z_2$ are said to be isomorphic, written 
$z_1\sim z_2$. References for this appendix are \cite{Mac, DR1, LR}
\paragraph{Functors and natural transformations :}
Consider two categories $\C_1$ and $\C_2$. A covariant 
functor $F:\C_1\rightarrow \C_2$ 
is said to be: \\[5pt]
\phantom{x}  \emph{faithful}, if $s,t\in(z_1,z_2)$ with $t\ne s$, then 
     $F(s)\ne F(t)$;\\[3pt] 
\phantom{x}  \emph{full}, if $F(z,z_1)= (F(z),F(z_1))$; \\[3pt]
\phantom{x}  \emph{dense}, if for any  $z_2\in \C_2$ there exists 
$z_1\in \C_1$ such that $F(z_1)\sim z_2$; \\[3pt]
\phantom{x}  \emph{involutive},   if  $\C_2=\C_1$ and 
$F\circ F = \mathrm{id}_{\C_1}$   
of $\C_1$,\\[5pt]
where $\mathrm{id}_{\C_1}$ is the identity functor of $\C_1$.\\[5pt] 
\indent A \emph{natural transformation} $u:F\rightarrow G$, 
between a pair of functors $F,G:\C_1\rightarrow \C_2$, 
is a map which assigns an arrow $u(z)$ of $\C_2$ to any object 
$z$ of $\C_1$ such that  
\[
\begin{array}{rll}
(i) &  u(z)\in (F(z), G(z)) & z\in \C_1 \\
(ii)&  F(f)\cdot u(z) = u(z_1)\cdot G(f) & f\in (z,z_1).
\end{array}
\]
$u$ is a natural \emph{isomorphism} whenever 
$u(z)$ is an isomorphism for any $z\in \C_1$.
In this case  we will say that $F$ and $G$ are  
\emph{isomorphic}, written as $F\sim G$.\\[5pt] 
\indent An  \emph{isomorphism} of categories, is functor 
$F:\C_1\rightarrow \C_2$ for which 
there exists another  functor $G:\C_2\rightarrow \C_1$ 
such that  
\[
 G\circ F  = 1_{\C_1}, \ \ F\circ G = 1_{\C_2}\ . 
\] 
Whenever  
\[
 G\circ F \sim 1_{\C_1}, \ \ \ F\circ G \sim 1_{\C_2}\ , 
\]
then $F$ is said to be an  \emph{equivalence}. 
If $F$ is an isomorphism (equivalence), then  
$\C_1$ and $\C_2$  are said to be \emph{isomorphic}
(\emph{equivalent}). It turns out that $F$ is equivalence if, and only if, 
it is a full, faithful and dense.
%**************************************
\paragraph{$\Crm^*-$categories :}
A category $\mathcal{C}$ is said to be a $\Crm^*$-category if
the set of the arrows $(z,z_1)$ between two objects $z,z_1$ is a complex
Banach space and the composition between arrows is bilinear 
a bilinear map $t,s\to t\cdot s$ with 
$\norm{t\cdot s}\leq\norm{t}\cdot \norm{s}$; 
there should be an adjoint, that is an involutive contravariant functor $*$
acting as the  identity on the objects and  the norm should satisfy the
$\mathrm{C}^*$-property,
namely  $\norm{r^{*} \cdot r} \ = \ \norm{r}^2$ for each
$r\in(z,z_1)$. Notice, that if $\mathcal{C}$ is a $\Crm^*$-category
then $(z,z)$ is a $\Crm^*$-algebra for
each $z$. \\[3pt]
\indent Assume that $\mathcal{C}$ is  a $\Crm^*$-category. An arrow
$v\in(z,z_1)$ is said to be  an isometry if $v^* \cdot v=1_z$;
a unitary, if it is an isometry and $v\cdot v^*=1_{z_1}$.
The property of  admitting  a unitary arrow, defines an equivalence
relation on the set of the objects of the category. We denote
by the symbol $[z]$ the unitary equivalence class of the object
$z$. An object $z$ is said to be irreducible if
$(z,z)=\mathbb{C}\cdot 1_z$. $\mathcal{C}$ is said to be closed
under  subobjects if for each orthogonal projection
$e\in(z,z)$, $e\ne 0$ there exists an isometry
$v\in(z_1,z)$ such that $v\cdot v^{*}  =  e$.
$\mathcal{C}$ is said to be closed under  direct sums,
if given $z_i \ i=1,2 $ there exists an
object $z$ and two isometries $w_i\in(z_i,z)$  such that
$w_1  \cdot w^{*}_1  +  w_2 \cdot w^{*}_2 \ = \ 1_z$. \\
\indent Given two $\Crm^*-$categories $\C_1$ and $\C_2$ 
a \emph{$^*-$functor} $F:\C_1\rightarrow \C_2$ is a functor which commute 
with the adjoint, and preserve the linear structure of arrows: 
\[
 F(\alpha\cdot t +\be\cdot s) = \alpha\cdot F(t) +\be\cdot F(s)
\] 
$\C_1$ and $\C_2$ are said to be  \emph{equivalent} 
(\emph{isomorphic}) if there exists a $^*-$functor 
$F:\C_1\rightarrow \C_2$ which is an equivalence (isomorphism). 
A natural isomorphism $u:F\rightarrow G$, with 
$F,G:\C_1\rightarrow \C_2$ $^*-$functors, is said to be 
\emph{natural unitary} whenever $u(z)$ is
a unitary arrow of $(F(z),G(z))$  for any object $z$ of $\C_1$. 
%******************************************** 
\paragraph{Symmetric tensor $\Crm^*-$categories :}
 A \emph{tensor $\Crm^*$-category} $\C$ is a
$\mathrm{C}^*$-category  which is equipped with a 
tensor product $\otimes$. This means that to each  pair 
of objects $z,z_1$ there is product  object $z\otimes z_1$,
and $\C$ has a unit object $\io$ such that 
$z\otimes \io= z = \io\otimes z$. Given two arrows 
$t\in (z,z_1)$ and $s\in(z_2,z_3)$  there is an arrow
$t\otimes s\in(z\otimes z_2, z_1\otimes z_3)$. The mapping 
$t,s\rightarrow t\otimes s$ is associative and bilinear, and 
\[
 1_\io\otimes t = t = t\otimes 1_\io, \ \ \ \ (t\otimes s)^*= t^*\otimes s^*,
\]
and the interchange law 
\[
(t\otimes s)\cdot (t_1\otimes s_1)  =  t\cdot t_1 \otimes s\cdot s_1
\]
holds whenever the right hand side is defined. \\[5pt]
\indent  From now on we will consider only tensor 
$\Crm^*$-categories,  with irreducible unit $\io$,   which are closed  under 
direct sums and subobjects.\\[5pt]
%*******************************************
%*******************************************
\indent $\C$ is said to be a \emph{symmetric} 
if it has a {\em permutation symmetry}. This means that there is 
a map  $\eps:\mathcal{C}\ni z_1,z_2\longrightarrow\eps(z_1,z_2)\in
(z_1\otimes z_2, z_2\otimes z_1)$
satisfying the relations:
\[
\begin{array}{rlrl}
(i)   & \eps(z_3,z_4)\cdot t\otimes s \ = \ s\otimes t\cdot\eps(z_1,z_2)\\
(ii)  &   \eps(z_1,z_2)^* \ = \ \eps(z_2,z_1)  \\
(iii) &  \eps(z_1,z_2\otimes z) \ = \ 1_{z_2}\otimes\eps(z_1,z)\cdot
                     \eps(z_1,z_2)\otimes 1_{z}  \qquad\qquad\qquad\qquad \\
(iv) &   \eps(z_1,z_2)\cdot\eps(z_2,z_1) \ = \ 1_{z_2\otimes z_1}
\end{array}
\]
where $t\in(z_2,z_4), s\in(z_1,z_3)$.
By $ii)- iv)$ it follows that
$\eps(z,\io)=\eps(\io,z) = 1_z$ for any $z$. \\[5pt]
%********************************************
%********************************************
\indent A {\em left inverse} of an object $z$ of $\C$ 
is a set of nonzero linear maps
$\phi^z  =  \{ \phi^z_{z_1,z_2}:
 (z\otimes z_1, z\otimes z_2)\longrightarrow(z_1,z_2) \}$ satisfying
\[
\begin{array}{rl}
(i) &  \phi^z_{z_3,z_4}(1_z\otimes t \cdot r \cdot 1_z\otimes s^*) =
  t\cdot  \phi^z_{z_1,z_2}(r)\cdot s^*, \qquad \qquad \qquad \\
(ii) &   \phi^z_{z_1\otimes z_3, z_2\otimes z_3} (r\otimes 1_{z_3})   =
   \phi^z_{z_1,z_2}(r) \otimes 1_{z_3},\\
(iii) &   \phi^z_{z_1,z_1}(s_1^*\cdot s_1) \geq 0,\\
(iv)  &  \phi^z_{\io,\io}(1_z) =\mathbbm{1}, 
\end{array}
\]
where $s\in(z_1,z_3)$, $t\in(z_2,z_4)$, $r\in(z\otimes z_1, z\otimes z_2)$ and 
$s_1\in (z\otimes z_1, z\otimes z_1)$.  $\C$ is said to 
\emph{have left inverses} if any object of $\C$ has left inverses.\\[5pt]
% \indent From now on we assume that 
% $\mathcal{C}$ has a symmetry $\eps$ and that any object of $\mathcal{C}$
% has left inverses. 
\indent Consider  two symmetric tensor $\Crm^*$--categories $\C_1$, $\C_2$.
Let $(\otimes_1,\eps_1)$ and $(\otimes_2,\eps_2)$ be the corresponding 
tensor products and permutation symmetries. A $^*-$functor $F: \C_1\rightarrow \C_2$ is said to be a 
\emph{symmetric tensor $^*-$functor}, if 
for any pair of objects $z_1,z_2$ of $\C_1$ and for any pair of arrows 
of $\C_1$, we have that  
\[
\begin{array}{lcll}
 F ( z\otimes_1 z_1)  & =  &F(z)\otimes_2 F(z_1)\ ,\\
 F ( t\otimes_1 s)    & =  & F(t)\otimes_2 (F(s)\ ,\\
 F(\eps_1(z,z_1))  & = & \eps_2(F(z), F(z_1)) \ . 
\end{array}
\]
% Assume that $F$ is a full and faithful symmetric tensor $^*-$functor.
% Let $r\in(\io,\con{z}\otimes_1 z)$ and  
% $\con{r}\in(\io, z\otimes_1\con{z})$ solve the conjugate equations
% with respect to $z$ and $\con{z}$. Then it turns out 
% that the pair $F(r)\in(\io,F(\con{z})\otimes_2 F(z))$
% and  $F(\con{r})\in(\io, F(z)\otimes_2F(\con{z}))$ 
% solve the conjugate equations in $\C_2$ with respect to $F(z)$ and 
% $F(\con{z})$. In particular,  $z$ is irreducible, if and only,   
% $F(z)$ is irreducible and they have same statistics 
% of $z$.\\ 
Two symmetric tensor $\Crm^*$-category $\C_1$ and $\C_2$ 
are said to be \emph{equivalent} (\emph{isomorphic}) 
if there exists a  symmetric tensor $^*-$functor
$F:\C_1\rightarrow\C_2$ which is an equivalence (isomorphism). 
A \emph{tensor natural transformation} 
$u:F\rightarrow G$ between two symmetric tensor $^*-$functors 
$G,F:\C_1\rightarrow \C_2$ is a natural transformation such that 
$u(z\otimes_1 z_1) = u(z)\otimes_2 u(z_1)$. 
It will be said to be a \emph{tensor natural isomorphism} 
(\emph{tensor natural unitary}) if it is a natural isomorphism 
(a natural unitary).
%*********************************************************
\paragraph{Statistics and conjugation :}
Let $\C$ be a symmetric tensor $\Crm^*-$category with left inverses. 
An object $z$ of $\mathcal{C}$ is said to have 
\emph{finite} statistics if it admits a \emph{standard left inverse}, 
that is a left inverse $\phi^z$ satisfying the relation   
\[
\phi^z_{z,z}(\eps(z,z))\cdot \phi^z_{z,z}(\eps(z,z)) = c\cdot 1_z 
 \mbox{ with } c>0
\]
The full subcategory $\mathcal{C}_\f$ of $\mathcal{C}$ whose objects 
\textit{have finite statistics}, is closed under direct sum, subobjects,
tensor products, and equivalence. Any object of $\mathcal{C}_\f$ is 
direct sums of irreducible objects. Given an irreducible 
object $z$ of $\C_\f$ and a  left inverse $\phi^z$ of $z$, then 
\[
\phi^z_{z,z}(\eps(z,z)) = \la(z)\cdot 1_z \ . 
\]
The number $\la(z)$ is an invariant of the  equivalence class of $z$, called 
the \emph{statistics parameter}.  It is the product of two invariants: 
\[
\la(z) = \chi(z)\cdot d(z)^{-1} \ \mbox{ where } \ \chi(z)\in\{1,-1\}, \ \ 
  d(z)\in\mathbb{N}
\] 
The possible statistics of $z$ are classified by the 
\emph{statistical phase} $\chi(z)$ distinguishing para-Bose $(1)$ 
and para-Fermi $(-1)$ statistics and by the \emph{statistical dimension}
$d(z)$ giving the order of the parastatistics. Ordinary 
Bose and Fermi statistics correspond to $d(z)=1$.\\[5pt]
\indent An object $z$ of $\C$ has \emph{conjugates} if there exists 
an object $\con{z}$ and a pair of arrows $r\in(\io,\con{z}\otimes z)$, 
$\con{r}\in(\io, z\otimes\con{z})$  satisfying the 
\emph{conjugate equations} 
\[
\con{r}^*\otimes 1_z\cdot 1_z\otimes r = 1_z, \ \ 
r^*\otimes 1_{\con{z}}\cdot 1_{\con{z}}\otimes \con{r} = 1_{\con{z}}.
\]
Conjugation is a property stable under subobjects, direct sums, tensor
products and, furthermore, it is stable under equivalence.  
It turns out that 
\[
 \mbox{if } z \mbox{ has conjugates, then } z \mbox{ has finite statistics.}
\] 
Consider two symmetric tensor $\Crm^*-$categories $\C_1$ and $\C_2$,
and let $F:\C_1\rightarrow \C_2$  be a full and faithful 
symmetric tensor $^*-$functor. Let $r\in(\io,\con{z}\otimes_1 z)$ and  
$\con{r}\in(\io, z\otimes_1\con{z})$ solve the conjugate equations
with respect to $z$ and $\con{z}$. Then it turns out 
that the pair $F(r)\in(\io,F(\con{z})\otimes_2 F(z))$
and  $F(\con{r})\in(\io, F(z)\otimes_2F(\con{z}))$ 
solve the conjugate equations in $\C_2$ with respect to $F(z)$ and 
$F(\con{z})$. In particular,  $z$ is irreducible, if and only,   
$F(z)$ is irreducible and has  statistics 
as $z$.\\[5pt]
\indent Finally, a symmetric tensor $\Crm^*-$category 
is said  \emph{to have conjugates} if \emph{any} object of the category 
has conjugates. Note that given a symmetric tensor 
$\Crm^*-$category $\C$, let  $\C_1$ the full subcategory of $\C$ 
whose objects have conjugates. Then $\C_1$ has conjugates. 
%****************************************************************
%****************************************************************
%****************************************************************
\paragraph{Doplicher-Roberts duality  theorem :} 
We recall some basic facts on the Doplicher-Roberts duality theorem 
of compact groups \cite{DR1} (DR-theorem) focusing 
on its functorial properties.\\ 
\indent Denote by $\overline{\Sym}$ the full subcategory 
of $\Sym$ (see Section \ref{Cbb}) whose objects 
are symmetric tensor $\Crm^*-$categories having conjugates.
The main technical result of the DR-theorem  is the embedding theorem: 
Any $\C\in\overline{\Sym}$ admits an embedding into a category 
of finite dimensional Hilbert spaces. To be precise, 
an \emph{embedding} of $\C$  is a pair $(H,\Hil)$, where 
$\Hil\in\overline{\Sym}$ is a category of finite dimensional Hilbert spaces,
while $H: \C  \rightarrow \Hil$ is a symmetric tensor $^*-$functor which 
is full and faithful. Given an embedding $(H,\Hil)$ of $\C$,  the set 
\[
 \mathrm{End}_\otimes(H)\defi\{ tensor \ natural \ unitaries  \ 
 u: H\rightarrow H  \}
\]
equipped with the composition $(u_1\circ u)(z)\defi u_1(z)\cdot u(z)$ is, 
with a suitable topology, a compact group. It turns out that 
$\C$ is the abstract dual of $\mathrm{End}_\otimes(H)$.  
This group is uniquely associated with  the category $\C$ only up to 
isomorphism. This is  because the embedding of the category $\C$ 
is not uniquely determined, so the compact group associated with 
$\C$ depends on the choice of the embedding. However for any other embedding
$(H',\Hil')$ of $\C$ there exists a tensor natural unitary equivalence
\begin{equation}
\label{X:1}
w:H\to H'
\end{equation}
which associates to any 
object $z\in\C$ a unitary operator $w(z)$  
from the Hilbert space $H(z)$ onto the Hilbert space  $H'(z)$. $w$ 
preserves the tensor products 
\[
w(z\otimes z_1) = w(z)\otimes' w(z_1) \ , \qquad z,z_1\in\C \ , 
\]
$\otimes'$ is the tensor product of $\Hil'$, and 
$w(z_1)\cdot H(t) = H'(t)\cdot w(z)$ for any $z,z_1\in\C$ and 
$t\in(z,z_1)$. It turns out that  the mapping 
\[
\mathrm{End}_\otimes(H)\ni u\rightarrow w\circ u \circ w^*\in 
\mathrm{End}_\otimes(H')
\]
is a group isomorphism, where 
\[
(w\circ u \circ w^*)(z)\defi w(z)\cdot u(z) \cdot w^*(z), \qquad z\in \C.
\]
We now want to see the functorial properties of the DR-theorem. 
Given two categories $\C_1,\C$, let 
$(H,\Hil)$ and $(H_1,\Hil)$ be a choice of embeddings. 
Consider a full and faithful, symmetric tensor $^*-$functor
$F_1:\C_1\rightarrow\C$. Then the pair 
$(H\circ F_1,\Hil)$ provides another embedding of $\C_1$. Let us denote by 
$w_{F_1}$ the tensor natural unitary equivalence 
$w_{F_1}: H_1\rightarrow H\circ F_1$ associated, by \ref{X:1}, 
with  the embeddings  $(H_1,\Hil)$ and $(H\circ F_1,\Hil)$. 
Given $u\in \mathrm{End}_\otimes(H)$,  define
\[
\alpha_{F_1}(u)(z_1)\defi w^*_{F_1}(z_1)  
 \circ  u(F(z_1))\circ w_{F_1}(z_1), \qquad z_1\in \C_1  
\]
It turns out that 
\[
\alpha_{F_1}: \mathrm{End}_\otimes(H)\rightarrow \mathrm{End}_\otimes(H_1) 
\]  
is a  group morphism.
Consider now a second category $\C_2$ and let $(H_2,\Hil_2)$ 
be an embedding of this category. If 
$F_2: \C_2\rightarrow \C_1$ is a functor, then by proceeding as above 
there is tensor natural unitary equivalence
$w_{F_1}: H_2\rightarrow H_1\circ F_2$ and a  group morphism 
\[
\alpha_{F_2}: \mathrm{End}_\otimes(H_1)\rightarrow \mathrm{End}_\otimes(H_2) 
\] 
defined, for any $u_1\in \mathrm{End}_\otimes(H_1)$,  as
\[
\alpha_{F_2}(u_1)(z_2) =  w^*_{F_2}(z_2)  
 \cdot  u_1(F_2(z_2))\cdot w_{F_2}(z_2) \ , \qquad  z_2\in \C_2
\] 
Now, observe that 
$\alpha_{F_2}\circ \alpha_{F_1}:
\mathrm{End}_\otimes(H)\rightarrow \mathrm{End}_\otimes(H_2)$. 
In particular  given $u\in \mathrm{End}_\otimes(H)$ and $z_2\in\C_2$
we have that 
\begin{align*}
(\alpha_{F_2}\circ \alpha_{F_1})&(u)(z_2)  = \\ 
& = w^*_{F_2}(z_2)\cdot \alpha_{F_1}(u)(F_2(z_2))\cdot w_{F_2}(z_2) \\
&  = w^*_{F_2}(z_2)\cdot w^*_{F_1}(F_2(z_2))\cdot 
u(F_1F_2(z_2))\cdot w_{F_1}(F_2(z_2))\cdot w_{F_2}(z_2).  
\end{align*}
Consider the tensor natural unitary equivalence 
$w_{F_1F_2}: H_2\rightarrow  H\circ F_1\circ F_2$, and the corresponding 
 group morphism 
$\alpha_{F_1F_2}: \mathrm{End}_\otimes(H)
\rightarrow \mathrm{End}_\otimes(H_2)$ 
defined, for any $u\in \mathrm{End}_\otimes(H)$,  as
\[
\alpha_{F_1F_2}(u)(z_2) =  
w_{F_1F_2}^*(z_2)\cdot  u(F_1F_2(z_2))\cdot w_{F_1F_2}(z_2) \ ,\qquad 
 z_2\in\C_2 \ . 
\]
Now, define 
\[
w_{F_1,F_2;F_1F_2}(z_2)\defi w_{F_2}^*(z_2)\cdot 
w_{F_1}^*(F_2(z_2))\cdot w_{F_1F_2}(z_2)\ , \qquad z_2\in\C_2\ . 
\]
Then it turns out that $w_{F_1,F_2;F_1F_2}$ is an element
of the group $\mathrm{End}_{\otimes}(H_2)$ and that 
\[
(\alpha_{F_2}\circ \alpha_{F_1})(u)(z_2)  =  
\mathrm{ad}_{w_{F_1,F_2;F_1F_2}(z_2)}\big(\alpha_{F_1F_2}(u)(z_2)\big) \ .  
\]
%%%%%%%%%%%%%%%%%%%%%

%****************************************************************
%****************************************************************
%****************************************************************

%****************************************************************
%***************** Biblbiografia ******************************
%****************************************************************

\end{document}